\documentclass[preprint, authoryear]{elsarticle} 
\usepackage[a4paper, total={6.5in, 9in}]{geometry}
\usepackage{lineno,hyperref}
\modulolinenumbers[5]

\usepackage{amsmath,amssymb,xcolor}
\usepackage{mathrsfs}
\makeatletter
\def\intkern@{\mkern-8mu\mathchoice{\mkern-4mu}{}{}{}}
\makeatother
\usepackage{graphicx}
\usepackage{dcolumn}
\usepackage{bm}
\usepackage{pdflscape}
\usepackage{afterpage}
\newcommand\michaelarraysettings{
	\setlength\arraycolsep{2pt} 
	\renewcommand\arraystretch{0} 
}

\journal{Journal of the Mechanics and Physics of Solids}

\begin{document}

\begin{frontmatter}

\title{Chain breaking in the statistical mechanical\\ constitutive theory of polymer networks}

\author{Michael R. Buche}

\author{Meredith N. Silberstein\corref{mycorrespondingauthor}}
\cortext[mycorrespondingauthor]{Corresponding author}
\ead{ms2682@cornell.edu}

\address{Theoretical and Applied Mechanics,\\ Sibley School of Mechanical and Aerospace Engineering,\\ Cornell University, Ithaca USA}

\begin{abstract}
	
	Elastomers are used in a wide range of applications because of their large strain to failure, low density, and tailorable stiffness and toughness.
	The mechanical behavior of elastomers derives mainly from the entropic elasticity of the underlying network of polymer chains.
	Elastomers under large deformation experience bonds breaking within the backbone chains that constitute the polymer network.
	This breaking of chains damages the network, can lead to material failure, and can be utilized as an energy dissipation mechanism.
	In the case of reversible bonds, broken chains may reform and heal the damage in the network.
	If the reversible bonds are dynamic, chains constantly break and reform and create a transient network.
	A fundamental constitutive theory is developed to model the mechanics of these polymer networks.
	A statistical mechanical derivation is conducted to yield a framework that takes in an arbitrary single-chain model (a Hamiltonian) and outputs the following: the single-chain mechanical response, the breaking and reforming kinetics, the equilibrium distribution of chains in the network, and the partial differential equations governing the deformation-coupled network evolution.
	This statistical mechanical framework is then brought into the continuum scale by using macroscopic thermodynamic constitutive theory to obtain a constitutive relation for the Cauchy stress.
	The potential-supplemented freely jointed chain ($u$FJC) model is introduced, and a parametric study of its mechanical response and breaking kinetics is provided.
	This single-chain model is then implemented within the constitutive framework, which we specialize and apply in two exemplary cases: the mechanical response and irreversible breakdown of a multinetwork elastomer, and the mechanical response of a dual crosslink gel.
	After providing a parametric study of the general constitutive model, we apply it to a hydrogel with reversible metal-coordination crosslinks.
	In several cases, we find that the breakdown of the network causes secondary physical mechanisms to become important and inhibit the accuracy of our model. 
	We then discuss these mechanisms and indicate how our existing framework can be adjusted to incorporate them in the future.

\end{abstract}

\begin{keyword}
statistical mechanics \sep constitutive theory \sep polymer networks \sep reversible cross-links
\end{keyword}

\end{frontmatter}


\section{\label{sec1}Introduction}

Bulk elastomer materials often consist of many single polymer chains crosslinked together to form a network of chains.
Elastomers tend to be soft, elastic, and highly stretchable, due to the entropic elasticity of the chains above the glass transition temperature \citep{Treloar1949physics}.
From large tires and small seals to soft robotics, elastomers are utilized in a wide variety of applications due to their resilience.
However, as the elastomer network is deformed more extensively, bonds begin to stretch and then chains begin to break.
While chain breaking can result in material failure for simpler elastomers, many advanced elastomers are cleverly designed to take advantage of chain breaking.
Firstly, elastomers may be strengthened, toughened, and made more stretchable through the incorporation of one or more sacrificial networks that begin to break down irreversibly when stretched, dissipating energy while secondary networks maintain the integrity of the material.
The sacrificial network is often embrittled as a swollen gel \citep{gong2003double}, but could also be pre-stretched using the secondary networks \citep{ducrot2016characterizing} or even designed without need for pre-stretching \citep{nakajima2019tough}.
The breaking in the sacrificial network may involve some additional functionality, such as mechanoluminescence \citep{ducrot2014toughening} and recently, chain-lengthening \citep{wangtoughening}.
Secondly, elastomers may utilize a range of reversible bonds in order to allow chains to reform after they have been broken.
This reversible breaking allows similar properties as the irreversibly-breaking cases, such as high stretchability and toughness, while also allowing new properties such as self-healing \citep{li2016highly}.
Alginate-based gels contain ionic crosslinking that breaks reversibly as the polymer is deformed, increasing toughness while enabling both self-healing and shape-memory \citep{sun2012highly}.
Metal-ligand interactions, which are inherently tunable \citep{khare2021transition}, when used as crosslinks provide a precise method to control polymer mechanical properties via the simple addition of neutral ligands \citep{vidavsky2020tuning}.
Dynamic reversible bonds may also be utilized: polymers with associative bond exchange reactions like vitrimers behave as an elastic solid at low temperatures while flowing more similarly to a viscous fluid at high temperatures, all the while maintaining the integrity of the network \citep{montarnal2011silica}.
Some of these covalent adaptable networks use light as a stimulus in order to trigger the dynamic bonds to permanently alter the material shape \citep{kloxin2013covalent}.
Utilizing a combination of interactions is also useful, such as the combination of permanent covalent bonds and transient physical bonds in dual-crosslink gels \citep{narita2013viscoelastic}.
Overall, the mechanical properties of these materials tend to be highly nonlinear, rate-dependent, and sensitive to changes in their chemistry.
Therefore, a truly physical constitutive model that accounts for the complexities of chains breaking in a network is desirable to maximize both predictive power and fundamental understanding for the wide range of available chemistries and combinations.

There are a variety of physically-based constitutive models for polymer networks that incorporate chain breaking, frequently using the freely-jointed chain (FJC) single-chain statistical mechanical model \citep{rubinstein2003polymer}.
A portion of these models are targeted towards the mechanical response of permanently-crosslinked elastomers, where chains or crosslinks are considered to break suddenly and irreversibly \citep{mao2017rupture,mao2018fracture}.
This approach has been successfully applied when modeling the irreversible damage or fracture of polymer networks \citep{talamini2018progressive,tehrani2017effect,li2020variational}.
Additionally, irreversible breaking has been incorporated into many models for multinetwork elastomers and gels \citep{lavoie2016rate,bacca2017model,morovati2019micro,lavoie2019continuum,zhong2020constitutive}, sometimes addressing a particular phenomenon such as necking instability \citep{zhao2012theory,vernerey2018statistical,morovati2020necking}.
Another portion of these physically-based constitutive models tends to be specialized for transient networks enabled by highly dynamic bonds.
Transient network theory is typically attributed to \citet{tanaka1992viscoelastic,tanaka1992viscoelasticALL3}, which is built upon foundational work from the 1940s to the 1990s \citep{green1946new,flory1960elasticity,thomas1966limitations,fricker1973theory}.
Recent development has been driven by Vernerey et al. (\citeyear{vernerey2017statistically,vernerey2018transient}), and has lead to successful application in fracture scenarios \citep{brighenti2017rate,shen2020rate}.
Other constitutive models for polymers with dynamic bonds combine physically-based insights with continuum-level constitutive laws, such those for the mechanics of dual-crosslink gels and of networks with temperature-sensitive dynamic covalent bonds \citep{meng2016stress,yu2018mechanics,lin2020constitutive,   hui2012constitutive,long2014time,guo2016mechanics,lu2020pseudo,   long2013modeling,long2014mechanics,sun2016thermomechanics}.

Although these existing physically-based constitutive models perform well for a range of materials, any one of them lacks widespread applicability.
In this manuscript we present a statistical mechanical derivation that can bridge these models, where these models are special cases of the general model.
This derivation will yield, from an arbitrary single-chain model Hamiltonian, (1) the single-chain mechanical response, (2) the equilibrium distribution of chains in the network, and (3) the mechanically-dependent kinetics of chain breaking and reforming.
While the first two connections have previously been established \citep{Buchestatistical2020}, the force-dependent kinetics have not yet been directly connected to the statistical mechanics of the single-chain model.
With limited additional assumptions, this statistical mechanical foundation will then be used to formulate macroscopic constitutive relations entirely informed by an arbitrary single-chain model.
This meticulous procedure carrying the underlying statistical mechanics through to the macroscale has many inherent benefits, such as consistency between the equilibrium configuration obtained by statistical thermodynamics and that obtained macroscopically, and the automatic satisfaction of the second law of thermodynamics.

\newpage

This manuscript is organized as follows: 
In Sec.~\ref{statmechsubsection}, beginning from the fundamentals of nonequilibrium statistical mechanics, we obtain evolution equations for the probability of finding an intact chain at a certain end-to-end vector within the network as a function of time. 
This derivation results in the single-chain mechanical behavior, equilibrium distributions of chains in the network, and chemical kinetics function of chain breaking/reforming all in terms of the single-chain partition functions.
In Sec.~\ref{macrosubsec}, our statistical theory is brought into the macroscale through the formulation of the Helmholtz free energy of the incompressible network.
After prescribing an affine deformation, a second-law analysis then results in the constitutive relation for the Cauchy stress entirely in terms of the intact chain distribution and single-chain mechanical response, where the residual inequality is shown to be arbitrarily satisfied.
With the general theory complete, in Sec.~\ref{singlechainimplsec} we introduce and implement the $u$FJC single-chain model: the freely jointed chain (FJC) model supplemented to have stiff, but flexible links with some potential energy $u$.
We utilize the Morse potential for $u$ and study various single-chain functions over a range of parameters.
Additionally, we present an original exact solution for the evolving intact chain probability distribution.
In Sec.~\ref{MARCOPOLO}, we consider several special cases from the limiting behavior of our model and apply them to exemplary polymers from the literature.
We then study the general behavior of the model, drawing conclusions in comparison to these simpler special cases and examining the results over a range of single-chain parameters.
Afterward we apply the general model to another polymers from the literature.
Finally, we discuss the successes and shortcomings of our model and propose improvements for future work.
We have implemented our model in \texttt{Python} and made it available on \texttt{github} to facilitate both adoption and adaption by interested readers \citep{Buchecodeforpaper}.

\section{\label{sec2}General theory}

\subsection{Statistical mechanics\label{statmechsubsection}}

We consider a classical, canonical statistical mechanical ensemble of noninteracting polymer chains that may break/reform via multiple reaction pathways. 
Beginning from the general nonequilibrium formalism of \citet{zwanzig2001nonequilibrium}, we derive a general evolution law for the probability distribution of intact chains at a certain end-to-end vector. 
We then apply the assumptions of transition state theory to obtain a simpler evolution law that does not require knowledge of the phase space distribution function.
After making some assumptions about the behavior of broken chains, we obtain conservation requirements for all chains in the network, as well as an evolution law for the probability of each broken chain species.

\subsubsection{Phase space principles\label{psp_sssec}}

\begin{figure*}[t]
\begin{center}
\includegraphics{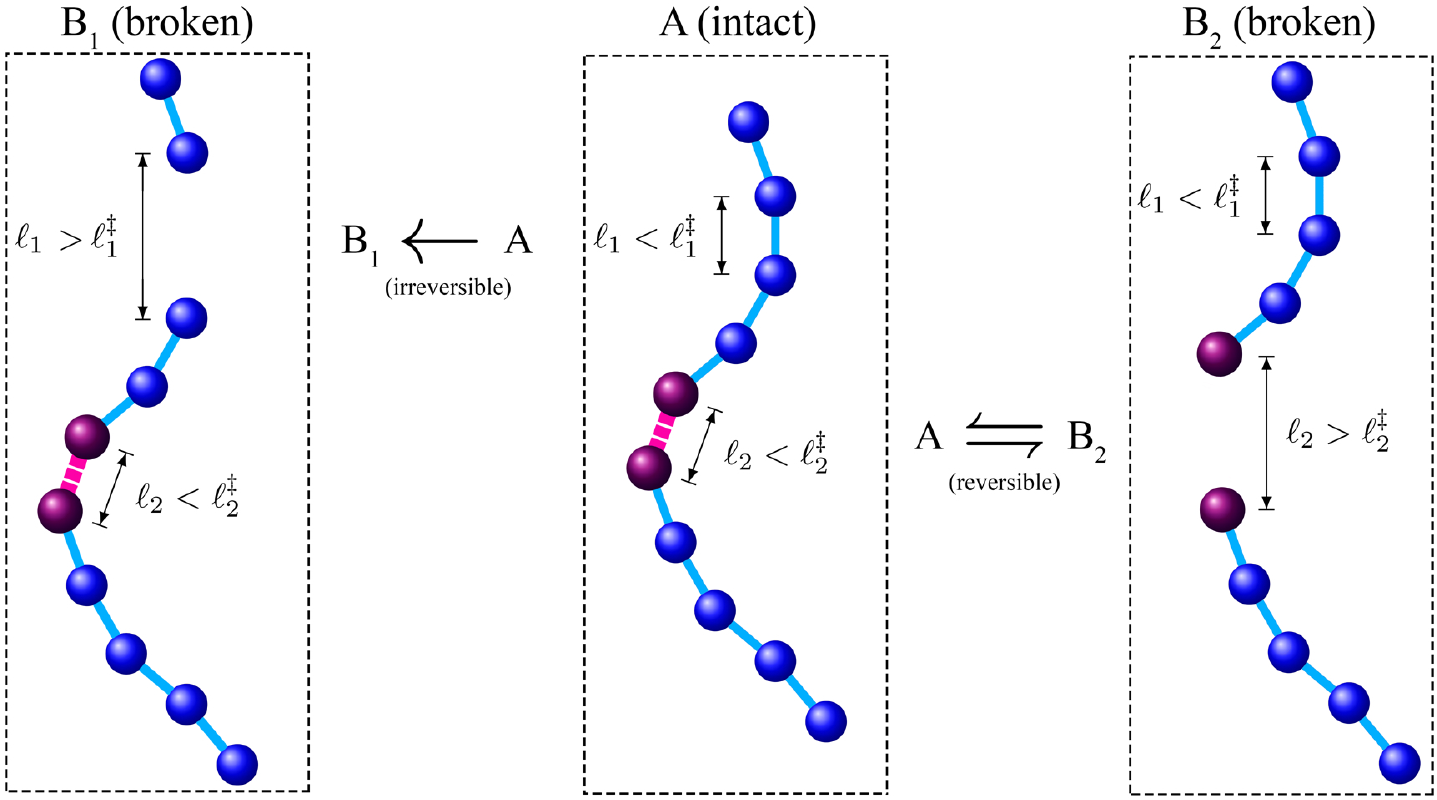}
\vspace{-5pt}
\end{center}
\caption{\label{fig1}
	Illustration of an intact chain (A) with various links of instantaneous length $\ell_i$ that may act as reaction coordinates. 
	In this illustration, the links may break irreversibly (see B$_1$) or reversibly (see B$_2$) when $\ell_i>\ell_i^\ddagger$.
}
\end{figure*}

In classical statistical mechanics \citep{mcq}, the phase space distribution function $f(\Gamma;t)$ provides the probability density at time $t$ that the system is in the state, denoted by $\Gamma$, with the atomic positions $\mathbf{q}$ and momenta $\mathbf{p}$. 
We may calculate the macroscopically observable value $\Phi(t)$ of some phase space function $\phi(\Gamma)$, which is the ensemble average of $\phi(\Gamma)$, or $\langle\phi\rangle$, as

\begin{equation}
	\Phi(t) = 
	\left\langle \phi \right\rangle \equiv 
	\int\cdots\int f(\Gamma;t) \phi (\Gamma) \,d\Gamma
	.
\end{equation}
In order to find $f(\Gamma;t)$, we integrate the evolution equation for $f(\Gamma;t)$, the Liouville equation

\begin{eqnarray}\label{Liouvilleeqn}
	\frac{\partial f}{\partial t} = 
	\left(-\mathscr{L}\right)f
	=
	\left(\frac{\partial H}{\partial\mathbf{q}}\cdot\frac{\partial}{\partial\mathbf{p}} - \frac{\partial H}{\partial\mathbf{p}}\cdot\frac{\partial}{\partial\mathbf{q}}\right)f
	,
\end{eqnarray}
with $\mathscr{L}$ being the Liouville operator. 
Since $f(\Gamma;t)$ does not evolve at equilibrium, $\mathscr{L}f^\mathrm{eq}=0$ for the equilibrium phase space distribution function $f^\mathrm{eq}(\Gamma)$.
This equilibrium distribution is the Boltzmann distribution

\begin{equation}\label{feq}
	f^\mathrm{eq}(\Gamma) = 
	\frac{e^{-\beta H(\Gamma)}}{\mathfrak{q}}
	,
\end{equation}
where $H(\Gamma)$ is the Hamiltonian of the system, $\beta=1/\mathfrak{b}T$ is the inverse temperature, $\mathfrak{b}$ is Boltzmann's constant, $T$ is the temperature, and 

\begin{equation}
	\mathfrak{q} = \int e^{-\beta H(\Gamma)} \,d\Gamma
\end{equation}
is the canonical partition function for the system. 
Note that we neglect the factors of Planck's constant $h$ that would nondimensionalize the partition functions, but this has no effect on our classically-obtained results.
The equilibrium ensemble average of some phase space function $\phi(\Gamma)$ is the time-independent average

\begin{equation}
	\Phi^\mathrm{eq} = 
	\left\langle \phi \right\rangle^\mathrm{eq} \equiv
	\frac{1}{\mathfrak{q}}\, \int\cdots\int e^{-\beta H(\Gamma)} \phi (\Gamma) \,d\Gamma
	.
\end{equation}
While solving for $f(\Gamma;t)$ would give us full knowledge of the system, it is impractical for our purposes since $\Gamma$ constitutes far too many state variables.
Fortunately, the macroscopic observables of interest in our case only require knowledge of a subset of the probability distribution of the phase space variables.
Specifically, we will only need the probability density distribution $P_\mathrm{A}(\boldsymbol{\xi};t)$ of intact chains with end-to-end vector $\boldsymbol{\xi}$ at time $t$ to calculate the macroscopic stress.
In order to track $P_\mathrm{A}(\boldsymbol{\xi};t)$, we will also need to consider the analogous distributions of broken chains $P_{\mathrm{B}_j}(\boldsymbol{\xi};t)$, where $j$ denotes that the chain has broken via the $j$th pathway; this is illustrated in Fig.~\ref{fig1}.
In the following section, we will write $P_\mathrm{A}(\boldsymbol{\xi};t)$ in terms of $f(\Gamma;t)$, and subsequently utilize this relation and the evolution equation for $f(\Gamma;t)$ to obtain the evolution equation for $P_\mathrm{A}(\boldsymbol{\xi};t)$.

\subsubsection{Evolution of intact chains\label{eic_sssec}}

The probability density distribution $P(\boldsymbol{\xi};t)$ of chains with end-to-end vector $\boldsymbol{\xi}$ at time $t$ is given by the ensemble average

\begin{eqnarray}\label{Pxi}
	P(\boldsymbol{\xi};t) = 
	\left\langle \delta^3\left[\mathbf{R}(\Gamma) - \boldsymbol{\xi}\right] \right\rangle
	,
\end{eqnarray}
where $\delta$ is the Dirac delta function and $\mathbf{R}(\Gamma)$ is the end-to-end vector of the chain as a function of phase space variables $\Gamma$.
This ensemble average effectively results in an integration of $f(\Gamma;t)$ over the portion of the phase space where the end-to-end vector of the chain is $\boldsymbol{\xi}$. 
A subset of the phase space variables $\Gamma$ are the reaction coordinates $\ell$ that determine whether a chain is intact or broken.
We consider $M$ of these reaction coordinates, where $\ell_i$ is then the $i$th reaction coordinate.
A chain is intact (A) if all $\ell_i<\ell^\ddagger_i $, while a chain is broken (B) if any $\ell_i>\ell^\ddagger_i$.
The reaction coordinates $\ell_i$ then create regions of the phase space where chains are intact ($\mathcal{A}$) or broken ($\mathcal{B}_i$).
The boundaries separating these two regions are the transition states with $\ell_i=\ell^\ddagger_i$.
Using Heaviside step functions $\Theta$, we may then write the probability density distribution $P_\mathrm{A}(\boldsymbol{\xi};t)$ that a polymer chain is both intact and at end-to-end vector $\boldsymbol{\xi}$ at time $t$ as

\begin{eqnarray}\label{PA}
	P_\mathrm{A}(\boldsymbol{\xi};t) = 
	\left\langle \delta^3\left[\mathbf{R}(\Gamma) - \boldsymbol{\xi}\right] \displaystyle\prod_{i=1}^{M} \Theta\left(\ell^\ddagger_i -\ell_i\right) \right\rangle
	.
\end{eqnarray}
We differentiate Eq.~\eqref{PA} with respect to time $t$ in order to produce the evolution equation for $P_\mathrm{A}(\boldsymbol{\xi};t)$.
Using Eq.~\eqref{Liouvilleeqn} and the properties of the Liouville operator $\mathscr{L}$ noted by \citet{zwanzig2001nonequilibrium}, we see that $\tfrac{d}{dt}\langle\phi\rangle=\langle\mathscr{L}\phi\rangle$ for some phase space function $\phi(\Gamma)$, so the evolution equation for $P_\mathrm{A}(\boldsymbol{\xi};t)$ is then

\begin{equation}\label{kjawefnkjn2r}
	\frac{\partial P_\mathrm{A}(\boldsymbol{\xi};t)}{\partial t} =
	\left\langle\mathscr{L}\left\{\delta^3\left[\mathbf{R}(\Gamma) - \boldsymbol{\xi}\right] \prod_{i=1}^{M} \Theta\left(\ell^\ddagger_i -\ell_i\right)\right\}\right\rangle
	.
\end{equation}
We will work from Eq.~\eqref{kjawefnkjn2r} to a readily usable form of the evolution equation for $P_\mathrm{A}(\boldsymbol{\xi};t)$, beginning with expansion using the product rule,
\begin{align}
	\frac{\partial P_\mathrm{A}(\boldsymbol{\xi};t)}{\partial t} = &
	\left\langle \mathscr{L}\left\{\delta^3\left[\mathbf{R}(\Gamma) - \boldsymbol{\xi}\right]\right\} \prod_{i=1}^{M} \Theta\left(\ell^\ddagger_i -\ell_i\right) \right\rangle 
	\nonumber \\ &+ 
	\sum_{j=1}^{M} \left\langle\delta^3\left[\mathbf{R}(\Gamma) - \boldsymbol{\xi}\right]\, \mathscr{L}\left\{\Theta\left(\ell^\ddagger_j -\ell_j\right)\right\} \prod_{\substack{i=1\\ i\neq j}}^{M} \Theta\left(\ell^\ddagger_i -\ell_i\right) \right\rangle
	.
\label{productrulez}
\end{align}
We consider the first term in Eq.~\eqref{productrulez} where the Liouville operator $\mathscr{L}$ acts on the delta function.
In effect, this term accounts for the evolution of $P_\mathrm{A}(\boldsymbol{\xi};t)$ due to $\dot{\boldsymbol{\xi}}_\mathrm{A}(\boldsymbol{\xi};t)$, the average rate of change of the end-to-end vector of an intact chain currently having end-to-end vector $\boldsymbol{\xi}$,

\begin{equation}
	\dot{\boldsymbol{\xi}}_\mathrm{A}(\boldsymbol{\xi};t) \equiv 
	\left\langle \dot{\mathbf{R}}(\Gamma)\, \delta^3\left[\mathbf{R}(\Gamma) - \boldsymbol{\xi}\right] \prod_{i=1}^{M} \Theta\left(\ell^\ddagger_i -\ell_i\right) \right\rangle
	.
\end{equation}
As detailed in general by \citet{zwanzig1961memory}, the first term in Eq.~\eqref{productrulez} can then be written as

\begin{eqnarray}\label{xidotevo}
	\left\langle \mathscr{L}\left\{\delta^3\left[\mathbf{R}(\Gamma) - \boldsymbol{\xi}\right]\right\} \prod_{i=1}^{M} \Theta\left(\ell^\ddagger_i -\ell_i\right) \right\rangle
	=
	-\frac{\partial}{\partial\boldsymbol{\xi}}\cdot\left[\dot{\boldsymbol{\xi}}_\mathrm{A}(\boldsymbol{\xi};t)P_\mathrm{A}(\boldsymbol{\xi};t)\right]
	.
\end{eqnarray}
We now consider the second set of terms in Eq.~\eqref{productrulez} where the Liouville operator $\mathscr{L}$ acts on each of the step functions $\Theta(\ell^\ddagger_j -\ell_j)$.
The derivative of the step function is the delta function, and $\mathscr{L}$ acting on the coordinates $\ell_j$ produces the velocities $p_j/m_j$, such that

\begin{equation}\label{lilguynosum}
	\mathscr{L}\left\{\Theta\left(\ell^\ddagger_j -\ell_j\right)\right\} = 
	-\frac{p_j}{m_j}\,\delta\left(\ell^\ddagger_j -\ell_j\right)
	.
\end{equation}
The summands of Eq.~\eqref{productrulez} are therefore the expected values of the velocity $-p_j/m_j$ along the $j$th reaction coordinate $\ell_j$ for a chain at end-to-end vector $\boldsymbol{\xi}$ in the $j$th transition state $\ell_j^\ddagger$.
These summands are understood as the evolution of $P_\mathrm{A}(\boldsymbol{\xi};t)$ due to flow across each transition state boundary.  
If we use step functions to split these flows into the forward $\mathcal{A}\to\mathcal{B}_j$ and reverse $\mathcal{B}_j\to\mathcal{A}$ reactions, with respective rates

\begin{align}
\label{dPABj}
	\mathcal{R}_j'(\boldsymbol{\xi};t) = &
	\left\langle \frac{p_j}{m_j}\,\Theta(p_j) \delta^3\left[\mathbf{R}(\Gamma) - \boldsymbol{\xi}\right] \delta\left(\ell^\ddagger_j -\ell_j\right) \prod_{\substack{i=1\\ i\neq j}}^{M} \Theta\left(\ell^\ddagger_i -\ell_i\right) \right\rangle
	,\\
	\mathcal{R}_j''(\boldsymbol{\xi};t) = &
	\left\langle -\frac{p_j}{m_j}\,\Theta(-p_j) \delta^3\left[\mathbf{R}(\Gamma) - \boldsymbol{\xi}\right] \delta\left(\ell^\ddagger_j -\ell_j\right) \prod_{\substack{i=1\\ i\neq j}}^{M} \Theta\left(\ell^\ddagger_i -\ell_i\right) \right\rangle
\label{dPBjA}
	,
\end{align}
then the final evolution law for the distribution of intact chains is

\begin{equation}
\label{PAevogeneral}
	\frac{\partial P_\mathrm{A}(\boldsymbol{\xi};t)}{\partial t} = 
	\sum_{j=1}^{M} \mathcal{R}_j''(\boldsymbol{\xi};t)
	- \sum_{j=1}^{M} \mathcal{R}_j'(\boldsymbol{\xi};t)
	- \frac{\partial}{\partial\boldsymbol{\xi}}\cdot\left[\dot{\boldsymbol{\xi}}_\mathrm{A}(\boldsymbol{\xi};t)P_\mathrm{A}(\boldsymbol{\xi};t)\right]
	.
\end{equation}
We have now obtained the general evolution law for the probability distribution of intact chains with a certain end-to-end vector $P_\mathrm{A}(\boldsymbol{\xi};t)$ in Eq.~\eqref{PAevogeneral}.
It remains an issue, however, that we require knowledge of the phase space distribution function $f(\Gamma;t)$, which is necessary to compute the ensemble averages for the reaction rates Eqs.~\eqref{dPABj} and \eqref{dPBjA}.
Consequently, we now utilize transition state theory in order to avoid computing these ensemble averages and therefore eliminate $f(\Gamma;t)$ from the evolution equation entirely.

\subsubsection{Transition state theory\label{tst_sssec}}

Our derivation so far is general for noninteracting chains, but we now make our first approximation.
Let the phase space function $f(\Gamma;t)$ maintain a local equilibrium in each species' region \citep{zwanzig2001nonequilibrium}, such that we may take $f(\Gamma;t)$ in the phase space region $\mathcal{A}$ to be approximately

\begin{equation}\label{floceqA}
	f(\Gamma; t) \approx
	\frac{P_\mathrm{A}(\boldsymbol{\xi};t)}{P_\mathrm{A}^\mathrm{eq}(\boldsymbol{\xi})}\, f^\mathrm{eq}(\Gamma)
	\quad\text{in region }\mathcal{A}
	,
\end{equation}
where $P_\mathrm{A}^\mathrm{eq}(\boldsymbol{\xi})$ is the equilibrium distribution of the end-to-end vectors of intact chains, and $f^\mathrm{eq}(\Gamma)=e^{-\beta H(\Gamma)}/\mathfrak{q}$ from Eq.~\eqref{feq}. 
This is representative of the fact that degrees of freedom not involved with breaking the chain (such as bond rotation) attain equilibrium much more quickly than those degrees of freedom associated with breaking the chain (i.e. bond breaking). 
Consequently, our transition state theory assumption now prevents us from considering cases where the timescales of bond breaking become close to those of intact chain dynamics.

The distribution of the end-to-end vectors of chains that have broken via the $j$th reaction pathway $P_{\mathrm{B}_j}(\boldsymbol{\xi};t)$ can be written by flipping the sign within the $j$th step function in Eq.~\eqref{PA},

\begin{equation}\label{PB}
	P_{\mathrm{B}_j}(\boldsymbol{\xi};t) = 
	\left\langle \delta^3\left[\mathbf{R}(\Gamma) - \boldsymbol{\xi}\right] \Theta\left(\ell_j - \ell^\ddagger_j\right) \displaystyle\prod_{\substack{i=1\\ i\neq j}}^{M} \Theta\left(\ell^\ddagger_i -\ell_i\right) \right\rangle
	.
\end{equation}
We may similarly approximate $f(\Gamma;t)$ in each broken chain phase space region as

\begin{equation}\label{floceqB}
	f(\Gamma; t) \approx
	\frac{P_{\mathrm{B}_j}(\boldsymbol{\xi};t)}{P_{\mathrm{B}_j}^\mathrm{eq}(\boldsymbol{\xi})}\, f^\mathrm{eq}(\Gamma)
	\quad\text{in each region }\mathcal{B}_j
	,
\end{equation}
where $P_{\mathrm{B}_j}^\mathrm{eq}(\boldsymbol{\xi})$ is the equilibrium distribution of the end-to-end vectors of chains that have broken via the $j$th reaction pathway. 
The equilibrium probabilities may be written using Eqs.~\eqref{PA} and \eqref{PB} in the equilibrium system, or by using the ratio of partition functions \citep{mcq,Buchestatistical2020} as

\begin{align}
\label{PAeqgen}
	P_\mathrm{A}^\mathrm{eq}(\boldsymbol{\xi}) = &
	\left\langle \delta^3\left[\mathbf{R}(\Gamma) - \boldsymbol{\xi}\right] \prod_{j=1}^{M}\Theta\left(\ell^\ddagger_j -\ell_j\right) \right\rangle^\mathrm{eq} = 
	\frac{\mathfrak{q}^*_\mathrm{A}(\boldsymbol{\xi})}{\mathfrak{q}}
	,\\
\label{PBjeqgen}
	P_{\mathrm{B}_j}^\mathrm{eq}(\boldsymbol{\xi}) = &
	\left\langle \delta^3\left[\mathbf{R}(\Gamma) - \boldsymbol{\xi}\right] \Theta\left(\ell_j - \ell^\ddagger_j\right) \displaystyle\prod_{\substack{i=1\\ i\neq j}}^{M} \Theta\left(\ell^\ddagger_i -\ell_i\right) \right\rangle^\mathrm{eq} = 
	\frac{\mathfrak{q}^*_{\mathrm{B}_j}(\boldsymbol{\xi})}{\mathfrak{q}}
	,
\end{align}
where the partition functions of an intact chain at end-to-end vector $\boldsymbol{\xi}$ and a chain that has broken via the $j$th reaction pathway at end-to-end vector $\boldsymbol{\xi}$ (the asterisk $*$ denotes the fixed $\boldsymbol{\xi}$) are given by

\begin{align}
\label{qA}
	\mathfrak{q}^*_\mathrm{A}(\boldsymbol{\xi}) = &
	\int\cdots\int e^{-\beta H(\Gamma)}\delta^3\left[\mathbf{R}(\Gamma) - \boldsymbol{\xi}\right] 
	\prod_{j=1}^{M}\Theta\left(\ell^\ddagger_j -\ell_j\right) \,d\Gamma 
	,
	\\
	\mathfrak{q}^*_{\mathrm{B}_j}(\boldsymbol{\xi}) = &
\label{qBj}
	\int\cdots\int e^{-\beta H(\Gamma)}\delta^3\left[\mathbf{R}(\Gamma) - \boldsymbol{\xi}\right] 
	\Theta\left(\ell_j - \ell^\ddagger_j\right) \displaystyle\prod_{\substack{i=1\\ i\neq j}}^{M} \Theta\left(\ell^\ddagger_i -\ell_i\right) \,d\Gamma 
	.
\end{align}
Similarly, the partition function of an intact chain at end-to-end vector $\boldsymbol{\xi}$ in the $j$th transition state is

\begin{equation}\label{qABjTS}
	\mathfrak{q}_{\ddagger_j}^* (\boldsymbol{\xi}) = 
	\int\cdots\int  e^{-\beta H(\Gamma)} \delta\left(p_j\right) \delta^3\left[\mathbf{R}(\Gamma) - \boldsymbol{\xi}\right]
	\delta\left(\ell^\ddagger_j -\ell_j\right) \displaystyle\prod_{\substack{i=1\\ i\neq j}}^{M} \Theta\left(\ell^\ddagger_i -\ell_i\right) \,d\Gamma 
	.
\end{equation}
Now, after utilizing Eqs.~\eqref{floceqA}--\eqref{qABjTS} (for detailed steps, see \ref{TSTmath}), the rates in Eqs.~\eqref{dPABj}--\eqref{dPBjA} become

\begin{align}
\label{Rjpkin}
	\mathcal{R}_j'(\boldsymbol{\xi};t) = & 
	k_j'(\boldsymbol{\xi})P_\mathrm{A}(\boldsymbol{\xi};t)
	,\\
\label{Rjppkin}
	\mathcal{R}_j''(\boldsymbol{\xi};t) = & 
	k_j''(\boldsymbol{\xi})P_{\mathrm{B}_j}(\boldsymbol{\xi};t)
	,
\end{align}
where the forward and reverse reaction rate coefficient functions are respectively given by

\begin{align}
\label{kBjA}
	k_j'(\boldsymbol{\xi}) = & 
	\frac{1}{\beta}\frac{\mathfrak{q}_{\ddagger_j}^* (\boldsymbol{\xi})}{\mathfrak{q}_\mathrm{A}^*(\boldsymbol{\xi})}
	,\\
\label{kABj}
	k_j''(\boldsymbol{\xi}) = & 
	\frac{1}{\beta}\frac{\mathfrak{q}_{\ddagger_j}^* (\boldsymbol{\xi})}{\mathfrak{q}_{\mathrm{B}_j}^*(\boldsymbol{\xi})}
	.
\end{align}
While extension-dependent rate coefficients have been previously considered for polymer networks \citep{green1946new,tanaka1992viscoelastic}, exact relations have not yet been discovered, leaving models to assume they are constant \citep{green1946new,vernerey2017statistically,brighenti2017rate,vernerey2018transient,guo2020mechanics}, or assume some other form \citep{tanaka1992viscoelasticALL3,lavoie2016rate,yu2018mechanics,shen2020rate,lin2020constitutive,lu2020pseudo,guo2021micromechanics} typically inspired by or in some way similar to the model of \citet{bell1978models}.
Eqs.~\eqref{kBjA} and \eqref{kABj} show that each forward and reverse reaction rate coefficient function is completely determined by the single-chain model via the partition functions.
These partition functions similarly determine the single-chain mechanical response and the equilibrium distribution of chain end-to-end vectors in the network \citep{Buchestatistical2020}.

After relating the ratio of the reaction rate coefficient functions to the equilibrium probabilities as

\begin{equation}\label{Kjeq}
	\frac{k_j''(\boldsymbol{\xi})}{k_j'(\boldsymbol{\xi})} = 
	\frac{\mathfrak{q}_\mathrm{A}^*(\boldsymbol{\xi})}{\mathfrak{q}_{\mathrm{B}_j}^*(\boldsymbol{\xi})} = 
	\frac{P_\mathrm{A}^\mathrm{eq}(\boldsymbol{\xi})}{P_{\mathrm{B}_j}^\mathrm{eq}(\boldsymbol{\xi})}
	,
\end{equation}
we finally rewrite Eq.~\eqref{PAevogeneral}, the evolution law for the probability distribution of intact chains as

\begin{equation}
\label{PAevokin}
	\frac{\partial P_\mathrm{A}(\boldsymbol{\xi};t)}{\partial t} = 
	-\sum_{j=1}^{M} k_j'(\boldsymbol{\xi}) \left[ P_\mathrm{A}(\boldsymbol{\xi};t) - 
	\frac{P_{\mathrm{B}_j}(\boldsymbol{\xi};t)}{P_{\mathrm{B}_j}^\mathrm{eq}(\boldsymbol{\xi})} \,P_\mathrm{A}^\mathrm{eq}(\boldsymbol{\xi})
	\right]
	- \frac{\partial}{\partial\boldsymbol{\xi}}\cdot\left[\dot{\boldsymbol{\xi}}_\mathrm{A}(\boldsymbol{\xi};t)P_\mathrm{A}(\boldsymbol{\xi};t)\right]
	.
\end{equation}
This evolution equation depends only upon the independent variables $\boldsymbol{\xi}$ and $t$.
We establish the evolution equation for the probability distribution of each species of broken chains $P_{\mathrm{B}_j}(\boldsymbol{\xi};t)$ in the following section.
Later in Sec.~\ref{constirelsubsubsec} we constitutively prescribe $\dot{\boldsymbol{\xi}}_\mathrm{A}(\boldsymbol{\xi};t)$ as a function of the deformation.

\subsubsection{Evolution of broken chains\label{disssubsubsec}}

In this section, we obtain simplified evolution laws for the probability density distribution of the $j$th broken chains $P_{\mathrm{B}_j}(\boldsymbol{\xi};t)$, and in the process further simplify Eq.~\eqref{PAevokin}.
To proceed, we first neglect the possibility that a chain may break via multiple pathways.
This is reasonable when all breaking pathways remain approximately inaccessible without considerable force acting on the chain, since broken chains will not support the force required to break again.
It is also reasonable when at most one breaking pathway is thermally accessible, such as the case with a chain with a single transient bond and many strong covalent bonds. 
This assumption inhibits our ability to model chains with many highly dynamic bonds, such as those within vitrimers.
Second, we neglect the possibility that broken chains of different reaction pathways may cross-reform, which prevents us from considering cases where groups of chains break and reform together resulting in altered contour lengths.
These two assumptions allow us to conclude that the total probability that a chain is either intact or is broken via a single reaction pathway is unity, yielding the conservation law

\begin{equation}\label{cons1}
	\iiint P_\mathrm{A}(\boldsymbol{\xi};t)\,d^3\boldsymbol{\xi} + \sum_{j=1}^M \iiint P_{\mathrm{B}_j}(\boldsymbol{\xi};t)\,d^3\boldsymbol{\xi} = 1
	.
\end{equation}
We evaluate Eq.~\eqref{cons1} at equilibrium and multiply it by the system partition function $\mathfrak{q}$, which with the equilibrium probabilities in Eqs.~\eqref{PAeqgen} and \eqref{PBjeqgen} then allows us to relate the partition functions as

\begin{equation}\label{qconserv}
	\mathfrak{q} = 
	\iiint \mathfrak{q}^*_\mathrm{A}(\boldsymbol{\xi}) \,d^3\boldsymbol{\xi} + \sum_{j=1}^M \iiint \mathfrak{q}^*_{\mathrm{B}_j}(\boldsymbol{\xi})\,d^3\boldsymbol{\xi} 
	\equiv
	\mathfrak{q}_{\mathrm{A}} + \sum_{j=1}^M \mathfrak{q}_{\mathrm{B}_j}
	.
\end{equation}
Once we specify a chain model, we can calculate $\mathfrak{q}^*_\mathrm{A}(\boldsymbol{\xi})$ and each $\mathfrak{q}^*_{\mathrm{B}_j}(\boldsymbol{\xi})$, and with Eq.~\eqref{qconserv} we may then calculate the equilibrium distributions in Eqs.~\eqref{PAeqgen} and \eqref{PBjeqgen}. 
Note that we have defined $\mathfrak{q}_{\mathrm{A}}$ and $\mathfrak{q}_{\mathrm{B}_j}$ in Eq.~\eqref{qconserv}, which are the partition functions of unconstrained intact chains and broken chains, respectively.
$\mathfrak{q}_{\mathrm{A}}$ and $\mathfrak{q}_{\mathrm{B}_j}$ are equivalently the integrals of $\mathfrak{q}^*_\mathrm{A}(\boldsymbol{\xi})$ and $\mathfrak{q}^*_{\mathrm{B}_j}(\boldsymbol{\xi})$ over all end-to-end vectors $\boldsymbol{\xi}$.
By the principal thermodynamic connection formula \citep{mcq} for the Helmholtz free energy $\psi = -\mathfrak{b}T\ln\mathfrak{q}$, we obtain 

\begin{equation}\label{qqexpyay}
	\Delta\Psi_{0_j} \equiv
	-\mathfrak{b}T \ln\left(\frac{ \mathfrak{q}_{\mathrm{B}_j} }{ \mathfrak{q}_{\mathrm{A}} }\right)
	,
\end{equation}
where $\Delta\Psi_{0_j}$ is then the net Helmholtz free energy change for the $j$th breaking reaction alone (only having to do with the reaction coordinate, not the rest of the chain).
We now approximate the two ends of any broken chain as effectively noninteracting, which allows the partition function $\mathfrak{q}^*_{\mathrm{B}_j}(\boldsymbol{\xi})$ to be constant in $\boldsymbol{\xi}$,

\begin{equation}\label{ewfrgdfgs3}
	\mathfrak{q}_{\mathrm{B}_j} \approx
	V_{\mathrm{B}_j} \mathfrak{q}^*_{\mathrm{B}_j}
	,
\end{equation}
where each $V_{\mathrm{B}_j}$ has units of volume.
Using Eq.~\eqref{qqexpyay}, we can define $V_{\mathrm{B}_j}$ in terms of $\mathfrak{q}_{\mathrm{A}}$, $\mathfrak{q}^*_{\mathrm{B}_j}$, and $\beta\Delta\Psi_{0_j}$ as

\begin{equation}\label{VBjeqn}
	V_{\mathrm{B}_j}\equiv (\mathfrak{q}_{\mathrm{A}}/\mathfrak{q}^*_{\mathrm{B}_j}) e^{-\beta\Delta\Psi_{0_j}}
	.
\end{equation}
Though we have treated the broken chain mechanics as independent of $\boldsymbol{\xi}$, the reforming reaction rate coefficient function $k_j''(\boldsymbol{\xi})$ in Eq.~\eqref{kABj} is still a strong function of $\boldsymbol{\xi}$ due to $\mathfrak{q}_{\ddagger_j}^*(\boldsymbol{\xi})$.
Our approximation in Eq.~\eqref{ewfrgdfgs3} causes all broken chain end-to-end vectors to be equally probable and therefore allows us to equate the probability of broken chains at end-to-end vector $\boldsymbol{\xi}$ to the average of broken chains at any end-to-end vector,

\begin{equation}\label{PBjxiapprox}
	P_{\mathrm{B}_j}(\boldsymbol{\xi};t) \approx
	\frac{P^\mathrm{tot}_{\mathrm{B}_j}(t)}{V_{\mathrm{B}_j}}
	.
\end{equation}
Since we do not track broken chains by end-to-end vector, we rewrite the conservation law from Eq.~\eqref{cons1} as

\begin{equation}\label{conservassumpd}
	\iiint P_\mathrm{A}(\boldsymbol{\xi};t)\,d^3\boldsymbol{\xi} + \sum_{j=1}^M P^\mathrm{tot}_{\mathrm{B}_j}(t) = 1
	,
\end{equation}
and, after taking the time derivative,

\begin{equation}\label{conservassumpdDT}
	\iiint \frac{\partial P_\mathrm{A}(\boldsymbol{\xi};t)}{\partial t}\,d^3\boldsymbol{\xi} + \sum_{j=1}^M \frac{\partial P^\mathrm{tot}_{\mathrm{B}_j}(t)}{\partial t} = 0
	.
\end{equation}
Applying the conservation requirement given by Eq.~\eqref{qconserv} and the relation between the partition functions given by Eq.~\eqref{ewfrgdfgs3}, we may now rewrite the equilibrium probabilities from Eqs.~\eqref{PAeqgen} and \eqref{PBjeqgen} as

\begin{align}
\label{PAeq}
	P_\mathrm{A}^\mathrm{eq}(\boldsymbol{\xi}) = &
	\frac{1}{1 + \sum_{\ell=1}^M e^{-\beta\Delta\Psi_{0_\ell}}}
	\left(\frac{\mathfrak{q}^*_\mathrm{A}(\boldsymbol{\xi})}{\iiint \mathfrak{q}^*_\mathrm{A}(\tilde{\boldsymbol{\xi}})\,d^3\tilde{\boldsymbol{\xi}}}\right)
	,\\
\label{PBjeq}
	P_{\mathrm{B}_j}^\mathrm{tot,eq} = &
	\frac{e^{-\beta\Delta\Psi_{0_j}}}{1 + \sum_{\ell=1}^M e^{-\beta\Delta\Psi_{0_\ell}}}
	,
\end{align}
and the evolution of the intact chains from Eq.~\eqref{PAevokin} as

\begin{equation}
\label{PAevokinVB}
	\frac{\partial P_\mathrm{A}(\boldsymbol{\xi};t)}{\partial t} = 
	-\sum_{j=1}^{M} k_j'(\boldsymbol{\xi}) \left[ P_\mathrm{A}(\boldsymbol{\xi};t) - \dfrac{P_{\mathrm{B}_j}^\mathrm{tot}(t)}{P_{\mathrm{B}_j}^\mathrm{tot,eq}}\,P^\mathrm{eq}_\mathrm{A}(\boldsymbol{\xi}) \right]
	- \frac{\partial}{\partial\boldsymbol{\xi}}\cdot\left[\dot{\boldsymbol{\xi}}_\mathrm{A}(\boldsymbol{\xi};t)P_\mathrm{A}(\boldsymbol{\xi};t)\right]
	.
\end{equation}
After writing the analogous evolution law for $P_{\mathrm{B}_j}(\boldsymbol{\xi};t)$, integrating for $P^\mathrm{tot}_{\mathrm{B}_j}(t)$, and using Eq.~\eqref{PBjxiapprox}, the evolution equation for $P^\mathrm{tot}_{\mathrm{B}_j}(t)$ is given by

\begin{equation}
	\frac{\partial P^\mathrm{tot}_{\mathrm{B}_j}(t)}{\partial t} = 
	\iiint k_j'(\boldsymbol{\xi}) \left[ P_\mathrm{A}(\boldsymbol{\xi};t) - \dfrac{P_{\mathrm{B}_j}^\mathrm{tot}(t)}{P_{\mathrm{B}_j}^\mathrm{tot,eq}}\,P^\mathrm{eq}_\mathrm{A}(\boldsymbol{\xi}) \right] d^3\boldsymbol{\xi}
	- \frac{P^\mathrm{tot}_{\mathrm{B}_j}(t)}{V_{\mathrm{B}_j}} \iiint \left[\frac{\partial}{\partial\boldsymbol{\xi}}\cdot\dot{\boldsymbol{\xi}}_\mathrm{B}(\boldsymbol{\xi};t)\right] d^3\boldsymbol{\xi}
\label{PBjtotevokinprelim}
	.
\end{equation}
The last term in Eq.~\eqref{PBjtotevokinprelim} is found equal to zero as follows:
Eq.~\eqref{PAevokinVB} is integrated over all $\boldsymbol{\xi}$ and the conservation requirement in Eq.~\eqref{conservassumpdDT} is applied in substituting in for $P_{\mathrm{B}_j}^\mathrm{tot}(t)$, where all the reaction-related terms then cancel.
We then apply the divergence theorem to this integral, which produces a balance law that we satisfy by requiring that the integrand is zero for all $\boldsymbol{\xi}$, which is

\begin{equation}\label{r3wr4entdf}
	\left[\dot{\boldsymbol{\xi}}_\mathrm{A}(\boldsymbol{\xi};t)P_\mathrm{A}(\boldsymbol{\xi};t) + \sum_{j=1}^{M} \dot{\boldsymbol{\xi}}_\mathrm{B}(\boldsymbol{\xi};t)\,\frac{P^\mathrm{tot}_{\mathrm{B}_j}(t)}{V_{\mathrm{B}_j}}\right]_{\partial_{\boldsymbol{\xi}}\mathcal{A}}
	= 0
	.
\end{equation}
Here $\partial_{\boldsymbol{\xi}}\mathcal{A}$ is the boundary of the region $\mathcal{A}$ with outward-pointing unit normal vector $\hat{\mathbf{n}}_{\partial_{\boldsymbol{\xi}}\mathcal{A}}$, a surface beyond which no intact chain may exist. 
Eq.~\eqref{r3wr4entdf} is a balance law for intact chains that are instantaneously broken via $\dot{\boldsymbol{\xi}}_\mathrm{A}(\boldsymbol{\xi};t)$ carrying them across $\partial_{\boldsymbol{\xi}}\mathcal{A}$.
We will now assume that a negligible amount of intact chains become extended to this intact limit without first breaking via the chemical reaction, which is $P_\mathrm{A}(\boldsymbol{\xi};t)|_{\partial_{\boldsymbol{\xi}}\mathcal{A}} \approx 0$. 
We then have $\dot{\boldsymbol{\xi}}_\mathrm{B}(\boldsymbol{\xi};t) \approx 0$ via Eq.~\eqref{r3wr4entdf}, which causes the last term in Eq.~\eqref{PBjtotevokinprelim} to become zero after again using the divergence theorem.
Eq.~\eqref{PBjtotevokinprelim} is now

\begin{equation}
\label{PBjtotevokin}
	\frac{\partial P^\mathrm{tot}_{\mathrm{B}_j}(t)}{\partial t} = 
	\iiint k_j'(\boldsymbol{\xi}) \left[ P_\mathrm{A}(\boldsymbol{\xi};t) - \dfrac{P_{\mathrm{B}_j}^\mathrm{tot}(t)}{P_{\mathrm{B}_j}^\mathrm{tot,eq}}\,P^\mathrm{eq}_\mathrm{A}(\boldsymbol{\xi}) \right] d^3\boldsymbol{\xi}
	.
\end{equation}
Eqs.~\eqref{PAevokinVB} and \eqref{PBjtotevokin} together with a prescription for $\dot{\boldsymbol{\xi}}_\mathrm{A}(\boldsymbol{\xi};t)$ create a set of evolution equations that govern the polymer network, are reasonable to solve, and have a firm foundation in the principles of statistical mechanics. 
Equipped with this framework to evaluate the relevant probabilities of chains in the network, we now use statistical thermodynamics to formulate the Helmholtz free energy and subsequently use macroscopic constitutive theory to obtain the constitutive relation for the Cauchy stress.


\newpage

\subsection{Macroscopic theory\label{macrosubsec}}

With our statistical mechanical framework complete, we turn now to the macroscopic description of the network. 
We begin by formulating the Helmholtz free energy of the network using statistical thermodynamics, which preserves our statistical mechanical framework as we move into the continuum scale.
Knowledge of the Helmholtz free energy allows us to utilize the Coleman-Noll procedure \citep{coleman13noll,coleman1967thermodynamics} to obtain constitutive relations for the entropy density and Cauchy stress. 
We do so after assuming that the temperature ($T$), deformation gradient ($\mathbf{F}$), probability density distribution of attached chains ($P_\mathrm{A}$), and probability of each broken chain type ($P_{\mathrm{B}_j}^\mathrm{tot}$) form a complete set of thermodynamic state variables. 
We additionally assume that, on average, the deformation gradient acts affinely on the intact chain end-to-end vectors.
Lastly, we show that the residual inequality -- solely dissipation due to the breaking/reforming of chains -- is already arbitrarily satisfied for the evolution laws we have derived in Sec.~\ref{statmechsubsection}. 

\subsubsection{Network Helmholtz free energy\label{machelm_ssec}}

The Helmholtz free energy $\mathscr{A}(t)$ of the network is analogous to that of a system of noninteracting particles of different chemical species \citep{mcq}

\begin{equation}
	\mathscr{A}(t) = 
	N_\mathrm{A}(t) \mu_\mathrm{A}(t) + \sum_{j=1}^{M} N_{\mathrm{B}_j}(t) \mu_{\mathrm{B}_j}(t) - N\mathfrak{b}T(t)
	,
\end{equation}
with the chemical potentials $\mu_i(t)$ given by

\begin{equation}
	\mu_i(t) = 
	-\mathfrak{b}T\ln\left[\frac{\mathfrak{q}_i}{N_i(t)}\right]
	,\quad i = \mathrm{A},\mathrm{B}_1,\ldots,\mathrm{B}_M
	,
\end{equation}
where $N_i(t)$ is the number of either intact or broken chains, and $N$ is the constant total number of chains.
We use Gibbs' postulate \citep{mcq} to write $\mu_\mathrm{A}(t)$ as the time-dependent average

\begin{equation}
	\mu_\mathrm{A}(t) = 
	\frac{1}{P_\mathrm{A}^\mathrm{tot}(t)}\,\iiint P_\mathrm{A}(\boldsymbol{\xi};t) \mu^*_\mathrm{A}(\boldsymbol{\xi};t) \,d^3\boldsymbol{\xi}
	,
\end{equation}
where (using $N_i(t)=P_i^\mathrm{tot}(t)N$) the chemical potential of an intact chain at end-to-end vector $\boldsymbol{\xi}$ is

\begin{equation}
	\label{muA}
	\mu^*_\mathrm{A}(\boldsymbol{\xi};t) = 
	-\mathfrak{b}T\ln\left[\frac{\mathfrak{q}^*_\mathrm{A}(\boldsymbol{\xi})}{NP_\mathrm{A}(\boldsymbol{\xi};t)}\right]
	.
\end{equation}
The broken chains have been assumed to be insensitive to extension, so we similarly utilize their chemical potentials as independent of the end-to-end vector $\boldsymbol{\xi}$,

\begin{equation}
	\label{muBj}
	\mu_{\mathrm{B}_j}(t) = 
	-\mathfrak{b}T\ln\left[\frac{V_{\mathrm{B}_j}\mathfrak{q}_{\mathrm{B}_j}^*}{NP^\mathrm{tot}_{\mathrm{B}_j}(t)}\right]
	.
\end{equation}
We may now write the Helmholtz free energy density $a(t)=\mathscr{A}(t)/V$ of our incompressible network of noninteracting polymer chains as

\begin{equation}\label{ayo}
	a(t) = 
	n\iiint P_\mathrm{A}(\boldsymbol{\xi};t) \mu^*_\mathrm{A}(\boldsymbol{\xi};t)\,d^3\boldsymbol{\xi} + n\sum_{j=1}^{M} P^\mathrm{tot}_{\mathrm{B}_j}(t) \mu_{\mathrm{B}_j}(t)
	-n\mathfrak{b}T(t) - p[J(t) - 1]
	,
\end{equation}
where $n=N/V$ is the constant total number density of chains, and $p$ is the pressure acting as a Lagrange multiplier enforcing the incompressibility constraint that $J=\det(\mathbf{F})=1$.
This specific formulation of $a(t)$ in Eq.~\eqref{ayo} is essential to our approach: in \ref{appmacroeq}, we show that it allows the equilibrium probabilities obtained from minimizing $a(t)$ with respect to each probability to exactly match those obtained beforehand from the statistical mechanical derivation (Eqs.~\eqref{PAeq} and \eqref{PBjeq}). 

\subsubsection{Constitutive relations\label{constirelsubsubsec}}

The time derivative of the Helmholtz free energy density is

\begin{equation}
	\dot{a}(t) = 
	n\iiint \dot{P}_\mathrm{A}(\boldsymbol{\xi};t) \mu^*_\mathrm{A}(\boldsymbol{\xi};t)\,d^3\boldsymbol{\xi} + n\sum_{j=1}^{M}\dot{P}^\mathrm{tot}_{\mathrm{B}_j}(t) \mu_{\mathrm{B}_j}(t)
	-nk\dot{T}(t) - p(t)\left[\mathbf{1}:\mathbf{L}(t)\right]
	,
\end{equation}
where $\mathbf{L}=\dot{\mathbf{F}}\cdot\mathbf{F}^{-1}$ is the velocity gradient.
We have factored out the time derivatives of the chemical potentials after utilizing the conservation requirement from Eq.~\eqref{conservassumpdDT}. 
Thermodynamically admissible processes must satisfy the second law of thermodynamics regarding irreversible entropy production, which is embodied in the Clausius-Duhem inequality

\begin{equation}\label{CD1}
	\dot{a} + s\dot{T} - \boldsymbol{\sigma}:\mathbf{L} \leq 0
	,
\end{equation}
where $s(t)$ is the entropy density and $\boldsymbol{\sigma}(t)$ is the Cauchy stress tensor \citep{truesdell2004non}. 
Note that this simplified form of the inequality already assumes incompressibility, neglects nonmechanical work, and assumes the classical constitutive relations for the entropy flux, the entropy source, and the heat flux \citep{paolucci2016continuum}. 
Further, we will not impose hyperbolicity requirements that would guarantee finite speeds of propagation \citep{muller2013rational}.

We will assume that $T(t)$, $\mathbf{F}(t)$, $P_\mathrm{A}(\boldsymbol{\xi};t)$, and $P^\mathrm{tot}_{\mathrm{B}_j}(t)$ together create a complete set of thermodynamic state variables.
This allows us to treat time derivatives of constitutive variables like $a(t)$ as fully implicit, where we may expand those time derivatives as partial derivatives with respect to the state variables. 
The three independent thermodynamic processes accounted for through our state variables are temperature change, deformation, and the chain breaking/reforming chemical reactions (denoted as rxn). 
The evolution of $a(t)$ is then expanded as

\begin{eqnarray}
	\dot{a} =  
	\left(\frac{\partial a}{\partial T}\right)_{\mathbf{F},\mathrm{rxn}}\dot{T}
	+ \left(\frac{\partial a}{\partial \mathbf{F}}\right)_{T,\mathrm{rxn}}:\dot{\mathbf{F}}
	-\mathcal{D}_\mathrm{rxn}
	,
\end{eqnarray}
where $\mathcal{D}_\mathrm{rxn}$ is the chemical dissipation per unit volume, calculated from the rate of change of Helmholtz free energy density over all breaking/reforming reactions

\begin{equation}\label{Drxndef}
	\mathcal{D}_\mathrm{rxn} \equiv -\left(\frac{\partial a}{\partial t}\right)_{\mathbf{F},T}
	.
\end{equation}
Substitution of our expansion into the Clausius-Duhem inequality in Eq.~\eqref{CD1} yields

\begin{equation}\label{CD2}
\left[\left(\frac{\partial a}{\partial T}\right)_{\mathbf{F},\mathrm{rxn}} + s\right]\dot{T} - \mathcal{D}_\mathrm{rxn}
+ \left[\left(\frac{\partial a}{\partial \mathbf{F}}\right)_{T,\mathrm{rxn}}\cdot\mathbf{F}^T - \boldsymbol{\sigma}\right]:\mathbf{L}
\leq 0
.
\end{equation}
We first consider the set of processes where the temperature varies arbitrarily, the deformation is held fixed, and the reactions do not proceed. 
In order to arbitrarily satisfy the inequality, we must then have

\begin{equation}
	s = 
	-\left(\frac{\partial a}{\partial T}\right)_{\mathbf{F},\mathrm{rxn}}
	,
\end{equation}
which is the expected constitutive relation for the entropy density. 
The inequality in Eq.~\eqref{CD2} then becomes

\begin{equation}\label{CD3}
	\left[\left(\frac{\partial a}{\partial \mathbf{F}}\right)_{T,\mathrm{rxn}}\cdot\mathbf{F}^T - \boldsymbol{\sigma}\right]:\mathbf{L} - \mathcal{D}_\mathrm{rxn} \leq 0
	.
\end{equation}
We next consider processes where the motion varies arbitrarily via $\mathbf{L}(t)$ and the reactions do not proceed. 
Since we have assumed to have a complete set of thermodynamic state variables, and that set does not include time derivatives of the deformation gradient, we have already ruled out dissipative stresses \citep{paolucci2016continuum}.
In order to arbitrarily satisfy the inequality in Eq.~\eqref{CD3}, we must then have

\begin{equation}\label{hyperelastic}
	\boldsymbol{\sigma} = \left(\frac{\partial a}{\partial \mathbf{F}}\right)_{T,\mathrm{rxn}}\cdot\mathbf{F}^T
	,
\end{equation}
which is the form of the stress for a hyperelastic material \citep{truesdell2004non}.
We will now assume that, on average, the end-to-end vectors $\boldsymbol{\xi}$ are affinely deformed by the deformation gradient $\mathbf{F}(t)$, which can be expanded as $\dot{\boldsymbol{\xi}}_\mathrm{A}(\boldsymbol{\xi};t)=\mathbf{L}(t)\cdot\boldsymbol{\xi}$. 
After applying this assumption and simplifying (see \ref{appstressretrieve}), the evolution equation for intact chains Eq.~\eqref{PAevokinVB} becomes

\begin{equation}\label{PAevokinVBfinal}
	\frac{\partial P_\mathrm{A}(\boldsymbol{\xi};t)}{\partial t} = 
	- \left[\frac{\partial P_\mathrm{A}(\boldsymbol{\xi};t)}{\partial\boldsymbol{\xi}}\,\boldsymbol{\xi}\right]:\mathbf{L}(t)
	-\sum_{j=1}^{M} k_j'(\boldsymbol{\xi}) \left[ P_\mathrm{A}(\boldsymbol{\xi};t) - \dfrac{P_{\mathrm{B}_j}^\mathrm{tot}(t)}{P_{\mathrm{B}_j}^\mathrm{tot,eq}}\,P^\mathrm{eq}_\mathrm{A}(\boldsymbol{\xi}) \right]
	,
\end{equation}
and the stress in Eq.~\eqref{hyperelastic} becomes

\begin{equation}\label{sigmafinal}
	\boldsymbol{\sigma}(t) = 
	n\iiint P_\mathrm{A}(\boldsymbol{\xi};t) \,\frac{\partial\psi^*_\mathrm{A}(\boldsymbol{\xi})}{\partial\boldsymbol{\xi}} \,\boldsymbol{\xi} \,d^3\boldsymbol{\xi}
	- p(t)\mathbf{1}
	.
\end{equation}
This general form of the stress has been obtained previously \citep{Buchestatistical2020}, but the evolution of $P_\mathrm{A}(\boldsymbol{\xi};t)$ is now more complicated here due to the breaking and reforming of chains.

\subsubsection{Residual inequality\label{resineq_ssec}}

Now that we have established each constitutive relation, we are left with the residual portion of the Clausius-Duhem inequality due to the dissipation $\mathcal{D}_\mathrm{rxn}(t)$.
Showing that $\mathcal{D}_\mathrm{rxn}(t)\geq 0$ is similar to the procedure for a reacting system of a finite number of discrete chemical species \citep{prigogine1967introduction,powers2016combustion}, but here we have the additional complication of having the reactions (chains breaking and reforming) occurring over the continuous variable $\boldsymbol{\xi}$ (see \ref{appgettheD} for details).
We find that the dissipation may be written succinctly as

\begin{equation}\label{Drxnttt}
	\mathcal{D}_\mathrm{rxn}(t) = 
	\sum_{j=1}^{M} \iiint \mathcal{D}_j^*(\boldsymbol{\xi};t) \,d^3\boldsymbol{\xi}
	,
\end{equation}
where the dissipation density for the $j$th reaction occurring at the end-to-end vector $\boldsymbol{\xi}$ is defined as

\begin{equation}\label{DjSrxnttt}
	\mathcal{D}_j^*(\boldsymbol{\xi};t) \equiv 
	n\mathfrak{b}T\left[\mathcal{R}_j'(\boldsymbol{\xi};t) - \mathcal{R}_j''(\boldsymbol{\xi};t)\right]\ln\left[\frac{\mathcal{R}_j'(\boldsymbol{\xi};t)}{\mathcal{R}_j''(\boldsymbol{\xi};t)}\right]
	.
\end{equation}
Since $[\mathcal{R}_j'(\boldsymbol{\xi};t)-\mathcal{R}_j''(\boldsymbol{\xi};t)]\ln[\mathcal{R}_j'(\boldsymbol{\xi};t)/\mathcal{R}_j''(\boldsymbol{\xi};t)]\geq 0$ and $n\mathfrak{b}T>0$, we are able to conclude

\begin{equation}\label{QEDyo}
	\mathcal{D}_j^*(\boldsymbol{\xi};t) \geq 0 \quad\text{for all}~j
	,\text{ and therefore}\quad
	\mathcal{D}_\mathrm{rxn}(t) \geq 0
	.
\end{equation}
This result means that not only is the residual inequality satisfied, but each chain breaking/reforming reaction at every end-to-end vector has a positive semi-definite dissipation.
Further, this is found without any additional restrictions on the thermodynamic state variables or the obtained constitutive relations, which can be directly attributed to the strong statistical mechanical foundation we have incorporated.

\subsection{General theory summary}

\begin{figure*}[t]
\begin{center}
\includegraphics{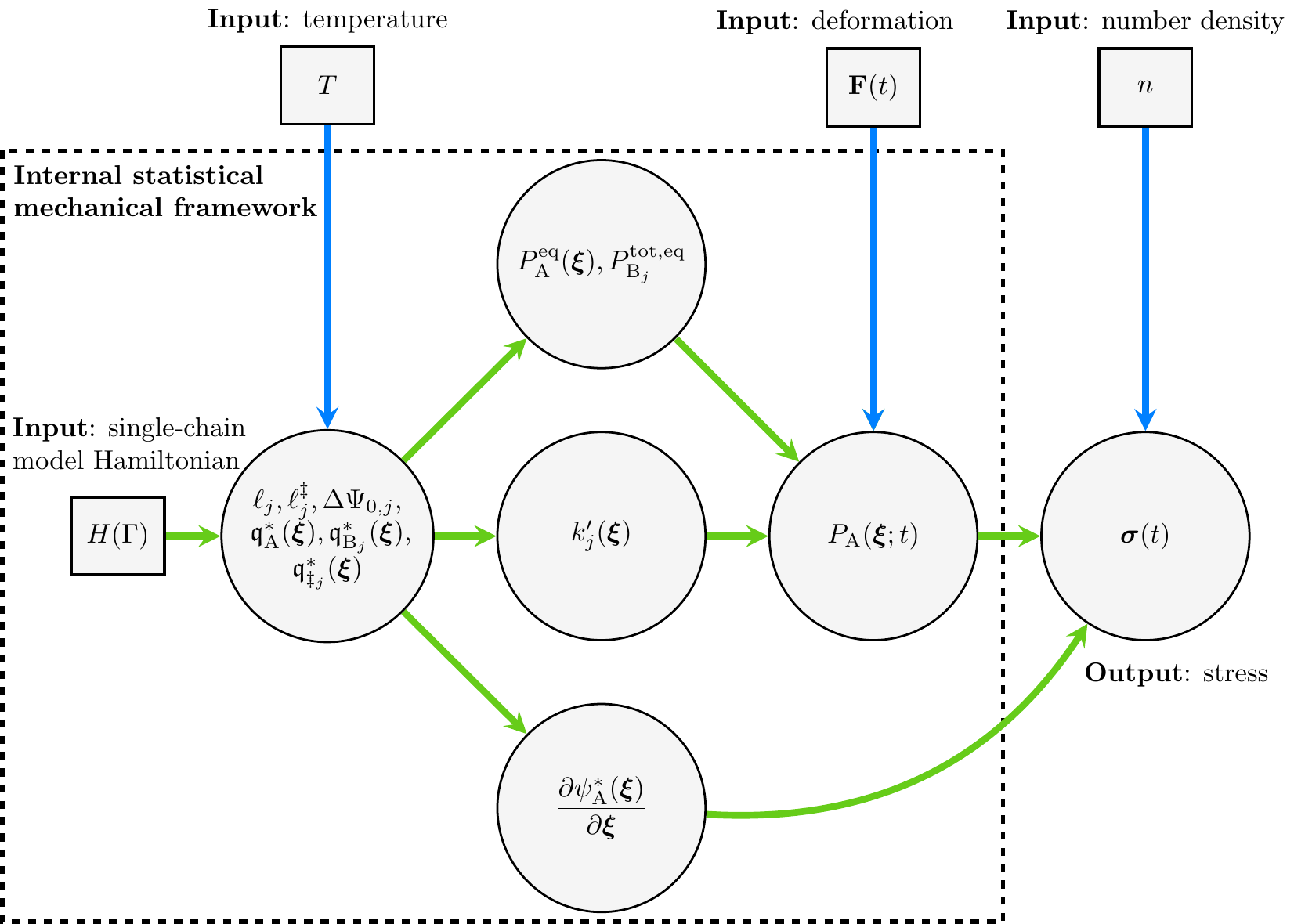}
\end{center}
\caption{\label{scheme1}
  Diagram describing the utilization of the general theory developed in Sec.~\ref{sec2}.
  After specifying the single-chain model Hamiltonian $H(\Gamma)$, the temperature $T$, the deformation gradient $\mathbf{F}(t)$, and the total number density of chains $n$, all quantities of interest may be systematically computed, ultimately resulting in the stress $\boldsymbol{\sigma}(t)$. 
}
\end{figure*}

Our general theory is now complete and can be utilized as illustrated by Fig.~\ref{scheme1}.
Two inputs are needed to constitutively define the polymer -- the single-chain model and the total number density of chains $n$. 
Two external conditions are also prescribed as inputs -- the temperature $T$ and the deformation gradient $\mathbf{F}(t)$.
A single-chain model is chosen through specification of a Hamiltonian $H(\Gamma)$, which contains $M$ transition state coordinates $\ell_j$ and locations $\ell_j^\ddagger$. 
Next, we calculate the partition function at the end-to-end vector $\boldsymbol{\xi}$ of the intact chains $\mathfrak{q}_\mathrm{A}^*(\boldsymbol{\xi})$ using Eq.~\eqref{qA}, of each broken chain species $\mathfrak{q}_{\mathrm{B}_j}^*$ (at some large $\boldsymbol{\xi}$ where the two broken ends do not interact) using Eq.~\eqref{qBj}, and of each transition state $\mathfrak{q}_{\ddagger_j}^*(\boldsymbol{\xi})$ using Eq.~\eqref{qABjTS}. 
We also compute the net free energy changes $\Delta\Psi_{0_j}$ using $H(\Gamma)$ or otherwise specify them as parameters. 
Equipped with these quantities, we are able to compute each reaction rate coefficient function $k_j'(\boldsymbol{\xi})$ using Eq.~\eqref{kBjA}, the equilibrium probability density distribution of intact chains $P^\mathrm{eq}_\mathrm{A}(\boldsymbol{\xi})$ using Eq.~\eqref{PAeq}, and the equilibrium probability of each broken chain species $P^\mathrm{tot,eq}_{\mathrm{B}_j}$ using Eq.~\eqref{PBjeq}.
We also compute $\psi_\mathrm{A}^*(\boldsymbol{\xi})$, the Helmholtz free energy of an intact chain at the end-to-end vector $\boldsymbol{\xi}$, using Eq.~\eqref{psinegktlnq}.
With a prescribed incompressible deformation history and assuming initial conditions, we have all the necessary information to formulate the evolution law for the probability density distribution of intact chains $P_\mathrm{A}(\boldsymbol{\xi};t)$ using Eq.~\eqref{PAevokinVBfinal} and those for the probability of each broken chain species $P^\mathrm{tot}_{\mathrm{B}_j}(t)$ using Eq.~\eqref{PBjtotevokin}.
The stress $\boldsymbol{\sigma}(t)$ is then computed using Eq.~\eqref{sigmafinal}, where the pressure $p(t)$ is solved for using the traction boundary conditions.

\section{Single-chain model specification\label{singlechainimplsec}}

We are now ready to specify a single-chain model and push it through our general framework as shown Fig.~\ref{scheme1}, computing each quantity of interest to ultimately obtain the stress as a function of deformation.
For the materials we will model, we require a single-chain model that incorporates force-sensitive reversible bond breaking.
Force-sensitive irreversible and force-insensitive (transient) reversible bonds are special cases of this broader class.
We propose the $u$FJC for our single-chain model: a freely jointed chain of flexible links, each with a potential $u$ that depends on the difference between the link length and its rest-length.
The Morse potential \citep{morse1929diatomic} is used for each link in order to allow the links to break and reform.
In the following section, we compute the functions related to the mechanical behavior, equilibrium distribution, and kinetics of breaking/reforming for the $u$FJC single-chain model and provide results with the Morse potential.
We then exactly solve the evolution equation for the distribution of intact chains in the network for all single-chain models with reaction pathways that are all equivalent.
Then we present how small adjustments can be made to the framework to account for when some links in the $u$FJC are weaker than the rest.

\subsection{The \textit{u}FJC model\label{subsecuFJC}}

The $u$FJC model is a freely jointed chain of $N_b$ flexible links, each with potential $u(\ell)$ that depends on the difference between the link length $\ell$ and its rest-length $\ell_b$.
This is similar to the freely jointed chain or FJC model \citep{Treloar1949physics,rubinstein2003polymer}, but with the rigid links replaced by these flexible ones.
If the potential is strictly harmonic, we retrieve the extensible freely jointed chain or EFJC model \citep{fiasconaro2019analytical}.
The links of the $u$FJC are considered broken if $\ell>\ell_\ddagger$ and intact otherwise. 
Each of the $N_b+1$ hinges are considered to be point masses with a mass of $m$, while the links are massless.
The single-chain Hamiltonian of this model is

\begin{equation}
	H(\Gamma) = 
	\sum_{i=1}^{N_b+1} \frac{p_i^2}{2m} + \sum_{i=1}^{N_b} u\left(\ell_i\right)
	.
\end{equation}
We substitute this Hamiltonian into Eq.~\eqref{qA} to compute the partition function $\mathfrak{q}_\mathrm{A}^*(\boldsymbol{\xi})=\mathfrak{q}^*_\mathrm{A,mom}\mathfrak{q}_\mathrm{A,con}^*(\boldsymbol{\xi})$ of an intact chain with end-to-end vector $\boldsymbol{\xi}$, where ``mom'' and ``con'' denote the momentum and configuration contributions.
Since the link potential only gives information about the intact state of the links, we must treat the net Helmholtz free energy change when breaking a link, $\Delta\Psi_0$, as an independent parameter.
While we may compute $\mathfrak{q}^*_\mathrm{A,mom}$ exactly without trouble \citep{mcq}, $\mathfrak{q}_\mathrm{A,con}^*(\boldsymbol{\xi})$ cannot be computed analytically in general and is difficult to evaluate numerically. 
This issue is typically resolved through computation of the Gibbs (isotensional) ensemble partition function and transforming back to the Helmholtz (isometric; canonical) partition function we desire.
This transformation is accomplished using an inverse Fourier transform, which is often approximated using a Legendre transformation when chains are sufficiently long \citep{manca2012elasticity,manca2014equivalence}, the so-called Gibbs-Legendre method \citep{Buchestatistical2020} of obtaining $\mathfrak{q}_\mathrm{A,con}^*(\boldsymbol{\xi})$.
After asymptotically approximating the mechanical response of the $u$FJC in the Gibbs ensemble, we will integrate it and use the Gibbs-Legendre method to obtain the Helmholtz free energy and then compute the desired partition function.

We obtain an asymptotic approximation for the single-chain mechanical response as

\begin{equation}\label{gammasimplest}
	\gamma(\eta)\sim\mathcal{L}(\eta) + \lambda(\eta) - 1
	\quad\text{for }\kappa\gg 1
	,
\end{equation}
where $\gamma\equiv\xi/N_b\ell_b$ is the nondimensional end-to-end length, $\mathcal{L}(\eta)=\coth(\eta)-1/\eta$ is the Langevin function, $\lambda(\eta)$ is the link stretch $\ell/\ell_b$ under the nondimensional force $\eta\equiv\beta f\ell_b$, and

\begin{equation}
	\kappa \equiv
	\beta\ell_b^2 \left.\frac{\partial^2 u(\ell)}{\partial \ell^2}\right|_{\ell=\ell_b}
\end{equation}
is the nondimensional link stiffness.
The full derivation of Eq.~\eqref{gammasimplest} is in \ref{appasymp}, where it is shown that the entropically-based mechanical response of the FJC (given by the Langevin function) may be approximated as decoupled from the link stretching for sufficiently stiff links.
There has been recent interest in similarly supplementing entropic polymer chain models with potential energy contributions \citep{mao2018fracture,talamini2018progressive,lavoie2019modeling,li2020variational,guo2021micromechanics} by a method where a chain free energy is minimized with respect to the potential degrees of freedom \citep{mao2017rupture}.
Although this current method performs well, we recommend that the asymptotic approach be used instead for both practical and physical reasons.
Practically, asymptotically-correct formulas such as those we provide here for the $u$FJC are easier to use than the current method, which requires implicitly solving nonlinear algebraic equations during the minimization process.
Physically, the minimization of thermodynamic free energies may only involve macroscopic thermodynamic state variables, not phase space degrees of freedom such as link length, which would approach their potential energy minima as thermal energy becomes scarce.
The apparent success of the current method can be attributed to the dominance of potential energy over the free energy minimization in the same limit, which in effect produces similar results.

\begin{figure*}[t]
\begin{center}
$
\michaelarraysettings
\begin{array}{cc}
\includegraphics{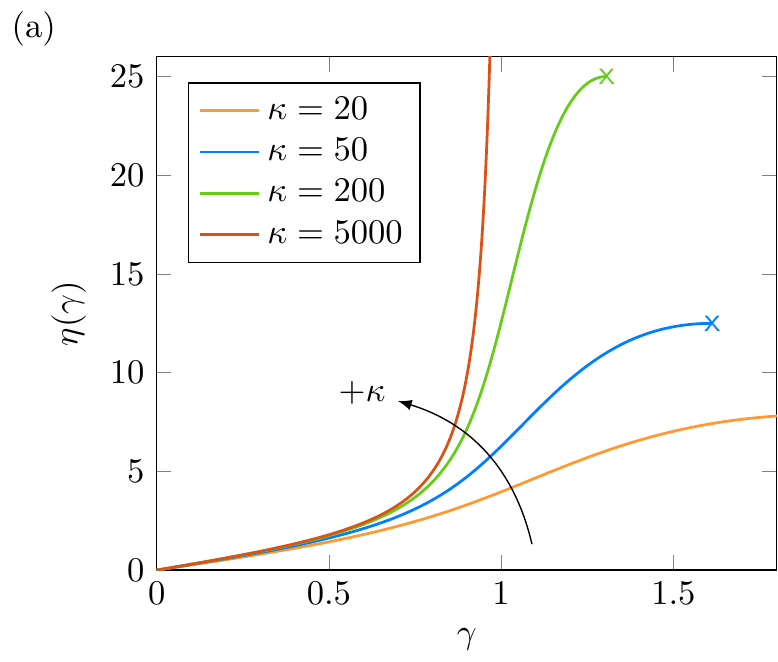}
&
\includegraphics{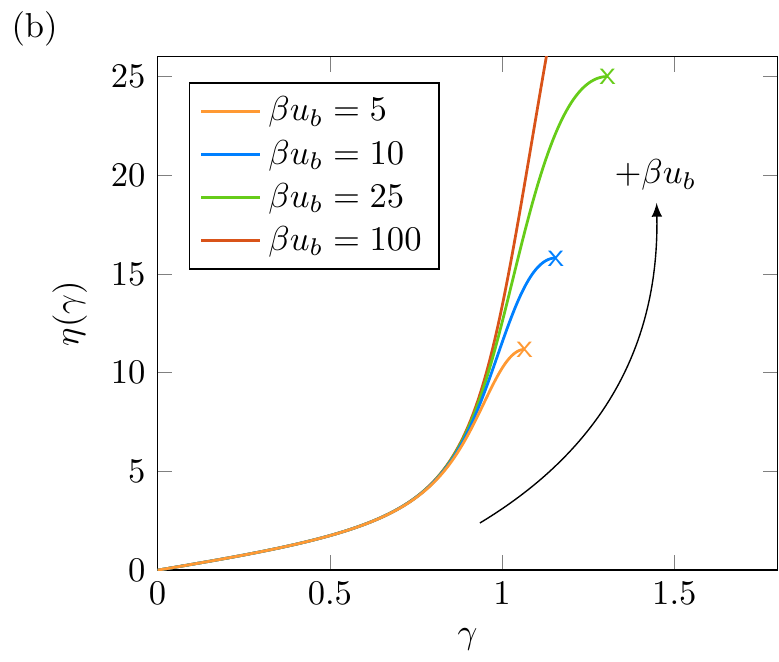}
\end{array}
$
\vspace{-5pt}
\end{center}
\caption{\label{figmech}
	Single-chain mechanical response when using the Morse-FJC model.
	(a) The nondimensional force $\eta=\beta f\ell_b$ as a function of the nondimensional end-to-end length $\gamma=\xi/N_b\ell_b$ for $\beta u_b = 25$ and varying $\kappa$.
	(b) $\eta$ as a function of $\gamma$ for $\kappa=200$ and varying $\beta u_b$.
}
\end{figure*}

We select the Morse potential \citep{morse1929diatomic} as the specific form of the $u$FJC link potential,

\begin{equation}\label{umorse}
	u(\ell) = 
	u_b\left[1 - e^{-\sqrt{k_b/2u_b}(\ell-\ell_b)}\right]^2
	,
\end{equation}
where $u_b$ is the depth of the Morse potential energy well and $k_b$ is the curvature near the bottom of the well (the stiffness).
The nondimensional parameters here are the nondimensional energy $\beta u_b$, the nondimensional stiffness $\kappa\equiv\beta k_b\ell_b^2$, and the link stretch $\lambda\equiv \ell/\ell_b$.
The stretch $\lambda_\ddagger\equiv 1 + \ln(2)\sqrt{2\beta u_b/\kappa}$ is the transition state stretch since this is where the nondimensional force $\eta$ will reach its maximum possible value of $\eta_\mathrm{max} = \sqrt{\kappa\beta u_b/8}$. 
The derivative of the Morse potential gives the force as a function of link length, which is then inverted and rescaled to obtain the stretch of a single link under force, 

\begin{equation}\label{Morselam}
	\lambda(\eta) = 
	1 + \sqrt{\frac{2\beta u_b}{\kappa}}\,\ln\left[\frac{2}{1+\displaystyle\sqrt{1 - \eta/\eta_\mathrm{max}}}\right]
	\quad\text{for } \eta \leq \eta_\mathrm{max} = \sqrt{\frac{\kappa\beta u_b}{8}}
	.
\end{equation}
Utilizing Eq.~\eqref{Morselam} with Eq.~\eqref{gammasimplest}, we plot the mechanical response of the Morse-FJC in Fig.~\ref{figmech}, varying $\kappa$ in Fig.~\ref{figmech}(a) and varying $\beta u_b$ in Fig.~\ref{figmech}(b).
As $\kappa$ increases, we observe a more dramatic transition near $\gamma=1$ as the increasingly-stiff links begin to be stretched.
Increasing $\kappa$ directly increases the maximum force $\eta_\mathrm{max}$ and causes it to be reached at lower stretches, thereby causing the maximum nondimensional end-to-end length to decrease with increasing $\kappa$ $(\gamma_\mathrm{max} \sim \lambda_\ddagger)$.
As we increase $\beta u_b$, we see an increase in both $\eta_\mathrm{max}$ and $\gamma_\mathrm{max}$; the overall mechanical response away from $\eta_\mathrm{max}$ is unchanged.
When varying either $\kappa$ or $\beta u_b$, the mechanical response at low $\gamma$ is unchanged since this regime is dominated by the initially-compliant entropic behavior of the Langevin function $\mathcal{L}(\eta)$.

We now compute the other thermodynamic functions of interest: the equilibrium distribution $P_\mathrm{A}^\mathrm{eq}(\boldsymbol{\xi})$ and the reaction rate coefficient function $k'(\boldsymbol{\xi})$.
For this single-chain model it is most practical to work in terms of the nondimensional chain end-to-end vector $\boldsymbol{\gamma} = \boldsymbol{\xi}/N_b\ell_b$.
Further, our single-chain functions and equilibrium distributions depend only on $\gamma=\sqrt{\boldsymbol{\gamma}\cdot\boldsymbol{\gamma}}$. We therefore introduce the nondimensional configurational Helmholtz free energy per link

\begin{equation}
	\vartheta_\mathrm{A,con}^*({\gamma}) \equiv 
	\frac{\beta\psi_\mathrm{A,con}^*({\gamma})}{N_b} 
	.
\end{equation}
We will assume that the number of links in our chain $N_b$ remains high enough to utilize the Gibbs-Legendre method \citep{Buchestatistical2020} to approximate $\psi_\mathrm{A,con}^*({\gamma})$, which in this case causes $\vartheta_\mathrm{A,con}^*({\gamma})$ to be independent of $N_b$.
While the Gibbs-Legendre method is invalid under sufficiently small forces \citep{neumann1985nonequivalence}, the regime of end-to-end lengths where this matters essentially vanishes with increasing $N_b$ \citep{manca2012elasticity,manca2014equivalence}.
Further, these tiny forces contribute little when integrating over all end-to-end lengths in Eq.~\eqref{sigmafinal} for the stress, causing the Gibbs-Legendre method to become correct as $N_b$ increases \citep{Buchestatistical2020}.
As detailed in \ref{appuFJCsimp}, we obtain 

\begin{equation}\label{psiconufjc}
	\vartheta_\mathrm{A,con}^*({\gamma}) = 
	\ln\left\{\frac{\eta\exp[\eta\mathcal{L}(\eta)]}{\sinh(\eta)}\right\} 
	+\beta u(\eta)
	,
\end{equation}
where $\eta = \eta(\gamma)$ is implied, which allows us to compute the nondimensional equilibrium distribution

\begin{equation}\label{PeqnicenoVb}
	\mathscr{P}_\mathrm{A}^\mathrm{eq}({\gamma}) = 
	\frac{1}{1 + N_b e^{-\beta\Delta\Psi_{0}}} \left(\frac{e^{-N_b\vartheta_\mathrm{A,con}^*({\gamma})}}{\iiint e^{-N_b\vartheta_\mathrm{A,con}^*(\tilde{{\gamma}})}\,d^3\tilde{\boldsymbol{\gamma}}}\right)
	.
\end{equation}
In \ref{appuFJCsimp} we also obtain the forward reaction rate coefficient function

\begin{equation}\label{kBAfinal}
	k'({\gamma}) = 
	\frac{\omega_\ddagger}{2\pi}\,e^{-\beta\Delta\Psi_\ddagger^*({\gamma})}
	.
\end{equation}
Here $\omega_\ddagger \equiv \sqrt{2\kappa/m\beta\ell_b^2}$ is the attempt frequency, and $\Delta\Psi_\ddagger^*({\gamma})$ is the Helmholtz free energy barrier of a single link to its transition state as a function of chain extension,

\begin{equation}\label{DeltaPsiTS}
	\Delta\Psi_\ddagger^*({\gamma}) \equiv 
	u(\ell_\ddagger) - u(\eta)
	- \mathfrak{b}T\ln\left\{\frac{\lambda_\ddagger\sinh(\lambda_\ddagger\eta)\exp[\eta\mathcal{L}(\eta)]}{\sinh(\eta)\exp[\lambda_\ddagger\eta\mathcal{L}(\lambda_\ddagger\eta)]}\right\} 
	.
\end{equation}
Note that we often use the initial rate $k'({0})$ in place of the attempt frequency $\omega_\ddagger$ or mass $m$ as a more convenient but equivalent parameterization.
The Helmholtz free energy barrier $\Delta\Psi_\ddagger^*({\gamma})$ consists of a positive contribution from the potential energy difference and a negative contribution from the entropy difference.
The initial nondimensional barrier is $\beta\Delta\Psi_\ddagger^*({0}) = \beta u_b/4 - 2\ln\lambda_\ddagger$, where recall $\lambda_\ddagger=1+\ln(2)\sqrt{2\beta u_b/\kappa}$.
Due to our requirement $\kappa\gg 1$, the nondimensional potential energy barrier $\beta u_b$ will tend to dominate the scale of $\beta\Delta\Psi_\ddagger^*({\gamma})$ and therefore $k'({\gamma})$.
If we assume $\lambda_\ddagger\approx 1$, we ignore the entropic term and take $u(\eta)\approx \eta\lambda(\eta)/2$, which then causes $k'({\gamma})$ in Eq.~\eqref{kBAfinal} to resemble the short-distance force-modified-barrier transition state theories \citep{kauzmann1940viscous,bell1978models} that have been applied to polymer chains \citep{zhurkov1984kinetic,tanaka1992viscoelasticALL3,silberstein2013modeling,silberstein2014modeling,meng2016stress,tehrani2017effect}.
The general case behavior of $k'({\gamma})$ here in Eq.~\eqref{kBAfinal} is more similar to the model of \citet{dudko2006intrinsic} than these short-distance approximated models, especially since it accounts for both entropic and potential energy effects.
Dudko's model is based on Kramers' theory of diffusive barrier crossing \citep{kramers1940brownian,zwanzig2001nonequilibrium} and has proved useful both in polymer chain AFM experiments \citep{schwaderer2008single} and polymer network constitutive models \citep{lavoie2019modeling}.
Our formulation for $k'({\gamma})$ has an advantage over Dudko's model: our $k'({\gamma})$ is directly connected to the statistical mechanics of the single-chain model, which provides guarantees such as dissipation requirements and solution existence.

\begin{figure*}[t]
\begin{center}
\includegraphics{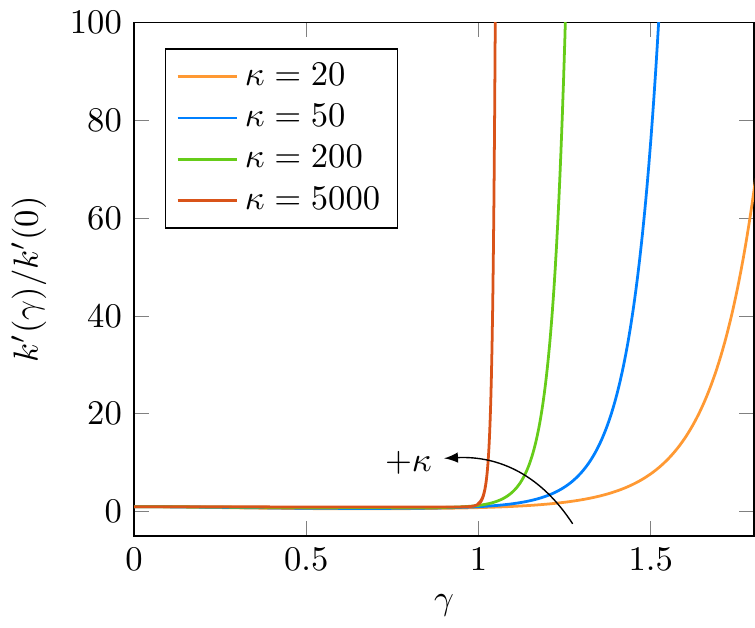}
\vspace{-5pt}
\end{center}
\caption{\label{figk}
	The reaction rate coefficient function $k'(\gamma)$ scaled by its initial value $k'(0)$ as a function of the nondimensional end-to-end length $\gamma=\xi/N_b\ell_b$, using the Morse-FJC model for $\beta u_b = 25$, and varying $\kappa$. 
}
\end{figure*}

The reaction rate coefficient $k'({\gamma})$ is plotted as a function of the nondimensional end-to-end length $\gamma$ in Fig.~\ref{figk} for $\beta u_b = 25$ and varying $\kappa$.
We find that $k'({\gamma})$ decreases slightly from $k'({0})$ as the chain is extended, which is due to the increasing entropy of the links.
After a critical chain extension just above unity, we find that $k'({\gamma})$ increases dramatically due to the potential energy barrier to the transition state being rapidly reduced.
As $\kappa$ is increased, this trend becomes even more dramatic since the potential energy barrier is proportional to $\kappa$, and the observed critical extension approaches unity.
Varying $\beta u_b$ provides little change to the shape of $k'({\gamma})/k'({0})$ and effectively varies the maximum allowable extension (not shown).
Eqs.~\eqref{kBAfinal}--\eqref{DeltaPsiTS} show that $k'({0})$ decays exponentially fast as $\beta u_b$ becomes large, so varying $\beta u_b$ changes the scale of $k'({\gamma})$ but not its shape.
Having both $\kappa\gg 1$ and $\beta u_b\gg 1$ simultaneously tends to result in reaction rate coefficient function $k'({\gamma})$ that is essentially constant before rapidly becoming effectively infinite at and above some critical extension $\gamma_c\gtrsim 1$, similar to the $\kappa=5000$ case in Fig.~\ref{figk}.

\subsection{Distribution evolution and solution\label{Psolsubsubsec}}

Now that the single-chain mechanics and kinetics for the $u$FJC model have been formulated, we next determine the evolution equation for the distribution of intact chains in the network and calculate the stress using Eq.~\eqref{sigmafinal}.
We will continue writing our results in terms of the nondimensional chain end-to-end vector $\boldsymbol{\gamma} = \boldsymbol{\xi}/N_b\ell_b$, and we correspondingly utilize the nondimensional distribution $\mathscr{P}_\mathrm{A}(\boldsymbol{\gamma};t)\equiv (N_b\ell_b)^3 P_\mathrm{A}(\boldsymbol{\xi};t)$.
It is equally probable that any given link will break, since each behaves the same under the Gibbs-Legendre method, so when we begin at equilibrium, we will have the same $P_{\mathrm{B}_j}^\mathrm{tot}(t) = P_{\mathrm{B}}^\mathrm{tot}(t)/M$ for each $j$ of the $M=N_b$ paths, and therefore we may apply conservation and simplify Eq.~\eqref{PAevokinVBfinal} using

\begin{equation}\label{q3rewrgt3wedf}
	{\rho}(t)\equiv
	\frac{P_{\mathrm{B}_j}^\mathrm{tot}(t)}{P_{\mathrm{B}_j}^\mathrm{tot,eq}} = 
	\frac{P_{\mathrm{B}}^\mathrm{tot}(t)}{P_{\mathrm{B}}^\mathrm{tot,eq}} = 
	\dfrac{1 - \iiint {P}_\mathrm{A}(\boldsymbol{\xi};t)\,d^3\boldsymbol{\xi}}{1 - \iiint {P}_\mathrm{A}^\mathrm{eq}(\boldsymbol{\xi})\,d^3\boldsymbol{\xi}} = 
	\dfrac{1 - \iiint \mathscr{P}_\mathrm{A}(\boldsymbol{\gamma};t)\,d^3\boldsymbol{\gamma}}{1 - \iiint \mathscr{P}_\mathrm{A}^\mathrm{eq}(\boldsymbol{\gamma})\,d^3\boldsymbol{\gamma}}
	.
\end{equation}
Eqs.~\eqref{PAevokinVBfinal} and \eqref{q3rewrgt3wedf} allows us to retrieve a linear first order integro-partial differential equation for $\mathscr{P}_\mathrm{A}(\boldsymbol{\gamma};t)$, 

\begin{equation}\label{PAevokinVBfinalsinglek}
	\dfrac{\partial \mathscr{P}_\mathrm{A}(\boldsymbol{\gamma};t)}{\partial t} = 
	- \left[\dfrac{\partial \mathscr{P}_\mathrm{A}(\boldsymbol{\gamma};t)}{\partial\boldsymbol{\gamma}}\,\boldsymbol{\gamma}\right]:\mathbf{L}(t) 
	-k(\boldsymbol{\gamma})\left\{\mathscr{P}_\mathrm{A}(\boldsymbol{\gamma};t) - \dfrac{\mathscr{P}^\mathrm{eq}_\mathrm{A}(\boldsymbol{\gamma})}{P_{\mathrm{B}}^\mathrm{tot,eq}}\left[1 - \iiint \mathscr{P}_\mathrm{A}(\tilde{\boldsymbol{\gamma}};t)\,d^3\tilde{\boldsymbol{\gamma}}\right]\right\}
	,
\end{equation}
where $k(\boldsymbol{\gamma})\equiv N_b k'(\boldsymbol{\gamma})$ is the net reaction rate coefficient function.
The proportionality of the net rate of breaking $k(\boldsymbol{\gamma})$ to the number of links $N_b$ is the effect predicted by \citet{lake1967strength} and has been used by recent models \citep{lavoie2016rate}.
This Lake-Thomas effect is also observed in the manner of our Gibbs-Legendre approximation, which causes an equal force to be experienced across all links and the total energy to scale with $N_b$. 
Eq.~\eqref{PAevokinVBfinalsinglek} appears similar to several from the literature \citep{tanaka1992viscoelastic,vernerey2017statistically,brighenti2017rate,vernerey2018transient,guo2020mechanics}, but there are two fundamental differences. 
First, two evolution equations are typically written -- one for the normalized probability $\mathscr{P}_\mathrm{A}(\boldsymbol{\gamma};t)/P_{\mathrm{A}}^\mathrm{tot}(t)$ and another for the concentration $n P_{\mathrm{A}}^\mathrm{tot}(t)$ -- rather than the single evolution equation for $\mathscr{P}_\mathrm{A}(\boldsymbol{\gamma};t)$ in Eq.~\eqref{PAevokinVBfinalsinglek}.
Second, it is often assumed that one may prescribe both forward and reverse reaction rate coefficients independently, where the forward rate would be the same $k(\boldsymbol{\gamma})$ here, but the reverse rate would be $k_a(\boldsymbol{\gamma})\mathscr{P}^\mathrm{eq}_\mathrm{A}(\boldsymbol{\gamma})/P_{\mathrm{B}}^\mathrm{tot,eq}$.
Not only would any $k_a(\boldsymbol{\gamma})\neq k(\boldsymbol{\gamma})$ violate the statistical mechanical derivation we have outlined in Sec.~\ref{statmechsubsection}, namely Eq.~\eqref{Kjeq}, but it would also cause the equilibrium distribution of the evolution equation $[k_a(\boldsymbol{\gamma})/k(\boldsymbol{\gamma})]\mathscr{P}^\mathrm{eq}_\mathrm{A}(\boldsymbol{\gamma})$ to differ from the equilibrium distribution from statistical mechanics $\mathscr{P}^\mathrm{eq}_\mathrm{A}(\boldsymbol{\gamma})$.
Separately specifying the forward and reverse reaction rate coefficient functions then causes a thermodynamic inconsistency regardless of the single-chain model used.
Conceptually, the kinetic rate(s) at which chemical systems approach equilibrium should not change the equilibrium configuration, since statistical thermodynamics allows equilibrium configurations to be obtained independently of the kinetics.

We present the path to obtain the exact solution to Eq.~\eqref{PAevokinVBfinalsinglek} in \ref{appexact}. 
For $\mathscr{P}_\mathrm{A}(\boldsymbol{\gamma};t\leq 0)=\mathscr{P}_\mathrm{A}^\mathrm{eq}(\boldsymbol{\gamma})$ and $\mathbf{F}(t\leq 0)=\mathbf{1}$, the solution can be written as

\begin{equation}\label{PAsol}
	\mathscr{P}_\mathrm{A}(\boldsymbol{\gamma};t) = 
	\int_{-\infty}^t \mathscr{P}_\mathrm{A}^\mathrm{eq}\left[{}_{(t)}\mathbf{F}(\tau)\cdot\boldsymbol{\gamma}\right] k\left[{}_{(t)}\mathbf{F}(\tau)\cdot\boldsymbol{\gamma}\right] \exp\left\{-\int_\tau^t k\left[{}_{(t)}\mathbf{F}(s)\cdot\boldsymbol{\gamma}\right]\,ds\right\} {\rho}(\tau)\,d\tau
	,
\end{equation}
where the relative deformation \citep{paolucci2016continuum} is defined as ${}_{(t)}\mathbf{F}(\tau) \equiv \mathbf{F}(\tau)\cdot\mathbf{F}^{-1}(t)$, and where the solution for ${\rho}(t)$, consistent with its definition in Eq.~\eqref{q3rewrgt3wedf} and the solution for $\mathscr{P}_\mathrm{A}(\boldsymbol{\gamma};t)$ in Eq.~\eqref{PAsol}, is given in \ref{appexact}.
The stress from Eq.~\eqref{sigmafinal}, which in nondimensional form is

\begin{equation}\label{sigmanondim}
	\frac{\boldsymbol{\sigma}(t) + p(t)\mathbf{1}}{n/\beta} = 
	N_b\iiint \mathscr{P}_\mathrm{A}(\boldsymbol{\gamma}; t) \,\eta(\gamma) \left(\frac{\boldsymbol{\gamma}\boldsymbol{\gamma}}{\gamma}\right) \,d^3\boldsymbol{\gamma}
	,
\end{equation}
can now be evaluated at any time $t$.
For $k(\boldsymbol{\gamma})\propto k'(\boldsymbol{\gamma})$ and $\mathscr{P}_\mathrm{A}^\mathrm{eq}(\boldsymbol{\gamma})$ derived from a single-chain model, such as the Morse-FJC, our solution is guaranteed to converge.
If one wishes to instead prescribe positive semidefinite functions $k'(\boldsymbol{\gamma})$ and $\mathscr{P}^\mathrm{eq}(\boldsymbol{\gamma})$ independently of a chain model, our solution still holds as long as 

\begin{equation}\label{ineqkPintegral}
	\iiint k'(\boldsymbol{\gamma}) \mathscr{P}_\mathrm{A}^\mathrm{eq}(\boldsymbol{\gamma})\,d^3\boldsymbol{\gamma} < \infty
	.
\end{equation}
If this condition is not met, not only does our solution not hold, but there is no solution in general, which means numerical methods will also fail.
This condition in Eq.~\eqref{ineqkPintegral} also keeps the dissipation $\mathcal{D}_\mathrm{rxn}(t)$ in Eq.~\eqref{Drxnttt} finite.
Thus, given admissible $k'(\boldsymbol{\gamma})$, $\mathscr{P}_\mathrm{A}^\mathrm{eq}(\boldsymbol{\gamma})$, and $\mathbf{F}(t)$, we may evaluate $\mathscr{P}_\mathrm{A}(\boldsymbol{\gamma};t)$ in Eq.~\eqref{PAsol} when integrating for the stress in Eq.~\eqref{sigmanondim}.
Our results here in Sec.~\ref{Psolsubsubsec} can be applied to other chains that have all identical reaction pathways and are long enough to utilize the Gibbs-Legendre method, or to chains of any length that instead have only a single reaction pathway (i.e. a chain with a single weak link).

\subsection{Adjustments for inhomogeneous chains\label{inhomo_ssec}}

\begin{figure*}[t]
\begin{center}
$
\michaelarraysettings
\begin{array}{cc}
\includegraphics{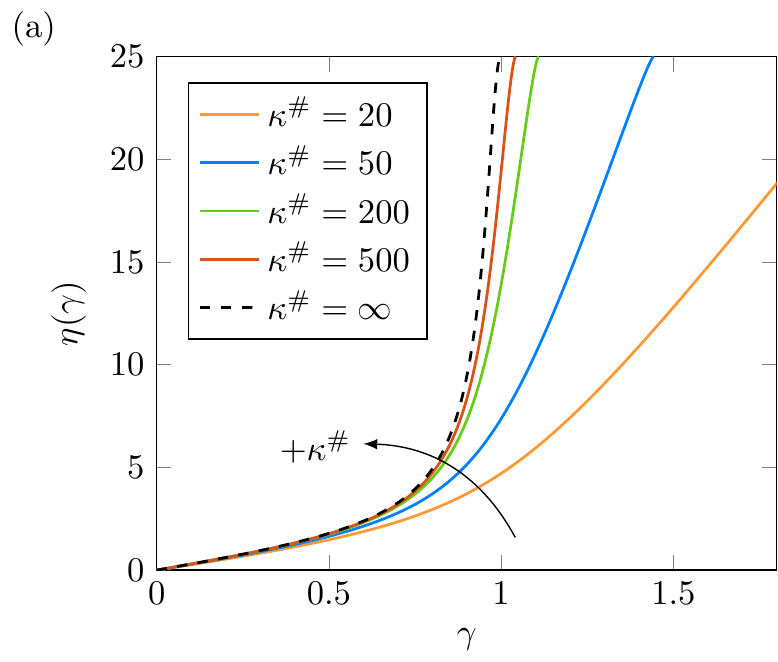}
& 
\includegraphics{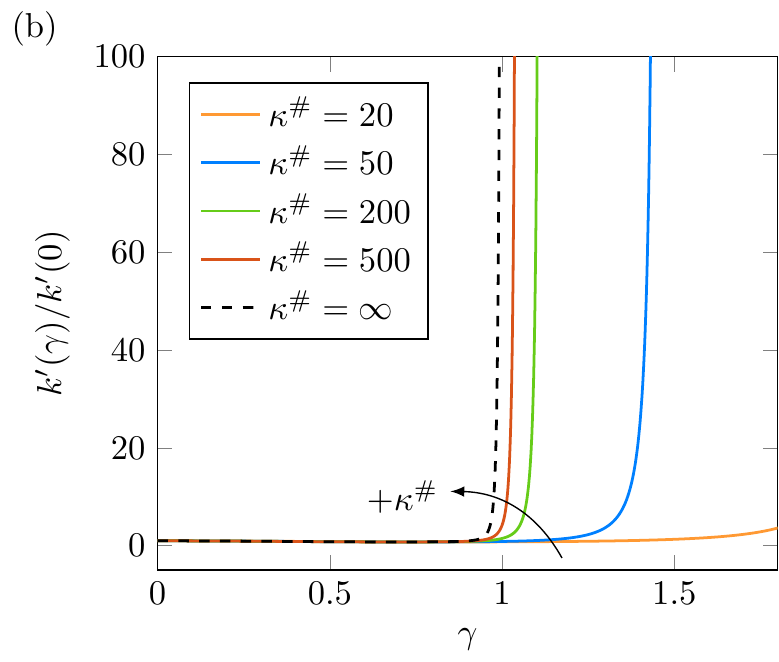}
\end{array}
$
\vspace{-5pt}
\end{center}
\caption{\label{figkappaH}
	Single-chain functions for the inhomogeneous Morse-FJC model with $N_b = 1$, $N_b^\# = 8$, $\kappa = 200$, $\beta u_b = 25$, and $\varsigma = 1$, while varying $\kappa^\#$.
	(a) The nondimensional force $\eta=\beta f\ell_b$ as a function of the nondimensional end-to-end length $\gamma=\xi/\ell_b(N_b+\varsigma N_b^\#)$.
	(b) The reaction rate coefficient function $k'(\gamma)$ scaled by its initial value $k'(0)$ as a function of $\gamma$.
}
\end{figure*}

Polymer networks are often synthesized to contain sacrificial bonds that are designed to break or activate before the rest of the bonds in the network \citep{ducrot2014toughening,clough2016covalent,wangtoughening}.
We can adjust our previous relations to accommodate these cases -- we consider the same $u$FJC model, but now with $N_b$ breakable links and $N_b^\#$ unbreakable links.
The unbreakable links are assumed to remain in the harmonic region (effectively EFJC), have rest-length $\ell_b^\#$, and nondimensional stiffness $\kappa^\#$. 
Due to the nature of the Gibbs-Legendre method, we may simply add these links onto our asymptotic approximations for the mechanical response and Helmholtz free energy.
Beginning with the mechanical response in Eq.~\eqref{gammasimplest}, for an inhomogeneous chain we now have

\begin{equation}\label{mechwithvarsigma}
	\gamma(\eta) \sim 
	\frac{N_b}{N_b + \varsigma N_b^\#}\left[\mathcal{L}(\eta) + \lambda(\eta)\right]
	+ \frac{\varsigma N_b^\#}{N_b + \varsigma N_b^\#}\left[\mathcal{L}(\varsigma\eta) + \frac{\varsigma\eta}{\kappa^\#}\right]
	,
\end{equation}
where $\varsigma\equiv\ell_b^\#/\ell_b$ is the ratio of rest-lengths of the two link types.
Note that the contour length of the chain that scales $\xi$ for $\gamma$ is now $N_b\ell_b+N_b^\#\ell_b^\# = N_b\ell_b(1+\varsigma N_b^\#/N_b)$.
Similarly, we adjust Eq.~\eqref{psiconufjc} for the nondimensional configurational Helmholtz free energy

\begin{equation}\label{psiconufjcinhomo}
	\left(N_b + N_b^\#\right)\vartheta_\mathrm{A,con}^*(\gamma) \sim 
	N_b\left[\ln\left\{\frac{\eta\exp[\eta\mathcal{L}(\eta)]}{\sinh(\eta)}\right\} + \beta u(\eta)\right]
	+ N_b^\#\left[\ln\left\{\frac{\varsigma \eta\exp[\varsigma \eta\mathcal{L}(\varsigma \eta)]}{\sinh(\varsigma \eta)}\right\} + \frac{(\varsigma\eta)^2}{2\kappa^\#} \right]
	,
\end{equation}
The equilibrium distribution $\mathscr{P}_\mathrm{A}^\mathrm{eq}(\gamma)$ is still given by Eq.~\eqref{PeqnicenoVb} after taking $N_b\vartheta_\mathrm{A,con}^*(\gamma)\mapsto(N_b+N_b^\#)\vartheta_\mathrm{A,con}^*(\gamma)$ and using Eq.~\eqref{psiconufjcinhomo}.
The forward rate coefficient function ${k}'(\gamma)$ is still given by Eqs.~\eqref{kBAfinal}--\eqref{DeltaPsiTS}, and the net reaction rate coefficient function is still $k(\gamma) = N_b k'(\gamma)$. 
The nondimensional stress from Eq.~\eqref{sigmanondim} is now

\begin{equation}\label{sigmanondimvarsigma}
	\frac{\boldsymbol{\sigma}(t) + p(t)\mathbf{1}}{n/\beta} = 
	\left(N_b + \varsigma N_b^\#\right)\iiint \mathscr{P}_\mathrm{A}(\boldsymbol{\gamma}; t) \,\eta(\gamma) \left(\frac{\boldsymbol{\gamma}\boldsymbol{\gamma}}{\gamma}\right) \,d^3\boldsymbol{\gamma}
	.
\end{equation}
The inhomogeneous single-chain mechanical response and reaction rate coefficient function is studied in Fig.~\ref{figkappaH} for varying $\kappa^\#$.
For these results we use $N_b = 1$, $N_b^\# = 8$, $\kappa = 200$, $\beta u_b = 25$, and $\varsigma=1$.
For $\kappa^\# < \kappa$, the mechanical response of the chain is dominated by the stretching of the unbreakable links, and the stiffer breakable link is stretched slowly due to the smaller forces reached per overall extension.
This in turn causes the reaction rate coefficient function to more gradually increase with extension since the breakable link will require larger chain extensions to experience the forces necessary for it to begin breaking.
For $\kappa^\# > \kappa$, the mechanical response of the chain is dominated by the stretching of the breakable link, where for $\gamma>1$ the chain extension becomes localized almost entirely in stretching the breakable link.
Due to this stretch localization, the reaction rate coefficient function begins to spike almost instantaneously at $\gamma=1$.
We also observe that for $\kappa^\# \gg \kappa$ we may instead take $\kappa^\#\to\infty$ to obtain an accurate FJC-based approximation, which is equivalent (after switching the Morse potential to the relevant potential) to many recent models \citep{mao2017rupture,mao2018fracture,li2020variational}.
Interestingly, combining $\kappa^\#\to\infty$ with $\varsigma\to\infty$ results in the FJC model that fails instantaneously for some $\gamma_c\lesssim 1$, which is a simplified form of the approach utilized by \citet{vernerey2018statistical}.

\section{Macroscopic results\label{MARCOPOLO}}

Now that we have implemented the $u$FJC single-chain model within our general theory, the resulting constitutive model is applied to predict the mechanics of polymer networks with bond breaking.
First, we consider two special cases (1) where chains break rate-independently and irreversibly, and (2) where chains dynamically break and reform in a transient manner.
In either case, we compare the theoretical results from the specialized models to experimental results from exemplary polymers in the literature.
Next, we demonstrate the features of the general model and examine the results in several parametric studies.
We then apply the general model to another polymer that is considered to have force-sensitive reversible chain breaking.

\subsection{Rate-independent irreversible breaking\label{speccaseirrev}} 

The reaction rate coefficient function $k'(\gamma)$ often behaves as being nearly constant at its initial value $k'(0)$ before suddenly becoming effectively infinite beyond some critical extension $\gamma_c$, as observed in Figs.~\ref{figk}--\ref{figkappaH}.
Physically, this corresponds to stiff but breakable links remaining unstretched until the entire chain is extended past the contour length, where the high and rapidly increasing force required for further extension almost immediately breaks the chain.
When thermal energy alone is insufficient to break the links, we neglect the force-free breaking by letting $k'(0)\to 0$, which then also implicitly neglects reforming.
Mathematically, the capability to neglect $k'(0)$ results from the nondimensional link energy $\beta u_b$ being sufficiently large in order to cause $k'(0)\propto e^{-\beta u_b}$ to become negligible compared to the slowest rate $1/\mathcal{T}$, where $\mathcal{T}$ is the total time of testing.
As shown in \ref{appspeccases}, the solution for the distribution of intact chains $\mathscr{P}_\mathrm{A}(\boldsymbol{\gamma};t)$ in this special case of rate independent irreversible breaking simplifies to

\begin{equation}\label{PAsolspecirrev}
	\mathscr{P}_\mathrm{A}(\boldsymbol{\gamma};t) = 
	\mathscr{P}_\mathrm{A}^\mathrm{eq}\left[\mathbf{F}^{-1}(t)\cdot\boldsymbol{\gamma}\right] \Theta(\boldsymbol{\gamma};t,0)
	,
\end{equation}
where the yield function $\Theta(\boldsymbol{\gamma};t,\tau)$ is defined as

\begin{equation}
	{\Theta(\boldsymbol{\gamma};t,\tau)} \equiv 
	\begin{cases}
	1, & \left\|{}_{(t)}\mathbf{F}(s)\cdot\boldsymbol{\gamma}\right\|_2 \leq \gamma_c ~~\forall s\in[\tau,t]
	,\\
	0, & \text{otherwise}
	.
	\end{cases}
\end{equation}
Similar forms of this special case have been considered previously, sometimes accounting for variability in the value of $\gamma_c$ \citep{vernerey2018statistical}.
Here, if we take $\gamma_c\to\infty$, chains never break and we retrieve $\mathscr{P}_\mathrm{A}(\boldsymbol{\gamma};t) = \mathscr{P}_\mathrm{A}^\mathrm{eq}\left[\mathbf{F}^{-1}(t)\cdot\boldsymbol{\gamma}\right]$, the expected solution for a network of non-breaking chains \citep{Buchestatistical2020}.

\begin{figure*}[t]
\begin{center}
$
\michaelarraysettings
\begin{array}{cc}
\includegraphics{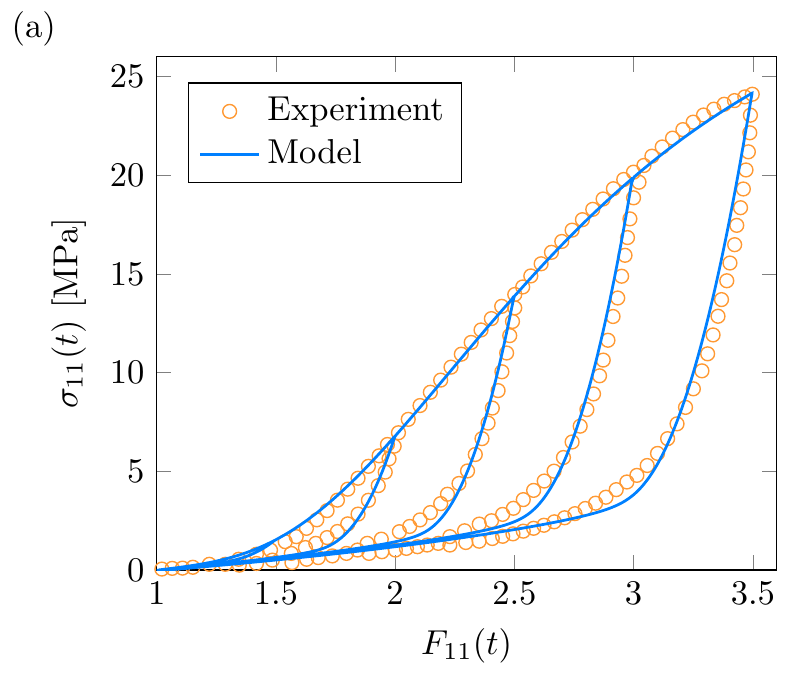}
& 
\includegraphics{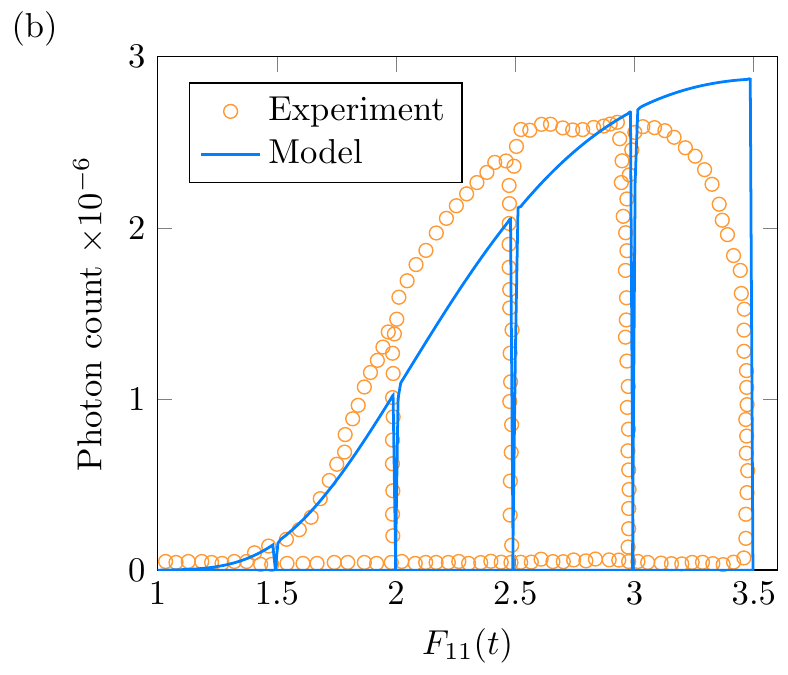}

\end{array}
$
\vspace{-5pt}
\end{center}
\caption{\label{figTN}
	(a) Mechanical response of the triple ethyl acrylate network (EA$_{0.5}$EAEA) of \citet{ducrot2014toughening} under cyclic uniaxial tension, and that predicted by the rate-independent irreversibly-breaking model.
	(b) Light emission from the sacrificial mechanoluminescent crosslinkers breaking within the first network in the same experiment, and that predicted by the model. 
}
\end{figure*}

This irreversible breaking is especially relevant when considering the sacrificial networks designed to break down within toughened elastomers.
These polymers experience noticeable hysteresis under cyclic deformation, exhibiting the Mullins effect \citep{mullins1948effect} and dissipating considerable amounts of energy \citep{webber2007large}.
Here, we consider the triple ethyl acrylate network (EA$_{0.5}$EAEA) of \citet{ducrot2014toughening}.
The first network (denoted EA$_{0.5}$) is synthesized using ethyl acrylate monomers and mechanoluminescent crosslinkers that are specifically weaker than the EA links, allowing the damage in the first network to be visualized.
The second network and third networks (each denoted EA) are synthesized by swelling the existing network in ethyl acrylate monomers and sparsely crosslinking them.
The resulting EA$_{0.5}$EAEA material, at a temperature of $T=20^\circ$C, was loaded in cyclic uniaxial tension while the stress and light emission were measured.
Repeat cycles showed negligible light emission and no change in stress, supporting the essential argument that chains in the network effectively break both irreversibly and rate-independently.
We model the first network as isotropically-swollen, with the volumetric swelling ratio $J=15.625$ known from the experiment.
The theory presented here may be quickly adjusted to account for this swelling: the equilibrium distribution under the swelling transforms as $\mathscr{P}_\mathrm{A}^\mathrm{eq}(\gamma)\mapsto\mathscr{P}_\mathrm{A}^\mathrm{eq}(J^{-1/3}\gamma)/J$ due to the isotropic swelling deformation $J^{1/3}\mathbf{1}$, where the factor of $J^{-1}$ preserves the total probability.
The number density of chains $n$ transforms under swelling as $n\mapsto n/J$, so if $n$ is known in the pre-swollen configuration, the nondimensional stress from Eq.~\eqref{sigmanondimvarsigma} under swelling is

\begin{equation}
	\frac{\boldsymbol{\sigma}(t) + p(t)\mathbf{1}}{n/\beta} = 
	\left(N_b + \varsigma N_b^\#\right) \iiint J^{-2} \mathscr{P}_\mathrm{A}(J^{-1/3}\boldsymbol{\gamma}; t) \,\eta(\gamma) \left(\frac{\boldsymbol{\gamma}\boldsymbol{\gamma}}{\gamma}\right) \,d^3\boldsymbol{\gamma}
	.
\end{equation}
The nondimensional modulus can be shown to be $3 J^{-1/3}$ for long chains, which is expected for the isotropic swelling of a network \citep{bacca2017model}.
The first network is modeled as a network of chains of $N_b=1$ irreversibly-breaking links and $N_b^\#$ unbreakable links.
The EA$_{0.5}$ material was reported to have a modulus of 0.6~MPa, which corresponds to $n/\beta=0.2$~MPa for the first network.
The second and third networks were over 100 times more sparsely crosslinked than the first, so we treat them as one effective filler network represented by the Neo-Hookean model, valid when chains are sufficiently long \citep{Buchestatistical2020}.
Since the EA$_{0.5}$EAEA material was reported to have a modulus of 1.5~MPa, we obtain $n/\beta=0.3$~MPa for the filler network.
The average number of monomers between mechanoluminescent crosslinkers in the first network was approximately 34, and the crosslinker itself offers some additional effective monomers, so we use $N_b^\#=38$.
We take $\beta u_b = 61.57$ corresponding to 150~kJ/mol, the zero-force bond energy of the mechanoluminescent crosslinker \citep{clough2016covalent}.
We find a good fit to the overall mechanical response for $\kappa = 9000$, $\kappa^\# = 6000$, and $\varsigma=4$, where $\varsigma>1$ represents that the bond breaking within the mechanoluminescent crosslinker is short compared to the monomer backbone length.
These large stiffness values and the small effective Kuhn length resulting from a link representing a single monomer are reasonable, as similar parameters have been used to fit the EFJC model to AFM experimental results for other acrylate chains \citep{grebikova2014single,grebikova2016recording}.
The critical chain extension $\gamma_c = 1.17$ results from the intact limit of the chain at $\eta_\mathrm{max} = 263.18$.
The predicted mechanical response of the material under cyclic uniaxial tension is shown in Fig.~\ref{figTN}(a) along with the experimental results.
We find a good overall agreement between the prediction of our theory and the mechanical response of the material but find some difficulty precisely capturing the unloading curve shape, a somewhat common issue when modeling this material \citep{bacca2017model,vernerey2018statistical,lavoie2019continuum}.
In addition to the mechanical response, we use our theory to predict the light intensity measured experimentally as the mechanoluminescent crosslinkers break by assuming that the intensity is proportional to the rate at which chains break in the first network \citep{lavoie2019continuum}.
The intensity would then be proportional to $-\tfrac{d}{dt}P_\mathrm{A}^\mathrm{tot}(t)$, and we find a good fit for a proportionality constant of $10^9$~photon$\cdot$seconds.
The theoretical prediction is shown in Fig.~\ref{figTN}(b) along with the experimental results, where we observe reasonable agreement but an important shape difference.
Specifically, our theory predicts a more gradual breakdown of the first network than is observed in the experiment.
There are several possible explanations for this discrepancy that we cannot distinguish among within the current framework.
First, it could be due to the polydispersity that is present within the first network which has been modeled here as effectively monodisperse.
Second, the assumption that the distribution of chains in the first network swell isotropically as the filler network is introduced could be invalid for this large degree of swelling.
Third, the breaking of chains in the first network could induce a more complicated damage mechanism that involves the filler networks \citep{morovati2020necking}.
Fourth, the discrepancy could be related to transfer reactions that create additional crosslinks during synthesis \citep{ducrot2014toughening}.

\subsection{Transient breaking} 

When the initial value of the net reaction rate coefficient function $k_0\equiv k(0)$ is appreciable and the critical extension $\gamma_c$ is large enough to be neglected, we obtain a constant reaction rate coefficient $k(\gamma) = k_0$ over all chain extensions.
This idea is traditionally referred to as the transient network model \citep{tanaka1992viscoelastic,vernerey2017statistically}.
As shown in \ref{appspeccases}, the solution for the distribution of intact chains $\mathscr{P}_\mathrm{A}(\boldsymbol{\gamma};t)$ in this special case simplifies to

\begin{equation}\label{PAsoltransient}
	\mathscr{P}_\mathrm{A}(\boldsymbol{\gamma};t) = 
	\int_{-\infty}^t \mathscr{P}_\mathrm{A}^\mathrm{eq}\left[{}_{(t)}\mathbf{F}(\tau)\cdot\boldsymbol{\gamma}\right] k_0 e^{-k_0(t-\tau)} \,d\tau
	.
\end{equation}
The transient network model has been considered previously in several forms, such as for a Gaussian distribution of freely-jointed chains \citep{vernerey2017statistically} or the Neo-Hookean representation retrieved when using ideal chains \citep{meng2019elasticity}.
The solution in Eq.~\eqref{PAsoltransient}, however, is more general since it is independent of single-chain model.
We would like to emphasize several vital features of this special transient case.
First, we are now limited to the utilization of single-chain models that are infinitely extensible since we have neglected $\gamma_c$.
Here we will simply replace the $u$FJC model with the EFJC model, which is infinitely extensible.
Second, note that we have taken $\rho(t)=1$, meaning that the total fraction of intact chains $P_\mathrm{A}^\mathrm{tot}(t)$ remains constant at $P_\mathrm{A}^\mathrm{tot,eq}$ for all time, which is true here (see \ref{appspeccases}).
The decoupling of single-chain kinetics from chain extension and the constant fraction of intact chains allows us to write the stress as 

\begin{equation}\label{stresswithstressy}
	\boldsymbol{\sigma}(t) = 
	\int_{-\infty}^t {}_{(t)}{\boldsymbol{\sigma}}(\tau) \, k_0 e^{-k_0(t-\tau)} \,d\tau
	,
\end{equation}
where ${}_{(t)}{\boldsymbol{\sigma}}(\tau)$ is the stress that the network, if permanent, experiences under the relative deformation ${}_{(\tau)}\mathbf{F}(t)$. 
Using Eq.~\eqref{sigmanondim}, we write this stress in nondimensional form as

\begin{equation}\label{stressy}
	\frac{{}_{(t)}{\boldsymbol{\sigma}}(\tau) + {}_{(t)}p(\tau)\mathbf{1}}{n/\beta} = 
	N_b \iiint \mathscr{P}_\mathrm{A}^\mathrm{eq}\left[{}_{(t)}\mathbf{F}(\tau)\cdot\boldsymbol{\gamma}\right] \,\eta(\gamma) \left(\frac{\boldsymbol{\gamma}\boldsymbol{\gamma}}{\gamma}\right) \,d^3\boldsymbol{\gamma}
	.
\end{equation}
This simplification allows us to eliminate considerable computational expense: for a given model, we may interpolate from tabulated values of ${}_{(t)}\mathbf{F}(0)$ and ${}_{(t)}{\boldsymbol{\sigma}}(0)$ in order to rapidly perform the integration over the history in Eq.~\eqref{stresswithstressy}.
In the ideal chain limit ($N_b\to\infty$) we obtain the Neo-Hookean model \citep{Buchestatistical2020}, where the right-hand side of Eq.~\eqref{stressy} becomes ${}_{(\tau)}\mathbf{F}(t) \cdot {}_{(\tau)}\mathbf{F}^T(t)$.

\begin{figure*}[t]
\begin{center}
$
\michaelarraysettings
\begin{array}{cc}
\includegraphics{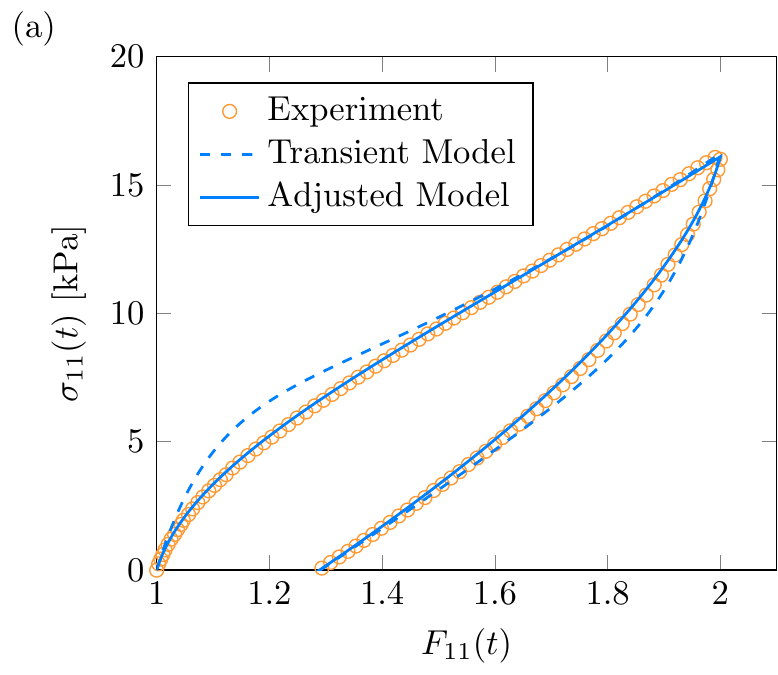}
& 
\includegraphics{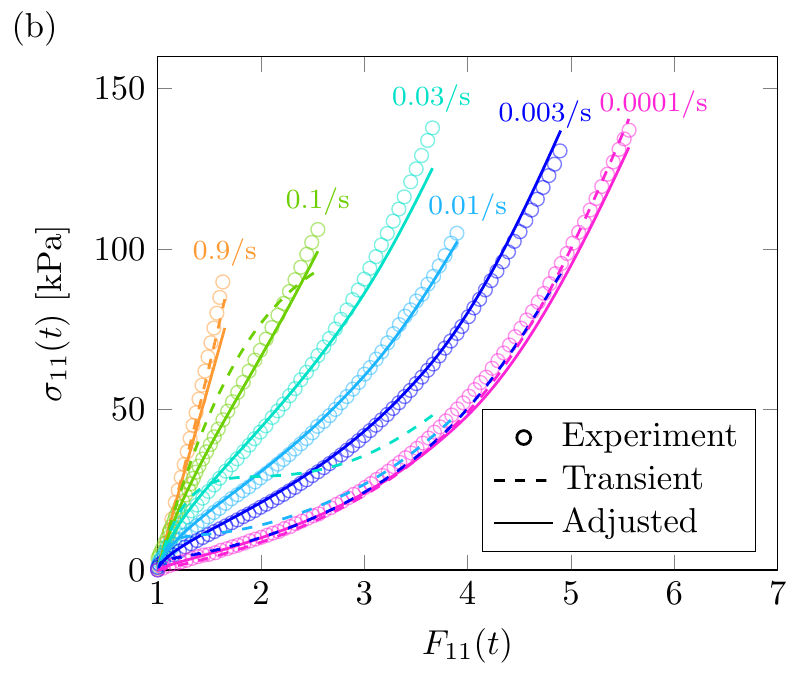}
\end{array}
$
\vspace{-5pt}
\end{center}
\caption{\label{figPVAlong}
	(a) Mechanical response of the PVA gel in \citet{long2014time} under cyclic uniaxial tension at a rate of 0.03/s, with predictions provided by the transient network model and the model adjusted to use a relaxation function. 
	(b) Mechanical response of the PVA gel of \citet{mayumi2013stress} under monotonic uniaxial tension at varying rates, with predictions provided again by both models.
}
\end{figure*}

It is common to consider a network that consists of both permanent and transiently-bonded chains \citep{hui2012constitutive}.
In this case, after introducing the fraction of permanent chains $0\leq x_p\leq 1$, Eq.~\eqref{stresswithstressy} becomes

\begin{equation}\label{transientxp}
	\boldsymbol{\sigma}(t) =
	x_p \, {}_{(t)}{\boldsymbol{\sigma}}(0) + (1 - x_p)\int_{-\infty}^t {}_{(t)}{\boldsymbol{\sigma}}(\tau) \, k_0 e^{-k_0(t-\tau)} \,d\tau
\end{equation}
We apply Eq.~\eqref{transientxp} under the ideal chain limit to the polyvinyl alcohol (PVA) gel with both permanent and transient crosslinks in \citet{long2014time} using $k_0 = 0.37$/s, $x_p=16$\%, and $n/\beta=24.15$~kPa.
Two cycles of uniaxial tension at a rate of 0.03/s are applied to the material with $\mathcal{T}_\mathrm{wait} = 30$~min between cycles.
Since $k_0\mathcal{T}_\mathrm{wait} \gg 1$, our transient model will fully relax between cycles and exactly repeat the mechanical response of the first cycle, as observed in experiment.
The results in Fig.~\ref{figPVAlong}(a) indicate show reasonable overall agreement, but there are discrepancies near the beginning of the loading and unloading potion of the cycles.
These results seem to indicate that the single-timescale approach of the transient network model is effective at short and long times, but that many timescales are required to capture the full material behavior.
In order to generalize for many timescales, we can adjust Eq.~\eqref{transientxp} to

\begin{equation}
	\boldsymbol{\sigma}(t) = 
	x_p \, {}_{(t)}{\boldsymbol{\sigma}}(0) + (1 - x_p)\int_{-\infty}^t {}_{(t)}{\boldsymbol{\sigma}}(\tau) \frac{\partial g(t -\tau)}{\partial\tau} \,d\tau
	,
\end{equation}
where $g(t)$ is any relaxation function. 
Taking $g(t)=e^{-k_0t}$ recovers the original transient network model.
We instead utilize a non-exponential relaxation function \citep{long2014time} which in effect represents many timescales,

\begin{equation}\label{gLongfun}
	g(t) = \left[1 + (\alpha - 1)\,\frac{t}{t_R}\right]^{1/(1 - \alpha)}
	,
\end{equation}
where $\alpha>1$ and $t_R$ is the characteristic bond breaking time. 
We utilize the parameters reported by \citet{guo2016mechanics} for our adjusted model, which in our case are $\alpha=2.6$, $t_R=0.6$~s, $x_p=10$\% and $n/\beta=24.15$~kPa.
The resulting mechanical response in Fig.~\ref{figPVAlong}(a) is a near-perfect fit and a substantial improvement upon the transient model.

The PVA gel we are considering was studied under large deformation and over 4~orders of magnitude of strain rates by \citet{mayumi2013stress}.
Our parameters change here since this material system is sensitive to synthesis conditions, often causing mterial parameters to change from batch-to-batch \citep{long2014time}.
These experiments were modeled by \citet{guo2016mechanics}, from which we obtain the relaxation function parameters $\alpha=1.99$, $t_R=3.23$~s, $x_p=4.68$\% and $n/\beta=37.78$~kPa.
We utilize Eqs.~\eqref{stresswithstressy} and \eqref{stressy} with the EFJC model, fitting $N_b=50$ and $\kappa = 40$ to the exponential hardening model used by \citet{guo2016mechanics}. 
For the transient network model we use the same $N_b=50$ and $\kappa = 40$, but find $x_p=5.85$\% and $k_0 = 0.12$/s by fitting to the 0.0001/s and 0.1/s rate results, respectively.
The modeling results in each case are shown in Fig.~\ref{figPVAlong}(b) with the experimental data.
Overall, the adjusted model provides a reasonable prediction of the mechanical response over all strain rates. 
We find that the transient network model tends to perform poorly here at intermediate rates and thus in modeling the mechanical response of more dynamic networks at intermediate timescales.
We attribute this to the insufficiency of the single timescale of the transient model in capturing the many timescales of the material observed in experiment.

\subsection{General behavior\label{genbessec}}

\begin{figure*}[t!]
\begin{center}
$
\michaelarraysettings
\begin{array}{cc}
\includegraphics{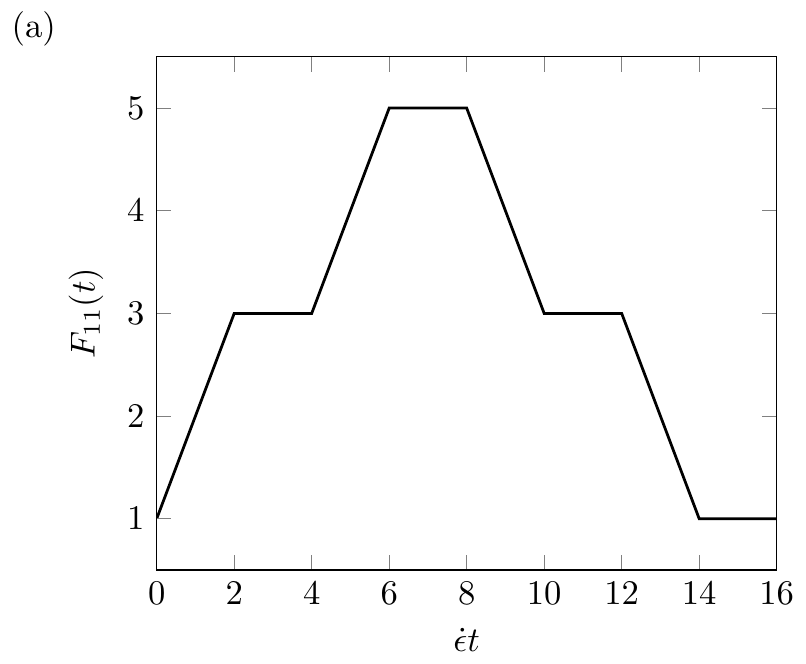}
& 
\includegraphics{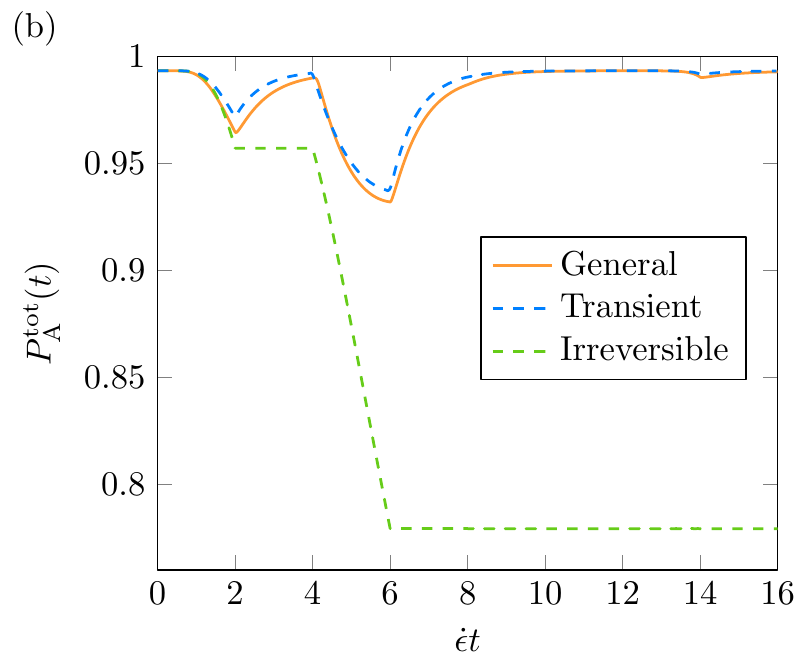}
\\
\includegraphics{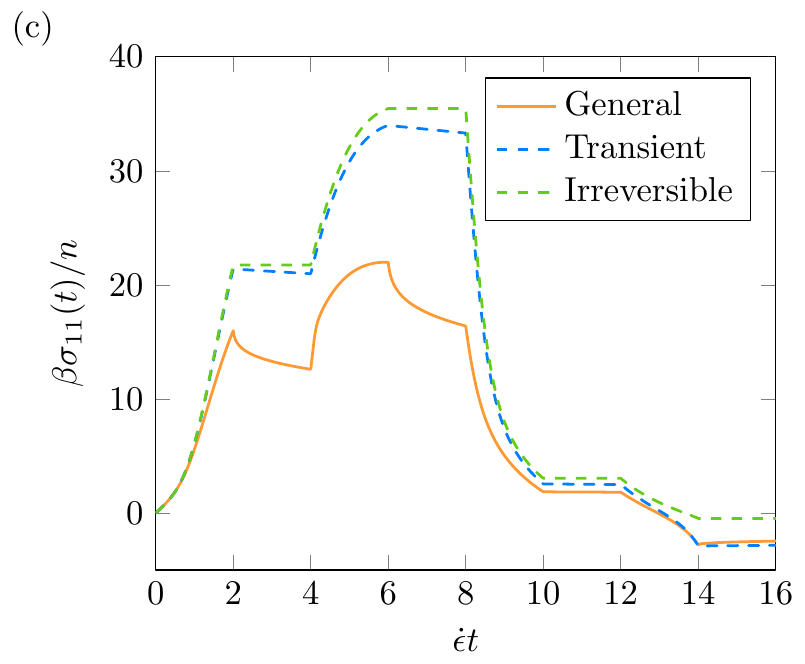}
&
\includegraphics{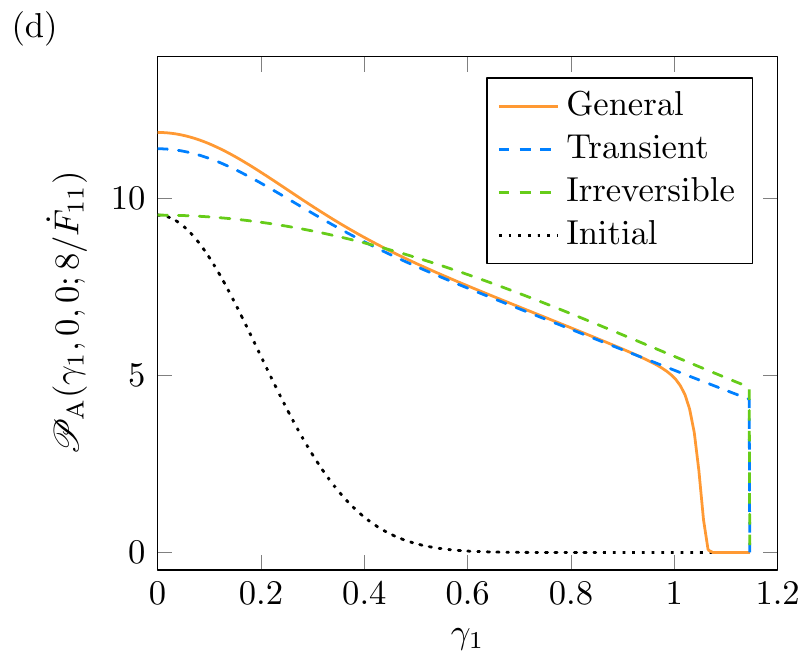}
\end{array}
$
\vspace{-5pt}
\end{center}
\caption{\label{figstepping}
	(a) The applied deformation $F_{11}(t)$ as a function of nondimensional time $\dot{\epsilon}t$, where $\dot{\epsilon}$ is a constant and the stress is uniaxial tension.
	(b) The total probability that a chain is intact, $P_\mathrm{A}^\mathrm{tot}(t)$, as function of nondimensional time for the general model and the two special cases. 
	(c) The applied nondimensional stress $\beta\sigma_{11}(t)/n$ as a function of nondimensional time for the same cases.
	(d) The distribution of intact chains $\mathscr{P}_\mathrm{A}(\boldsymbol{\gamma};t)$ aligned with the loading direction (along the $\gamma_1$-axis) at the halfway point $t=8/\dot{\epsilon}$.
	The nondimensional parameters are $N_b=1$, $N_b^\#=8$, $\kappa=200$, $\kappa^\#=500$, $\varsigma=1$, $\beta u_b=100$, $\beta\Delta\Psi_0=5$, and $k_0/\dot{\epsilon}=1/100$, with $\gamma_c=\gamma_\mathrm{max}=1.146$ for the special cases.
}
\end{figure*}

We now examine the behavior of the general model in comparison to the special cases we have just outlined, rate-independent reversible breaking and transient breaking.
The Morse-FJC model is used in each case with the same parameters, apart from the approximations made to the reaction rate coefficient function for the special cases.
The critical extension $\gamma_c$ for both special cases is taken to be $\gamma_\mathrm{max}=1.146$, the maximum extension where the chain remains intact.
For the version of the transient model that does not neglect $\gamma_c$, see \ref{appspeccases}.
Using an exemplary set of parameters, we apply a series of ramps of rate $\dot{\epsilon}$ and holds to the deformation gradient, as shown in Fig.~\ref{figstepping}(a) as a function of the nondimensional time $\dot{\epsilon}t$.
The total probability that a chain is intact, $P_\mathrm{A}^\mathrm{tot}(t)$, under this deformation is shown in Fig.~\ref{figstepping}(b).
While a similar fraction of chains break under the first loading period for each case, a significant amount of reforming occurs under the following holding period for the general and transient case, in contrast with the irreversible case where reforming is neglected.
In all cases, the unloading periods in the second half of the deformation history break a negligible fraction of chains.
Overall, the transient case seems to provide a reasonable approximation of the general case for $P_\mathrm{A}^\mathrm{tot}(t)$ here. 
The stress in Fig.~\ref{figstepping}(c), however, is substantially different for the two special cases versus the general case.
Due to their neglect of the nontrivial shape of $k(\gamma)$, either special case underestimates the fraction of high-extension chains being broken and therefore tends to overestimate the stress.
The reforming process negligibly affects the stress here since chains tend to reform towards $\mathscr{P}_\mathrm{A}^\mathrm{eq}(\boldsymbol{\gamma})$, i.e. the stress-free configuration, leading to a surprisingly similar stress response for the transient and irreversible models.
During the holding periods we observe much more substantial stress relaxation in the general case compared to little in the transient case, which is again due to $k(\gamma)$ providing a region where chains break quickly but not instantaneously.
This effect is clearly apparent from examining $\mathscr{P}_\mathrm{A}(\gamma_1,0,0;t)$ after the second hold period in Fig.~\ref{figstepping}(d), where $\gamma_1$ is the component of $\boldsymbol{\gamma}$ aligned with the loading direction.
Fig.~\ref{figstepping}(d) shows that the special cases overestimate the fraction of chains at larger extensions and therefore both the stress and the rate of breaking chains.
This higher rate of breaking in the general case additionally provides a higher rate of reforming, which is why the general case shows the highest probability of chains near $\gamma=0$.

\subsubsection*{Parametric studies}

Next we conduct parametric studies to understand the dependencies of the general model.
In each case we apply one cycle of uniaxial tension with strain rate $\dot{\epsilon}$ and maximum stretch 9, plotting the results as a function of the nondimensional time $\dot{\epsilon}t$.
The nondimensional base parameters are $N_b=1$, $N_b^\#=8$, $\kappa=200$, $\kappa^\#=500$, $\varsigma=1$, $\beta u_b=100$, $\beta\Delta\Psi_0=5$, and $k_0/\dot{\epsilon}=1/100$.
First, we vary the nondimensional initial reaction rate $k_0/\dot{\epsilon}$, or equivalently, the nondimensional strain rate $\dot{\epsilon}/k_0$.
Fig.~\ref{figmacroparam1}(a) shows the nondimensional stress $\beta\sigma_{11}(t)/n$ as a function of the nondimensional time $\dot{\epsilon}t$.
As $k_0$ increases relative to $\dot{\epsilon}$, we see a decrease in the time it takes for the network to yield as chains break more rapidly.
Increasing $k_0$ also causes broken chains to reform more rapidly, which causes an increasing amount of compressive stress when fully returning the network to zero strain.
This compressive stress results from chains reforming towards their stress-free equilibrium distribution while the deformation is still being applied to the network.
The rate-independent irreversibly-breaking special case ($k_0=0$) is shown for reference, which still differs considerably from the lowest $k_0$ case where $k_0/\dot{\epsilon}=10^{-4}$.
For additional insight, we plot the fraction of intact chains within the network $P_\mathrm{A}^\mathrm{tot}(t)$ as a function of the nondimensional time $\dot{\epsilon}t$ in Fig.~\ref{figmacroparam1}(b).
As $k_0$ increases relative to $\dot{\epsilon}$, we see that chains reform more rapidly and correspondingly, more chains are intact at any time.
When $k_0$ becomes sufficiently large it appears that $P_\mathrm{A}^\mathrm{tot}(t)$ remains approximately constant at its equilibrium value $P_\mathrm{A}^\mathrm{tot,eq}$, however, this is not exactly true (see \ref{appspeccases}).
We see that even when $k_0$ is quite small compared to $\dot{\epsilon}$, reforming still takes place after a sufficient amount of chains are broken.
This is because the total reforming rate is proportional to $\rho(t)=[1 - P_\mathrm{A}^\mathrm{tot}(t)]/P_\mathrm{B}^\mathrm{tot,eq}$, which spikes if appreciable amount chains break when $P_\mathrm{B}^\mathrm{tot,eq}$ is small, e.g. here where we have $P_\mathrm{B}^\mathrm{tot,eq} = 6.69\times 10^{-3}$.

Second, we examine the behavior of the general model while varying the nondimensional stiffness of the unbreakable links $\kappa^\#$ and keeping that of the breakable link constant, $\kappa=200$.
Referring back to the single-chain mechanical response and the reaction rate coefficient function in Fig.~\ref{figkappaH}, 
we recall that while increasing $\kappa^\#$ stiffens the chain near its full extension, it causes the chain to break more rapidly and reduces the maximum extensibility due to an increasing amount of strain localization in the breakable link.
Fig.~\ref{figmacroparam1}(c) shows the nondimensional stress $\beta\sigma_{11}(t)/n$, where we see that the more rapid breaking resulting from increasing $\kappa^\#$ is manifested as more rapid yielding of the network.
While larger $\kappa^\#$ causes the stress to increase at small deformations, smaller $\kappa^\#$ enables much higher stresses to be reached at large deformations due to increased maximum chain extensibility.
The fraction of intact chains, shown in Fig.~\ref{figmacroparam1}(d), verifies that as $\kappa^\#$ increases, chains break more rapidly as the deformation is applied.
Overall, these results illustrate that for a given breakable bond, maximizing single-chain extensibility may be much more effective than maximizing chain backbone stiffness when it comes to strengthening networks.
These results additionally illustrate that force-driven breaking of chains within the network is substantially increased after ensuring that the breakable bond is far less stiff than the rest of the chain.
Lastly, we find that for $\kappa^\#\gg \kappa$ utilizing the relevant rigid-constraint single-chain model ($\kappa^\#=\infty$, which is the FJC model here) captures both single-chain results and macroscopic-level results.

Third, we examine the behavior of the general model while varying the number of unbreakable links $N_b^\#$.
We keep the number of breakable links $N_b=1$ constant, so we effectively vary the chain length.
Our results are independent of whether the number density of chains, $n$, or number density of unbreakable links, $nN_b^\#$, is kept constant as $N_b^\#$ increases since we use the nondimensional stress.
As $N_b^\#$ increases, the nondimensional stress, shown in Fig.~\ref{figmacroparam1}(e), follows a similar trend we observed in Fig.~\ref{figmacroparam1}(c) when decreasing $\kappa^\#$.
Longer chains require less force to have the same end-to-end length as shorter chains, and additionally the average nondimensional end-to-end length at equilibrium decreases as chains become longer \citep{rubinstein2003polymer, Buchestatistical2020}.
Combined, these two effects allow a network of longer chains to deform more without breaking down and thus reach higher nondimensional stresses without yielding.
This is verified by examining the fraction of intact chains, shown in Fig.~\ref{figmacroparam1}(f), where we see less overall breaking as chains become longer.
The results in the ideal chain limit ($N_b^\#=\infty$) are also shown in both Figs.~\ref{figmacroparam1}(e) and \ref{figmacroparam1}(f), and are equivalent to the results obtained when using the Neo-Hookean model \citep{Buchestatistical2020}.
As $N_b^\#$ increases, the results of the general model matches the Neo-Hookean model for an increasing amount of time, but continued deformation of the network always causes the two to diverge as the finite-length chains in the general model stiffen and begin to break.

\afterpage{
	\begin{figure*}[th!]
	\begin{center}
	$
	\michaelarraysettings
	\begin{array}{cc}
	\includegraphics{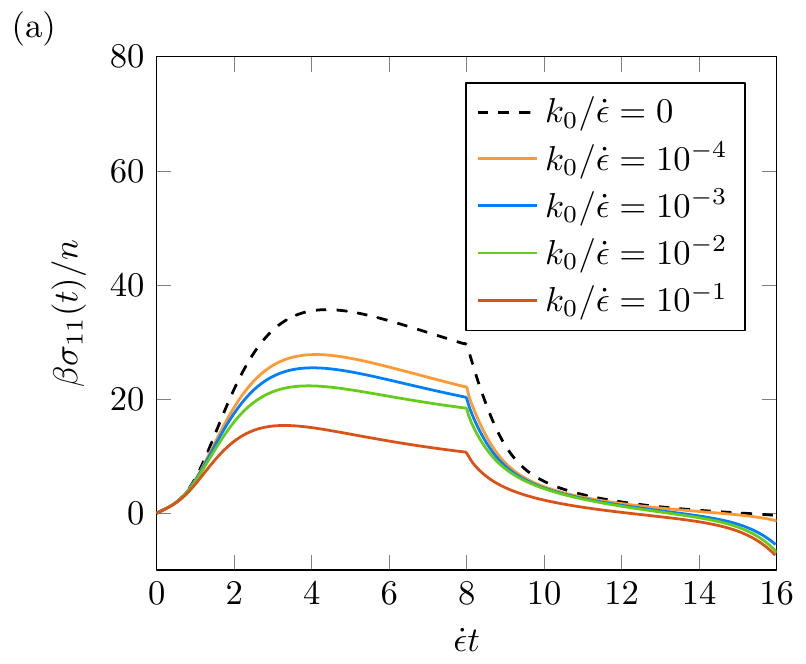}
	& 
	\includegraphics{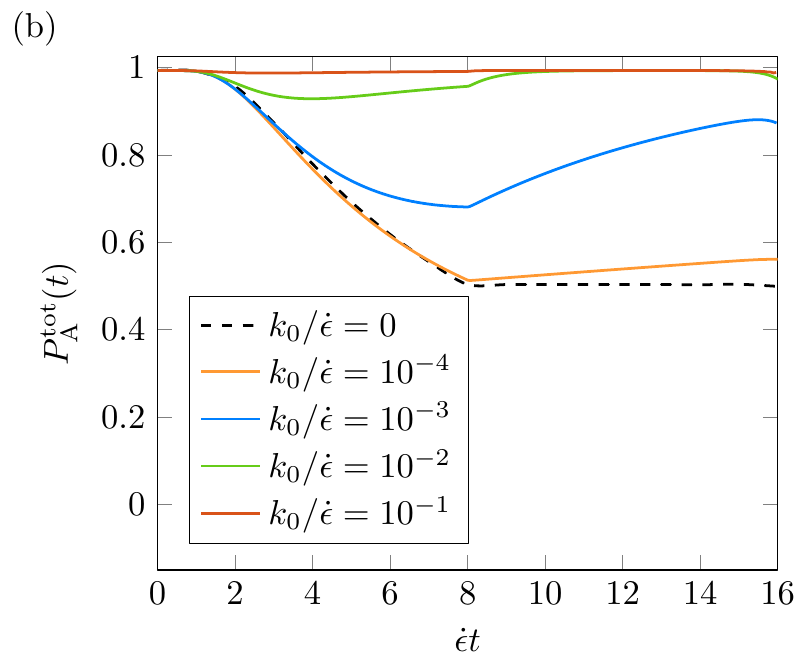}
	\\
	\includegraphics{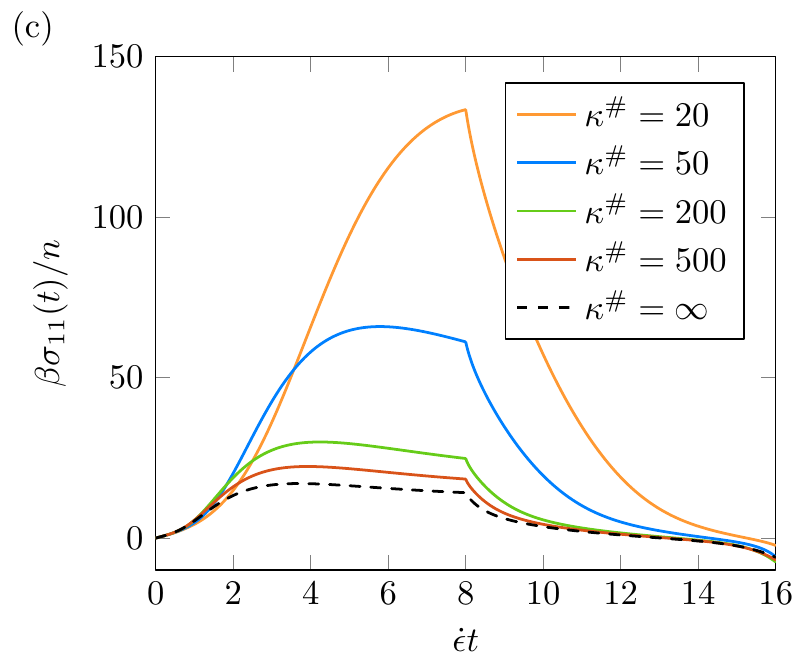}
	& 
	\includegraphics{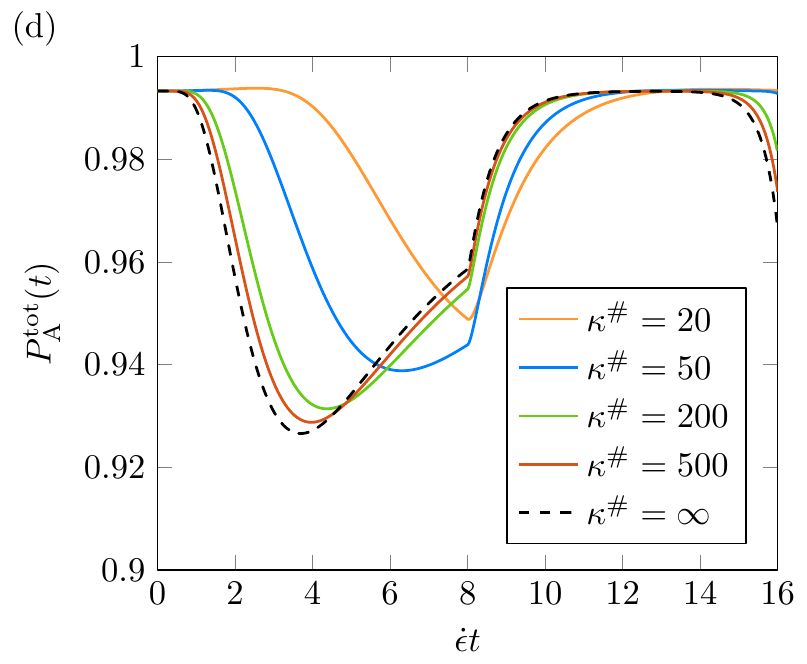}
	\\
	\includegraphics{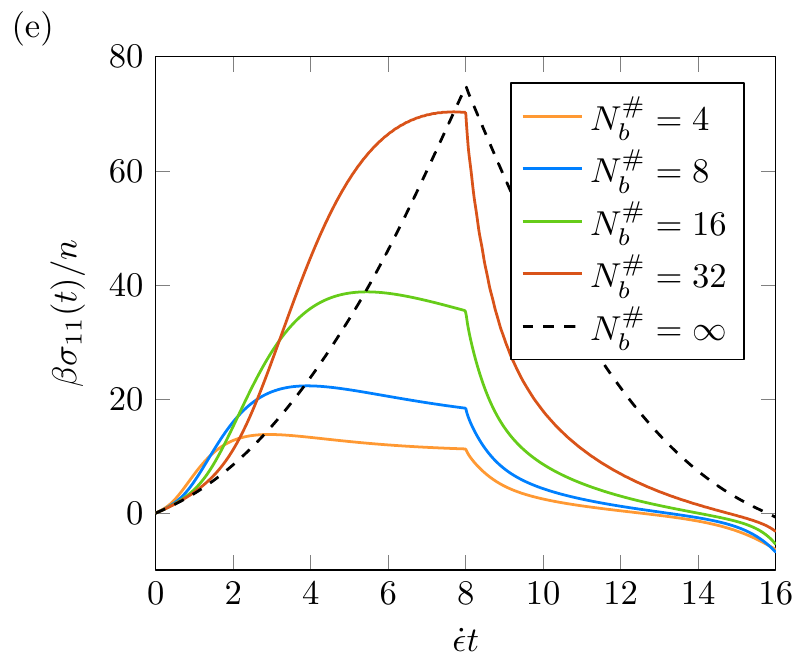}
	& 
	\includegraphics{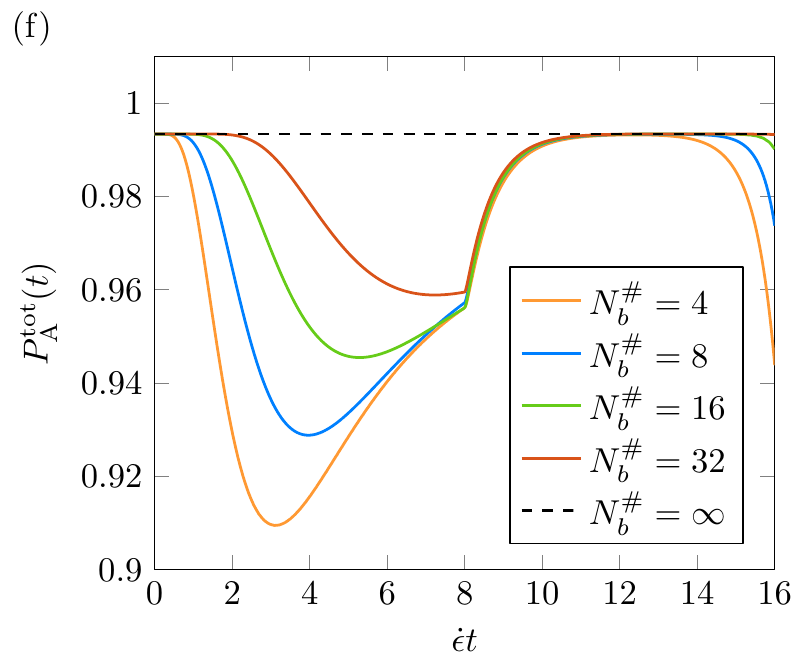}
	\end{array}
	$
	\vspace{-5pt}
	\end{center}
	\caption{\label{figmacroparam1}
		Parametric studies concerning rates and the unbreakable links, where we vary
		(a,b) $k_0/\dot{\epsilon}$, the nondimensional initial reaction rate,
		(c,d) $\kappa^\#$, the nondimensional stiffness of the unbreakable links, and
		(e,f) $N_b^\#$, the number of unbreakable links, while keeping the number of breakable links, $N_b=1$, constant.
		For one cycle of uniaxial monotonic tension, the nondimensional stress, $\beta\sigma_{11}(t)/n$, and total probability that a chain is intact, $P_\mathrm{A}^\mathrm{tot}(t)$, are plotted as a function of the nondimensional time $\dot{\epsilon}t$.
	}
	\end{figure*}
\newpage
	\begin{figure*}[th!]
	\begin{center}
	$
	\michaelarraysettings
	\begin{array}{cc}
	\includegraphics{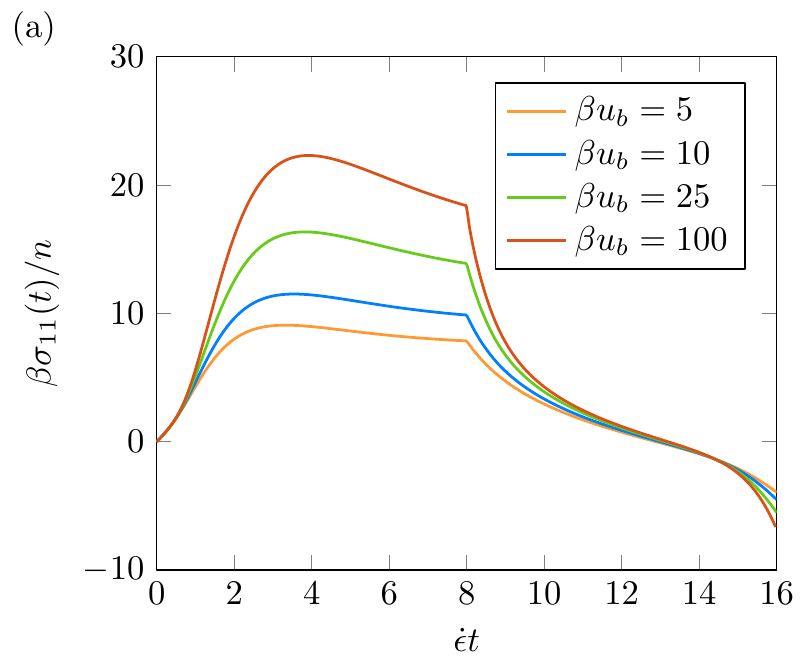}
	& 
	\includegraphics{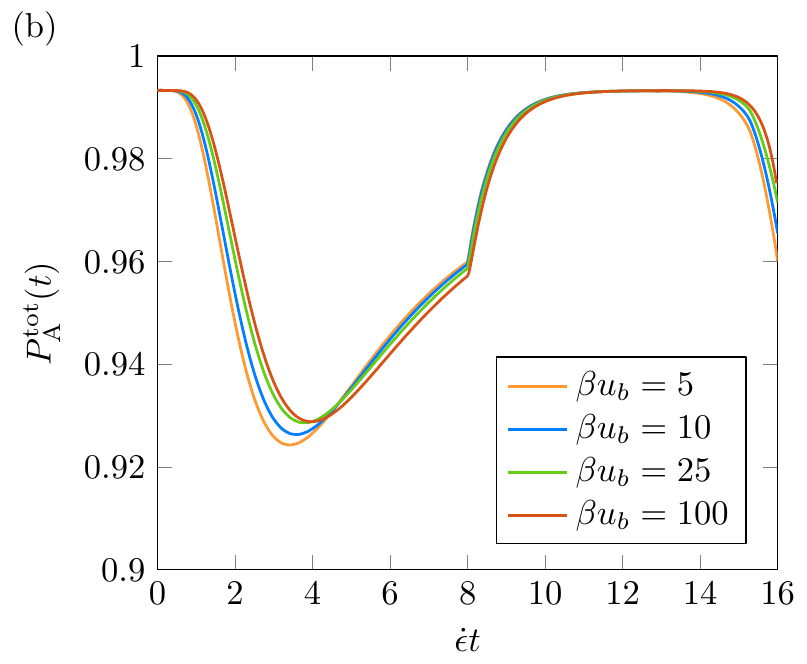}
	\\
	\includegraphics{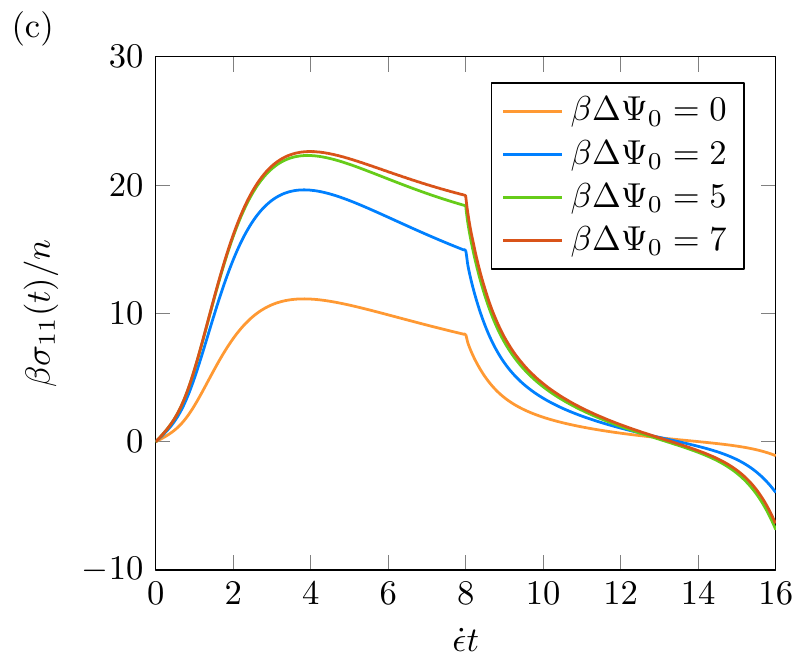}
	& 
	\includegraphics{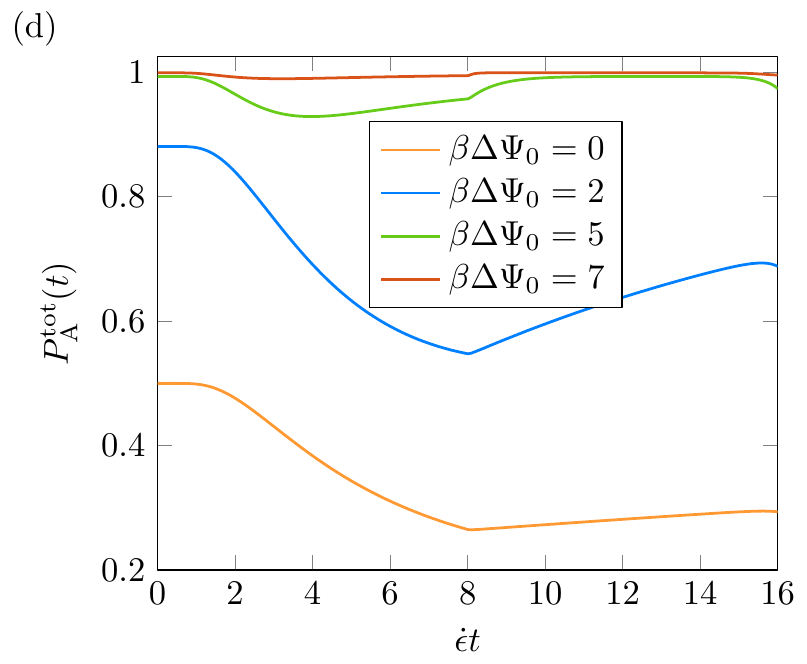}
	\\
	\includegraphics{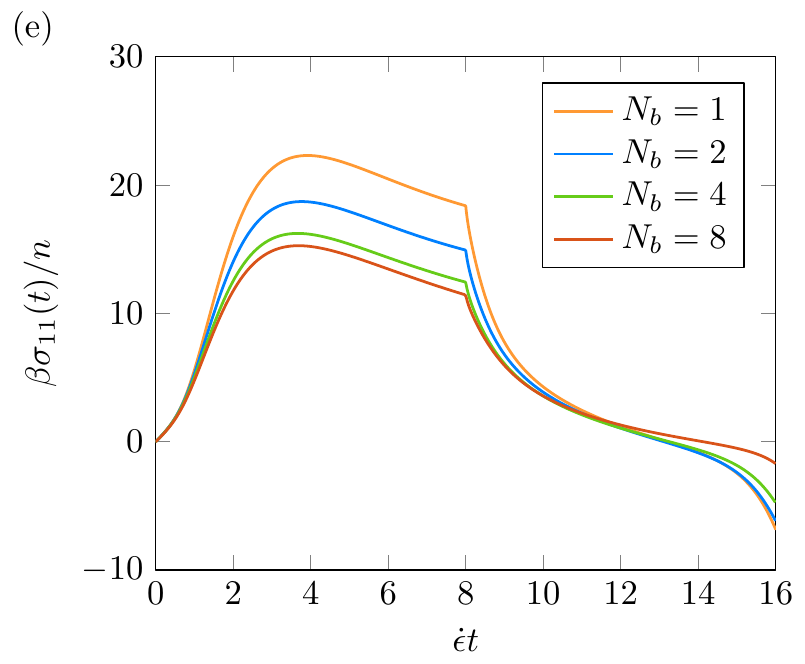}
	& 
	\includegraphics{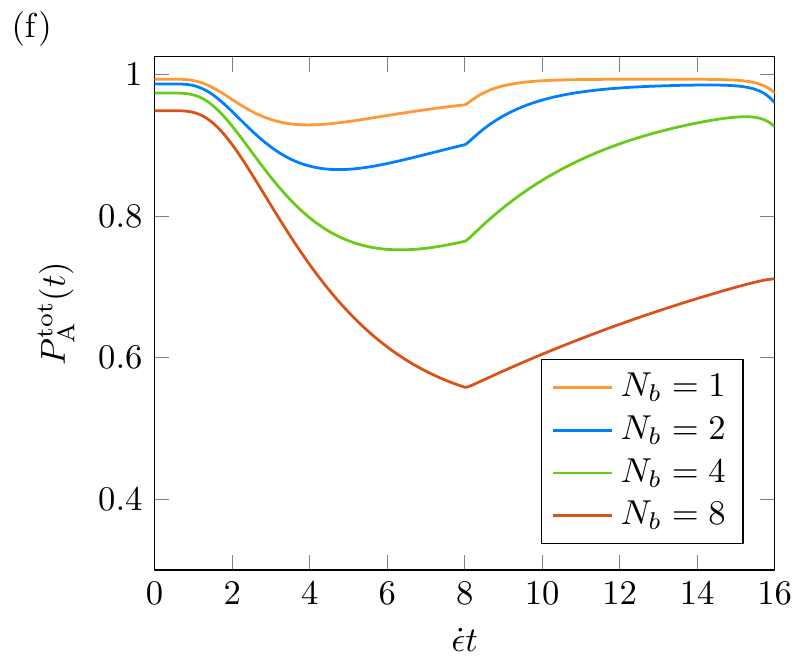}
	\end{array}
	$
	\vspace{-5pt}
	\end{center}
	\caption{\label{figmacroparam2}
		Parametric studies concerning the breakable link(s), where we vary
		(a,b) $\beta u_b$, the nondimensional breakable link energy,
		(c,d) $\beta\Delta\Psi_0$, the net Helmholtz free energy change when breaking a chain, and
		(e,f) $N_b$, the number of breakable links, while keeping the total number of links, $N_b+N_b^\#=9$, constant.
		For one cycle of uniaxial monotonic tension, the nondimensional stress, $\beta\sigma_{11}(t)/n$, and total probability that a chain is intact, $P_\mathrm{A}^\mathrm{tot}(t)$, are plotted as a function of the nondimensional time $\dot{\epsilon}t$.
	}
	\end{figure*}
}

Fourth, we examine the behavior of the general model while varying the nondimensional breakable link energy $\beta u_b$.
Referring back to the single-chain mechanical response in Fig.~\ref{figmech}(b), we recall that increasing $\beta u_b$ caused the maximum nondimensional force that a chain could support, $\eta_\mathrm{max}\equiv\sqrt{\kappa\beta u_b/8}$, to increase.
We also recall that the overall mechanical response away from $\eta_\mathrm{max}$ and the reaction rate coefficient function tended not to change when varying $\beta u_b$.
Fig.~\ref{figmacroparam2}(a) shows the nondimensional stress $\beta\sigma_{11}(t)/n$ as a function of the nondimensional time $\dot{\epsilon}t$.
As $\beta u_b$ increases, the shape of the curve remains relatively unchanged while the overall stress level increases due to increasing $\eta_\mathrm{max}$.
The fraction of intact chains within the network $P_\mathrm{A}^\mathrm{tot}(t)$ as a function of the nondimensional time $\dot{\epsilon}t$ is shown in Fig.~\ref{figmacroparam2}(b).
From $P_\mathrm{A}^\mathrm{tot}(t)$ we see that the breakdown of the network is only mildly lessened by increasing $\beta u_b$, which can be further understood after reconsidering the mechanical response in Fig.~\ref{figmech}(b) and observing that the maximum extensibility $\gamma_\mathrm{max}$ also only mildly increases with $\beta u_b$.
While these results show that the network is strengthened by increasing the energy of the breakable bond, it is unlikely that this energy would be controllable in the range of many factors of $\mathfrak{b}T=1/\beta$.
Correspondingly, the mechanics of the network are relatively insensitive to the breakable bond energy. 

Fifth, we examine the behavior of the general model while varying $\beta\Delta\Psi_0$, the net Helmholtz free energy change when breaking a chain.
This parameter directly controls the total probability at equilibrium that a chain is intact, $P_\mathrm{A}^\mathrm{tot,eq}$, or broken, $P_\mathrm{B}^\mathrm{tot,eq} = 1 - P_\mathrm{A}^\mathrm{tot,eq}$, via Eq.~\eqref{PeqnicenoVb}.
Fig.~\ref{figmacroparam2}(c) shows the nondimensional stress, where we see that increasing $\beta\Delta\Psi_0$ decreases the initial modulus (roughly $3nP_\mathrm{A}^\mathrm{tot,eq}/\beta$) as well as the overall stress. 
Both trends are almost entirely due to decreasing $P_\mathrm{A}^\mathrm{tot,eq}$: when scaling the nondimensional stress by $P_\mathrm{A}^\mathrm{tot,eq}$, the curves collapse on one another (not shown).
Although it is not observed in the stress, $\beta\Delta\Psi_0$ also has a strong effect on the rate of reforming.
The net reverse reaction rate coefficient function, $k(\boldsymbol{\gamma})\mathscr{P}_\mathrm{A}^\mathrm{eq}(\boldsymbol{\gamma})/P_\mathrm{B}^\mathrm{tot,eq}$, multiplies $P_\mathrm{B}^\mathrm{tot}(t)$ in Eq.~\eqref{PAevokinVBfinalsinglek}.
The rate of breaking relative to reforming is then $\mathscr{P}_\mathrm{A}^\mathrm{eq}(\boldsymbol{\gamma})/P_\mathrm{B}^\mathrm{tot,eq}$, which becomes large as $\beta\Delta\Psi_0$ increases.
The stress is mostly unaffected by the rate of reforming since chains reform towards the equilibrium distribution $\mathscr{P}_\mathrm{A}^\mathrm{eq}(\boldsymbol{\gamma})$ where they tend not to contribute to the overall stress.
Importantly, this reforming towards the stress-free equilibrium distribution does not appear to provide any appreciable stress reduction, although more rapid reforming does cause an increasing amount of compression when the strain is returned to zero.
The fraction of intact chains, shown in Fig.~\ref{figmacroparam2}(d), allows us to better examine the effects of the rate of reforming broken chains.
As $\beta\Delta\Psi_0$ increases, more rapid reforming prevents a large percentage of broken chains to be reached at any time, and the equilibrium value (which also increases) is more quickly recovered after the loading portion.
When $\beta\Delta\Psi_0$ is sufficiently large, the reforming of broken chains becomes so rapid that $P_\mathrm{A}^\mathrm{tot}(t)$ appears to remain constant at its equilibrium value.
This result might lead one to believe that for $\beta\Delta\Psi_0\gg 1$, $\rho(t)=P_\mathrm{B}^\mathrm{tot}(t)/P_\mathrm{B}^\mathrm{tot,eq}$ could be approximated as $\rho(t)\sim 1$ within the solution for $\mathscr{P}_\mathrm{A}(\boldsymbol{\gamma};t)$ in Eq.~\eqref{PAsol}, thus avoiding computing the solution for $\rho(t)$ in \ref{appexact}.
Such an approximation fails entirely: $\rho(t)$ spikes and decays rapidly as chains break and are immediately reformed, playing a crucial role in the solution for $\mathscr{P}_\mathrm{A}(\boldsymbol{\gamma};t)$.
Taking $\rho(t)\sim 1$ then retrieves entirely different results where reforming is vastly under-predicted and $P_\mathrm{A}^\mathrm{tot}(t)$ does not actually remain constant (not shown).
This is unfortunate, because cases where $\beta\Delta\Psi_0\gg 1$ tend to be the most computationally expensive and represent the cases where chains require considerable force to break but reform quickly afterward.

Sixth and lastly, we examine the behavior of the general model while varying the number of breakable links $N_b$.
We keep the total number of links $N_b+N_b^\#=9$ constant, so the chain length remains constant and the number of unbreakable links $N_b^\#$ varies accordingly.
In effect, this varying $N_b$ while keeping $N_b+N_b^\#$ constant varies the fraction of the chain that is breakable.
The stress in Fig.~\ref{figmacroparam2}(e) decreases mildly as $N_b$ increases while retaining the same overall shape. 
This stress decrease is mostly due to chains breaking more quickly as $N_b$ increases since the net reaction rate coefficient function is still $k(\gamma) = N_b k'(\gamma)$ scales directly with $N_b$.
It is also due to the equilibrium fraction of intact chains $P_\mathrm{A}^\mathrm{tot,eq}$ decreasing as $N_b$ increases, i.e. Eq.~\eqref{PeqnicenoVb}.
This is directly observed in Fig.~\ref{figmacroparam2}(f), where the evolution of the fraction of intact chains in the network, $P_\mathrm{A}^\mathrm{tot}(t)$, is shown.
As $N_b$ increases, chains break more rapidly and allow a larger fraction of chains to be broken overall.
The total rate of reforming seems to increase with $N_b$, but this is simply due to more chains being broken, driving faster reforming due to the system being further out of equilibrium.
Our results here verify that including more breakable bonds within a chain of a fixed contour length causes substantially increased bond breaking under a given deformation.

\subsubsection*{General model application}

\begin{figure*}[t]
\begin{center}
\includegraphics{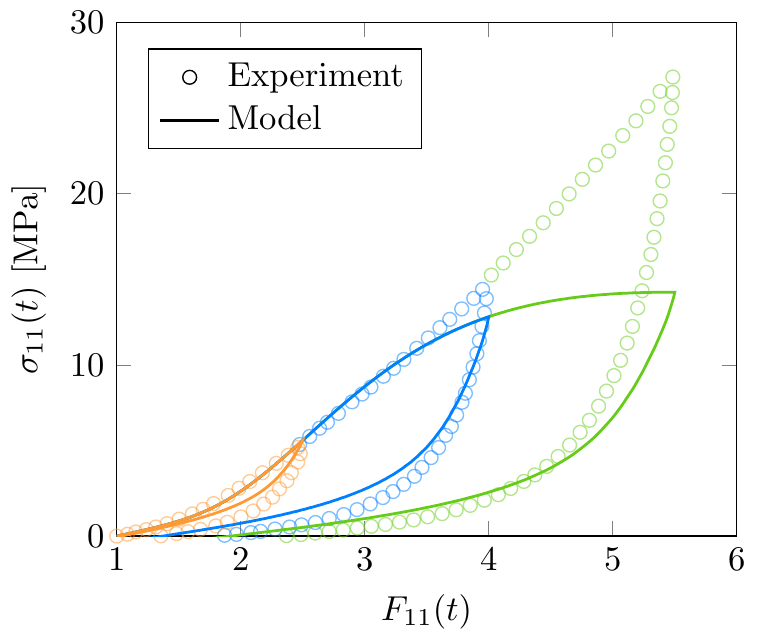}
\vspace{-5pt}
\end{center}
\caption{\label{figCu}
	Stress as a function of stretch for one cycle of uniaxial monotonic tension, repeated for different maximum stretches.
	The experimental results for the metal-coordinated gel of \citet{lin2020constitutive} are shown with those predicted by the general model.
}
\end{figure*}

Now that our parametric study is finished, we apply the general model to a material system from the literature with force-sensitive reversible crosslinks.
A tough, self-recovering hydrogel was synthesized by \citet{zheng2016metal} using metal-coordination complexes as reversible crosslinks.
This system was modeled by \citet{lin2020constitutive}, where the crosslink breaking rate was taken to increase as the network experienced more stress to account for force-sensitive breaking.
We obtain parameters for our model as follows.
The hydrogel was synthesized with 10\% mole fraction of crosslinking monomers, so when taking $N_b=1$ we obtain $N_b^\#=9$ after taking each link to represent a monomer.
We obtain $n/\beta=0.48$~MPa from half of the reported shear modulus, $0.96$~MPa.
We take $k_0=2\times 10^{-4}$/s from the force-free rate of breaking obtained for the model of \citet{lin2020constitutive}, and similarly take $\beta\Delta\Psi_0=8.55$ in order to match total reforming rate of the model, $\hat{K}=1$/s.
Otherwise, we find $\varsigma=1$, $\beta u_b=100$, $\kappa=200$, and $\kappa^\#=400$ provides the best fit. 
The stress as a function of applied stretch is shown in Fig.~\ref{figCu} for one cycle of uniaxial monotonic tension, where different tests are performed to different maximum stretches.
For the first test (to a stretch of 2.5), we find that our model accurately predicts the loading curve from the experiment but upon unloading overpredicts the recovery and thus underpredicts the dissipation.
For the second test (to a stretch of 4), the model begins to better predict the growing amount of dissipation, but still overpredicts the recovery and begins to yield.
For the third test (to a stretch of 5.5), the model continues to yield and deviates strongly from the experimental loading curve.
Overall, since our model cannot create large amounts of dissipation without significantly breaking down the network, it is unable to capture the mechanical response observed in experiment due to the accompanying significant yielding.
We attribute this to our model not accounting for the viscoplastic deformation resulting from broken portions of the network freely flowing before reforming.
This was not encountered earlier in Sec.~\ref{speccaseirrev} when modeling the multinetwork elastomer since the secondary networks provided integrity while the sacrificial network broke down.
This viscous flow was included in the model of \citet{lin2020constitutive} and allowed it to make more accurate predictions.
Here, the viscous flow would allow for increased dissipation without requiring the significant network breakdown that creates artificial yielding, and additionally would reduce the predicted amount of recovery.

\newpage

\section{Conclusion}

We have accomplished a fundamental derivation leading to a constitutive model for the stress-strain behavior of elastomers with chain breaking that properly incorporates the statistical mechanics of a general single-chain model. 
We have shown that the single-chain mechanical response, the breaking and reforming kinetics, and the equilibrium distribution of chains in the network are all inextricably determined by the single-chain model Hamiltonian.
Our meticulous formulation was seamlessly brought to the continuum scale, where we obtained the Cauchy stress entirely in terms of the applied deformation, the temperature, the network chain density, and the single-chain model.
We showed that the second law of thermodynamics was automatically arbitrarily satisfied as a consequence of our statistical mechanical treatment.
We introduced, developed, and studied the potential-supplemented freely jointed chain ($u$FJC) model.
We then accomplished a valuable exact solution for the evolving distribution of chains in the network, which is applicable beyond the $u$FJC model.

Next, we developed two special cases of our constitutive model and applied them to exemplary polymer materials from the literature.
In the case of the multinetwork elastomer, the rate-independent irreversible theory was mostly successful in modeling both the mechanical response of the material and the breakdown of the sacrificial network.
In the case of the dual crosslink gel, the single timescale of the transient network model was mostly unable to capture the rate-dependent mechanical response of the material.
We then examined the behavior of our constitutive model in the general case, where we contrasted it with the two special cases and performed several parametric studies to highlight how material performance could be tuned. 
Afterward we applied the general model to a hydrogel with force-sensitive reversible metal-coordinated crosslinks, where we found that the model was unable to capture the toughness of the material without artificially yielding.
Overall, we observed the following: while chain breaking is a dominant feature in the large deformation of many elastomeric systems, related phenomena often become similarly important as the network breaks down, limiting the success of our current approach.
For example, breaking bonds could increase the effective contour length of chains rather than simply reduce the amount of intact chains, similar to network alteration theory \citep{marckmann2002theory,chagnon2006development}.
This idea is supported by a recent molecular dynamics study, which determined that network breakdown in the same multinetwork elastomer we considered here was strongly tied to the evolving shortest contour length between crosslinks \citep{yin2020topological}.
Another example: significant network breakdown -- especially in the case of a single network material -- seems to be accompanied by additional viscoplastic flow not included in our model \citep{lin2020constitutive}.
Nonetheless, our existing theory provides a strong foundation to build upon and include additional physical mechanisms.
The presented approach is a robust method to obtain macroscopic constitutive functions in terms of molecular functions and parameters.

\section*{Acknowledgments}
	This material is based in part upon work supported by the National Science Foundation under Grant No. CAREER-1653059.
	Calculations in this work used the Extreme Science and Engineering Discovery Environment (XSEDE) \texttt{Stampede2} cluster, which is supported by National Science Foundation Grant ACI-1548562.

\appendix

\section{Simplification under transition state theory\label{TSTmath}}

Here we detail the mathematical operations involved with simplifying the reaction rates $\mathcal{R}_j'(\boldsymbol{\xi};t)$ and $\mathcal{R}_j''(\boldsymbol{\xi};t)$ in Eqs.~\eqref{dPABj} and \eqref{dPBjA} when applying the assumptions of transition state theory, resulting in Eqs.~\eqref{Rjpkin} and \eqref{Rjppkin}.
We first substitute the primary assumption of transition state theory Eq.~\eqref{floceqA} into Eq.~\eqref{dPABj} for

\begin{equation}
	\mathcal{R}_j'(\boldsymbol{\xi};t) = 
	\frac{P_\mathrm{A}(\boldsymbol{\xi};t)}{\mathfrak{q}P_\mathrm{A}^\mathrm{eq}(\boldsymbol{\xi})} \int\cdots\int e^{-\beta H(\Gamma)} \frac{p_j}{m_j}\,\Theta(p_j) \delta^3\left[\mathbf{R}(\Gamma) - \boldsymbol{\xi}\right] \delta\left(\ell^\ddagger_j -\ell_j\right) \prod_{\substack{i=1\\ i\neq j}}^{M} \Theta\left(\ell^\ddagger_i -\ell_i\right) \,d\Gamma
	.
\end{equation}
Similarly, we substitute Eq.~\eqref{floceqB} into Eq.~\eqref{dPBjA} for

\begin{equation}
	\mathcal{R}_j''(\boldsymbol{\xi};t) = 
	\frac{P_{\mathrm{B}_j}(\boldsymbol{\xi};t)}{\mathfrak{q}P_{\mathrm{B}_j}^\mathrm{eq}(\boldsymbol{\xi})} \int\cdots\int e^{-\beta H(\Gamma)} \left(-\frac{p_j}{m_j}\right)\Theta(-p_j) \delta^3\left[\mathbf{R}(\Gamma) - \boldsymbol{\xi}\right] \delta\left(\ell^\ddagger_j -\ell_j\right) \prod_{\substack{i=1\\ i\neq j}}^{M} \Theta\left(\ell^\ddagger_i -\ell_i\right) \,d\Gamma
	.
\end{equation}
Note that we have used Eq.~\eqref{feq}, which is $f^\mathrm{eq}(\Gamma) = e^{-\beta H(\Gamma)}/\mathfrak{q}$.
We may complete the portion of these integrals related to the momentum along the reaction coordinate, which contributes a factor of $\mathfrak{b}T=1/\beta$ in either case, and where we retain a delta function in order to keep the integration over the whole phase space:

\begin{align}
	\int\cdots\int e^{-\beta H(\Gamma)} \frac{p_j}{m_j}\,\Theta(p_j) \delta^3\left[\mathbf{R}(\Gamma) - \boldsymbol{\xi}\right] \delta\left(\ell^\ddagger_j -\ell_j\right) \prod_{\substack{i=1\\ i\neq j}}^{M} \Theta\left(\ell^\ddagger_i -\ell_i\right) \,d\Gamma = 
	\nonumber \\ 
	\int\cdots\int e^{-\beta H(\Gamma)} \left(-\frac{p_j}{m_j}\right)\Theta(-p_j) \delta^3\left[\mathbf{R}(\Gamma) - \boldsymbol{\xi}\right] \delta\left(\ell^\ddagger_j -\ell_j\right) \prod_{\substack{i=1\\ i\neq j}}^{M} \Theta\left(\ell^\ddagger_i -\ell_i\right) \,d\Gamma = 
	\nonumber \\ 
	\frac{1}{\beta}\int\cdots\int e^{-\beta H(\Gamma)} \delta\left(p_j\right) \delta^3\left[\mathbf{R}(\Gamma) - \boldsymbol{\xi}\right] \delta\left(\ell^\ddagger_j -\ell_j\right) \prod_{\substack{i=1\\ i\neq j}}^{M} \Theta\left(\ell^\ddagger_i -\ell_i\right) \,d\Gamma
	.
\end{align}
The remaining integral is the partition function of the transition state $\mathfrak{q}_{\ddagger_j}^* (\boldsymbol{\xi})$ in Eq.~\eqref{qABjTS}, and the reaction rates are now rewritten as

\begin{align}
	\mathcal{R}_j'(\boldsymbol{\xi};t) = &
	\frac{\mathfrak{q}_{\ddagger_j}^* (\boldsymbol{\xi})}{\beta\mathfrak{q}P_\mathrm{A}^\mathrm{eq}(\boldsymbol{\xi})}\,P_\mathrm{A}(\boldsymbol{\xi};t)
	,\\
	\mathcal{R}_j''(\boldsymbol{\xi};t) = &
	\frac{\mathfrak{q}_{\ddagger_j}^* (\boldsymbol{\xi})}{\beta\mathfrak{q}P_{\mathrm{B}_j}^\mathrm{eq}(\boldsymbol{\xi})}\,P_{\mathrm{B}_j}(\boldsymbol{\xi};t)
	.
\end{align}
We use Eq.~\eqref{PAeqgen} for $\mathfrak{q}P_\mathrm{A}^\mathrm{eq}(\boldsymbol{\xi}) = \mathfrak{q}^*_\mathrm{A}(\boldsymbol{\xi})$
and Eq.~\eqref{PBjeqgen} for $\mathfrak{q}P_{\mathrm{B}_j}^\mathrm{eq}(\boldsymbol{\xi}) = \mathfrak{q}^*_{\mathrm{B}_j}(\boldsymbol{\xi})$; with the reaction rate coefficient functions $k_j'(\boldsymbol{\xi})$ and $k_j''(\boldsymbol{\xi})$ defined in Eq.~\eqref{kBjA} and Eq.~\eqref{kABj}, this allows us to obtain the simplified reaction rates $\mathcal{R}_j'(\boldsymbol{\xi};t)$ and $\mathcal{R}_j''(\boldsymbol{\xi};t)$ in Eqs.~\eqref{Rjpkin}--\eqref{Rjppkin}.

\section{Extended derivations for the macroscopic theory\label{appendixmacrovarious}}

\subsection{Macroscopically-obtained equilibrium\label{appmacroeq}}

Here we show that the equilibrium probabilities $P_\mathrm{A}^\mathrm{eq}(\boldsymbol{\xi})$ and $P_{\mathrm{B}_j}^\mathrm{tot,eq}$ in Eqs.~\eqref{PAeq} and \eqref{PBjeq}, respectively, may be obtained from the Helmholtz free energy density $a(t)$ in Eq.~\eqref{ayo} through minimization.
We use the Lagrange multiplier $\Lambda$ to enforce the conservation requirement in Eq.~\eqref{conservassumpd} and write

\begin{equation}
	a_\Lambda(t) \equiv a(t) - \Lambda\left[\iiint P_\mathrm{A}(\boldsymbol{\xi};t)\,d^3\boldsymbol{\xi} + \sum_{j=1}^M P^\mathrm{tot}_{\mathrm{B}_j}(t) - 1\right]
	.
\end{equation}
We now take the functional derivative \citep{giaquinta2004calculus} of $a_\Lambda(t)$ with respect to the intact chain probability distribution $P_\mathrm{A}(\boldsymbol{\xi};t)$ and each broken chain probability $P^\mathrm{tot}_{\mathrm{B}_j}(t)$.
Evaluating the results at equilibrium and setting them equal to zero, we obtain the following:

\begin{align}
	\left.\left(\frac{\delta a_\Lambda}{\delta P_\mathrm{A}(\boldsymbol{\xi};t)}\right)_{P^\mathrm{tot}_{\mathrm{B}_j}}\right|^\mathrm{eq} = &~
	n\mu_\mathrm{A}^{*,\mathrm{eq}}(\boldsymbol{\xi}) - n\mathfrak{b}T - \Lambda 
	=0
	,\\
	\left.\left(\frac{\delta a_\Lambda}{\delta P^\mathrm{tot}_{\mathrm{B}_j}(t)}\right)_{P_\mathrm{A}}\right|^\mathrm{eq} = &~
	n\mu_{\mathrm{B}_j}^\mathrm{eq} - n\mathfrak{b}T - \Lambda
	=0
	.
\end{align}
We then see that all chemical potentials are equal at equilibrium, consistent with statistical thermodynamics \citep{mcq}.
Using Eq.~\eqref{muA}, we find the equilibrium probability distribution of intact chains to be

\begin{equation}\label{fe23rwefdfwe}
	P_\mathrm{A}^\mathrm{eq}(\boldsymbol{\xi}) = 
	\frac{e^{\beta\Lambda/n + 1}}{N}\,\mathfrak{q}^*_\mathrm{A}(\boldsymbol{\xi})
	,
\end{equation}
and using Eq.~\eqref{muBj} with Eq.~\eqref{qqexpyay} and Eq.~\eqref{ewfrgdfgs3}, we find each broken chain total probability to be

\begin{equation}\label{few423wres}
	P_{\mathrm{B}_j}^\mathrm{tot,eq} = 
	\frac{e^{\beta\Lambda/n + 1}}{N}\, \mathfrak{q}_{\mathrm{A}} e^{-\beta\Delta\Psi_{0_j}}
	.
\end{equation}
Recall that $\mathfrak{q}_{\mathrm{A}}$ is the integral of $\mathfrak{q}^*_\mathrm{A}(\boldsymbol{\xi})$ over all $\boldsymbol{\xi}$.
In order to solve for the Lagrange multiplier $\Lambda$ we apply the conservation requirement Eq.~\eqref{conservassumpd}, to obtain

\begin{equation}
	\frac{e^{\beta\Lambda/n + 1}}{N}\,\mathfrak{q}_\mathrm{A} + \sum_{j=1}^M \frac{e^{\beta\Lambda/n + 1}}{N}\, \mathfrak{q}_{\mathrm{A}} e^{-\beta\Delta\Psi_{0_j}}
	= 1
	,
\end{equation}
which is then rearranged to solve for the entire factor

\begin{equation}
	\frac{e^{\beta\Lambda/n + 1}}{N} = 
	\frac{1}{\mathfrak{q}_\mathrm{A}} \frac{1}{1 + \sum_{j=1}^M e^{-\beta\Delta\Psi_{0_j}}}
	.
\end{equation}
Substitution of this factor into Eq.~\eqref{fe23rwefdfwe} results in Eq.~\eqref{PAeq}, and into Eq.~\eqref{few423wres} results in Eq.~\eqref{PBjeq}.

\subsection{Retrieving the stress\label{appstressretrieve}}

Starting from the hyperelastic form of the stress given in Eq.~\eqref{hyperelastic}, here we retrieve the form of the stress in Eq.~\eqref{sigmafinal}.
To begin, we consider the evolution of the probability distributions due to the deformation from Eq.~\eqref{PAevokinVB}, 

\begin{equation}\label{defevoPA}
	\left.\frac{\partial P_\mathrm{A}(\boldsymbol{\xi};t)}{\partial t}\right|_\mathbf{F} = 
	- \frac{\partial}{\partial\boldsymbol{\xi}}\cdot\left[\dot{\boldsymbol{\xi}}_\mathrm{A}(\boldsymbol{\xi};t)P_\mathrm{A}(\boldsymbol{\xi};t)\right]
	,
\end{equation}
and from Eq.~\eqref{PBjtotevokin}, we have more simply

\begin{equation}
	\left.\frac{\partial P^\mathrm{tot}_{\mathrm{B}_j}(t)}{\partial t}\right|_\mathbf{F} = 
	0
	.
\end{equation}
We now integrate by parts using an extension of the divergence theorem, neglecting the probability of intact chains existing on the boundary $P_\mathrm{A}(\boldsymbol{\xi};t)|_{\partial_{\boldsymbol{\xi}}}\approx 0$ as we have previously discussed near the end of Sec.~\ref{disssubsubsec},

\begin{equation}
	\iiint \frac{\partial}{\partial\boldsymbol{\xi}}\cdot\left[\dot{\boldsymbol{\xi}}_\mathrm{A}(\boldsymbol{\xi};t)P_\mathrm{A}(\boldsymbol{\xi};t)\right] \mu^*_\mathrm{A}(\boldsymbol{\xi};t) \,d^3\boldsymbol{\xi}
	=
	-\iiint P_\mathrm{A}(\boldsymbol{\xi};t) \,\frac{\partial\mu^*_\mathrm{A}(\boldsymbol{\xi};t)}{\partial\boldsymbol{\xi}}\cdot \dot{\boldsymbol{\xi}}_\mathrm{A}(\boldsymbol{\xi};t) \,d^3\boldsymbol{\xi}
	.
\end{equation}
Next, we make the affine assumption $\dot{\boldsymbol{\xi}}_\mathrm{A}(\boldsymbol{\xi};t)=\mathbf{L}(t)\cdot\boldsymbol{\xi}$, which is that the end-to-end vectors $\boldsymbol{\xi}$ are affinely deformed by the deformation gradient $\mathbf{F}(t)$ on average.
Upon carrying out the derivatives involved in Eq.~\eqref{hyperelastic}, the stress is

\begin{equation}\label{stressintermsofmu}
	\boldsymbol{\sigma}(t) = 
	n\iiint P_\mathrm{A}(\boldsymbol{\xi};t) \,\frac{\partial\mu^*_\mathrm{A}(\boldsymbol{\xi};t)}{\partial\boldsymbol{\xi}}\,\boldsymbol{\xi} \,d^3\boldsymbol{\xi}
	- p(t)\mathbf{1}
	.
\end{equation}
The affine assumption allows us to expand Eq.~\eqref{defevoPA}, where the divergence of $\dot{\boldsymbol{\xi}}_\mathrm{A}(\boldsymbol{\xi};t)$ is zero since $\mathbf{1}:\mathbf{L}(t)=0$ due to incompressibility, leaving only the term containing the gradient of $P_\mathrm{A}(\boldsymbol{\xi};t)$.
We substitute this nonzero term into the evolution equation for $P_\mathrm{A}(\boldsymbol{\xi};t)$ in Eq.~\eqref{PAevokinVB} to obtain

\begin{equation}\label{appPAevokinVBfinal}
	\frac{\partial P_\mathrm{A}(\boldsymbol{\xi};t)}{\partial t} = 
	- \left[\frac{\partial P_\mathrm{A}(\boldsymbol{\xi};t)}{\partial\boldsymbol{\xi}}\,\boldsymbol{\xi}\right]:\mathbf{L}(t)
	-\sum_{j=1}^{M} k_j'(\boldsymbol{\xi}) \left[ P_\mathrm{A}(\boldsymbol{\xi};t) - \dfrac{P_{\mathrm{B}_j}^\mathrm{tot}(t)}{P_{\mathrm{B}_j}^\mathrm{tot,eq}}\,P^\mathrm{eq}_\mathrm{A}(\boldsymbol{\xi}) \right]
	,
\end{equation}
which is Eq.~\eqref{PAevokinVBfinal} in the manuscript.
The stress in Eq.~\eqref{stressintermsofmu} can be written as a function of the time-independent Helmholtz free energy of an intact chain $\psi^*_\mathrm{A}(\boldsymbol{\xi})$ rather than the time-dependent corresponding chemical potential $\mu^*_\mathrm{A}(\boldsymbol{\xi};t)$. 
The chemical potential of an intact chain $\mu^*_\mathrm{A}(\boldsymbol{\xi};t)$ is given by Eq.~\eqref{muA} in terms of the partition function $\mathfrak{q}^*_\mathrm{A}(\boldsymbol{\xi})$ and the probability distribution $P_\mathrm{A}(\boldsymbol{\xi};t)$.
The principal thermodynamic connection formula allows the partition function to be written as a function of the Helmholtz free energy,

\begin{equation}\label{psinegktlnq}
	\mathfrak{q}^*_\mathrm{A}(\boldsymbol{\xi}) = e^{-\beta\psi^*_\mathrm{A}(\boldsymbol{\xi})}
	,
\end{equation}
so we then expand the gradient of $\mu^*_\mathrm{A}(\boldsymbol{\xi};t)$ in Eq.~\eqref{muA} as

\begin{equation}\label{gradgradgrad}
	\frac{\partial\mu^*_\mathrm{A}(\boldsymbol{\xi};t)}{\partial\boldsymbol{\xi}} = 
	\frac{\partial\psi^*_\mathrm{A}(\boldsymbol{\xi})}{\partial\boldsymbol{\xi}} + \frac{\mathfrak{b}T}{P_\mathrm{A}(\boldsymbol{\xi};t)}\frac{\partial P_\mathrm{A}(\boldsymbol{\xi};t)}{\partial\boldsymbol{\xi}}
	.
\end{equation}
Substitution of the second term in Eq.~\eqref{gradgradgrad} into the stress Eq.~\eqref{stressintermsofmu} results in 

\begin{equation}\label{morestepsplz}
	\boldsymbol{\sigma}(t) = 
	n\iiint P_\mathrm{A}(\boldsymbol{\xi};t) \,\frac{\partial\psi^*_\mathrm{A}(\boldsymbol{\xi})}{\partial\boldsymbol{\xi}} \,\boldsymbol{\xi} \,d^3\boldsymbol{\xi}
	+ n\mathfrak{b}T\iiint \frac{\partial P_\mathrm{A}(\boldsymbol{\xi};t)}{\partial\boldsymbol{\xi}}\,\boldsymbol{\xi} \,d^3\boldsymbol{\xi}
	- p(t)\mathbf{1}
	.
\end{equation}
The second term in Eq.~\eqref{morestepsplz} is simplified by again using integration by parts and $P_\mathrm{A}(\boldsymbol{\xi};t)|_{\partial_{\boldsymbol{\xi}}\mathcal{A}} \approx 0$,

\begin{equation}\label{sphericalbyebye}
	n\mathfrak{b}T\iiint \frac{\partial P_\mathrm{A}(\boldsymbol{\xi};t)}{\partial\boldsymbol{\xi}}\,\boldsymbol{\xi} \,d^3\boldsymbol{\xi}
	=
	-n\mathfrak{b}TP^\mathrm{tot}_\mathrm{A}(t)\mathbf{1}
	.
\end{equation}
The pressure term in Eq.~\eqref{morestepsplz} and this term in Eq.~\eqref{sphericalbyebye} are both spherical.
Since the pressure $p(t)$ is merely a Lagrange multiplier enforcing incompressibility, we take $p(t)+n\mathfrak{b}TP^\mathrm{tot}_\mathrm{A}(t) \mapsto p(t)$ without loss of generality.
The stress in Eq.~\eqref{morestepsplz} is then

\begin{equation}\label{appsigmafinal}
	\boldsymbol{\sigma}(t) = 
	n\iiint P_\mathrm{A}(\boldsymbol{\xi};t) \,\frac{\partial\psi^*_\mathrm{A}(\boldsymbol{\xi})}{\partial\boldsymbol{\xi}} \,\boldsymbol{\xi} \,d^3\boldsymbol{\xi}
	- p(t)\mathbf{1}
	,
\end{equation}
which is Eq.~\eqref{sigmafinal} in the manuscript.
This stress is then nondimensionalized as follows: we first rearrange the pressure term and scale by $\beta/n$ for

\begin{equation}
	\frac{\boldsymbol{\sigma}(t) + p(t)\mathbf{1}}{\beta/n} = 
	\iiint P_\mathrm{A}(\boldsymbol{\xi};t) \,\frac{\partial\beta\psi^*_\mathrm{A}(\boldsymbol{\xi})}{\partial\boldsymbol{\xi}} \,\boldsymbol{\xi} \,d^3\boldsymbol{\xi}
	.
\end{equation}
Next, we substitute in $\boldsymbol{\xi} = N_b\ell_b \boldsymbol{\gamma}$, using $\mathscr{P}_\mathrm{A}(\boldsymbol{\gamma};t)\equiv (N_b\ell_b)^3 P_\mathrm{A}(\boldsymbol{\xi};t)$ and then $P_\mathrm{A}(\boldsymbol{\xi};t)\,d^3\boldsymbol{\xi} = \mathscr{P}_\mathrm{A}(\boldsymbol{\gamma};t)\,d^3\boldsymbol{\gamma}$,

\begin{equation}
	\frac{\boldsymbol{\sigma}(t) + p(t)\mathbf{1}}{\beta/n} = 
	\iiint \mathscr{P}_\mathrm{A}(\boldsymbol{\gamma};t) \,\frac{\partial\beta\psi^*_\mathrm{A}(\boldsymbol{\gamma})}{\partial{\gamma}} \left(\frac{\boldsymbol{\gamma}\boldsymbol{\gamma}}{\gamma}\right) \,d^3\boldsymbol{\gamma}
	.
\end{equation}
We then substitute in $\beta\psi^*_\mathrm{A}(\boldsymbol{\gamma}) = N_b\vartheta_\mathrm{A}^*({\gamma})$, where $\eta(\boldsymbol{\gamma}) = \partial\vartheta_\mathrm{A}^*({\gamma})/\partial\gamma$, to obtain Eq.~\eqref{sigmanondim}.
When an inhomogeneous chain (consists of both breakable and unbreakable links) is used for the $u$FJC model, the homogeneous chain contour length transforms, $N_b\ell_b\mapsto (N_b+\varsigma N_b^\#)\ell_b$.
To adjust our nondimensional representation of the stress in Eq.~\eqref{sigmanondim}, we must then transform $\boldsymbol{\gamma}$ using $N_b\gamma\mapsto (N_b+\varsigma N_b^\#)\gamma$, where $\mathscr{P}_\mathrm{A}(\boldsymbol{\gamma};t)\,d^3\boldsymbol{\gamma}$ is invariant.
The net result within Eq.~\eqref{sigmanondim} is effectively

\begin{equation}
	N_b\left(\frac{\boldsymbol{\gamma}\boldsymbol{\gamma}}{\gamma}\right) \mapsto
	\left(N_b + \varsigma N_b^\#\right)\left(\frac{\boldsymbol{\gamma}\boldsymbol{\gamma}}{\gamma}\right)
	,
\end{equation}
which taken within Eq.~\eqref{sigmanondim} produces Eq.~\eqref{sigmanondimvarsigma}.

\subsection{Expressing the chemical dissipation\label{appgettheD}}

Starting from Eq.~\eqref{Drxndef}, here we retrieve the form of the dissipation due to the chemical reactions $\mathcal{D}_\mathrm{rxn}(t)$ that allows us to conclude $\mathcal{D}_\mathrm{rxn}(t) \geq 0$.
We begin by expanding Eq.~\eqref{Drxndef}, 

\begin{equation}\label{appDrxnt}
	\mathcal{D}_\mathrm{rxn}(t) = 
	-n\iiint \left.\frac{\partial P_\mathrm{A}(\boldsymbol{\xi};t)}{\partial t}\right|_\mathrm{rxn} \mu^*_\mathrm{A}(\boldsymbol{\xi};t)\,d^3\boldsymbol{\xi} 
	- n\sum_{j=1}^{M} \left.\frac{\partial P^\mathrm{tot}_{\mathrm{B}_j}(t)}{\partial t}\right|_\mathrm{rxn} \mu_{\mathrm{B}_j}(t)
	.
\end{equation}
It is possible to write the time derivatives and chemical potentials in Eq.~\eqref{appDrxnt} strictly in terms of the original forward and reverse reactions rates, $\mathcal{R}_j'(\boldsymbol{\xi};t)$ and $\mathcal{R}_j''(\boldsymbol{\xi};t)$.
Referring back to Eq.~\eqref{PAevogeneral}, we first write

\begin{equation}\label{feawfwe234}
	\left.\frac{\partial P_\mathrm{A}(\boldsymbol{\xi};t)}{\partial t}\right|_\mathrm{rxn} = 
	-\sum_{j=1}^{M} \left[\mathcal{R}_j'(\boldsymbol{\xi};t) - \mathcal{R}_j''(\boldsymbol{\xi};t)\right]
	.
\end{equation}
Similarly with Eq.~\eqref{PBjtotevokin}, we use $\mathcal{R}_j(\boldsymbol{\xi};t) = k_j'(\boldsymbol{\xi})P_{\mathrm{A}}(\boldsymbol{\xi};t)$ from Eq.~\eqref{Rjpkin} to write

\begin{equation}
	\left.\frac{\partial P^\mathrm{tot}_{\mathrm{B}_j}(t)}{\partial t}\right|_\mathrm{rxn} = 
	\iiint \left[\mathcal{R}_j'(\boldsymbol{\xi};t) - k_j'(\boldsymbol{\xi})P^\mathrm{eq}_\mathrm{A}(\boldsymbol{\xi})\,\dfrac{P_{\mathrm{B}_j}^\mathrm{tot}(t)}{P_{\mathrm{B}_j}^\mathrm{tot,eq}} \right] d^3\boldsymbol{\xi}
	.
\end{equation}
Next, we use Eq.~\eqref{Kjeq} for $k_j'(\boldsymbol{\xi})P^\mathrm{eq}_\mathrm{A}(\boldsymbol{\xi}) = k_j''(\boldsymbol{\xi})P^\mathrm{eq}_{\mathrm{B}_j}(\boldsymbol{\xi})$.
We then use Eq.~\eqref{PBjxiapprox} to simplify

\begin{equation}
	P^\mathrm{eq}_{\mathrm{B}_j}(\boldsymbol{\xi})\dfrac{P_{\mathrm{B}_j}^\mathrm{tot}(t)}{P_{\mathrm{B}_j}^\mathrm{tot,eq}} =
	P^\mathrm{eq}_{\mathrm{B}_j}(\boldsymbol{\xi})\dfrac{V_{\mathrm{B}_j}P_{\mathrm{B}_j}(\boldsymbol{\xi};t)}{V_{\mathrm{B}_j}P_{\mathrm{B}_j}^\mathrm{eq}(\boldsymbol{\xi})} =
	P_{\mathrm{B}_j}(\boldsymbol{\xi};t)
	,
\end{equation}
which with $\mathcal{R}_j''(\boldsymbol{\xi};t) = k_j''(\boldsymbol{\xi})P_{\mathrm{B}_j}(\boldsymbol{\xi};t)$ from Eq.~\eqref{Rjppkin} allows us to obtain 

\begin{equation}\label{carefdun}
	\left.\frac{\partial P^\mathrm{tot}_{\mathrm{B}_j}(t)}{\partial t}\right|_\mathrm{rxn} = 
	\iiint \left[\mathcal{R}_j'(\boldsymbol{\xi};t) - \mathcal{R}_j''(\boldsymbol{\xi};t)\right] \,d^3\boldsymbol{\xi} 
	.
\end{equation}
With Eqs.~\eqref{feawfwe234} and \eqref{carefdun}, the dissipation in Eq.~\eqref{appDrxnt} now becomes

\begin{equation}\label{kkjnkwefsdf}
	\mathcal{D}_\mathrm{rxn}(t) = 
	n\sum_{j=1}^{M} \iiint \left[\mathcal{R}_j'(\boldsymbol{\xi};t) - \mathcal{R}_j''(\boldsymbol{\xi};t)\right] \left[ \mu^*_\mathrm{A}(\boldsymbol{\xi};t) - \mu_{\mathrm{B}_j}(t) \right] \,d^3\boldsymbol{\xi}
\end{equation}
Using both Eq.~\eqref{muA} and Eq.~\eqref{muBj}, we obtain the difference between the chemical potentials

\begin{equation}
	\mu_\mathrm{A}^*(\boldsymbol{\xi};t) - \mu_{\mathrm{B}_j}(t) = 
	\mathfrak{b}T\ln\left[\frac{\mathfrak{q}_{\mathrm{B}_j}^*}{\mathfrak{q}^*_\mathrm{A}(\boldsymbol{\xi})}\frac{P_\mathrm{A}(\boldsymbol{\xi};t)}{P^\mathrm{tot}_{\mathrm{B}_j}(t)/V_{\mathrm{B}_j}}\right]
	.
\end{equation}
After noting $\mathfrak{q}_{\mathrm{B}_j}^*/\mathfrak{q}^*_\mathrm{A}(\boldsymbol{\xi}) = k_j'(\boldsymbol{\xi})/k_j''(\boldsymbol{\xi})$ using Eq.~\eqref{Kjeq} and $P^\mathrm{tot}_{\mathrm{B}_j}(t)/V_{\mathrm{B}_j} = P_{\mathrm{B}_j}(\boldsymbol{\xi};t)$ using Eq.~\eqref{PBjxiapprox}, 

\begin{equation}
	\mu_\mathrm{A}^*(\boldsymbol{\xi};t) - \mu_{\mathrm{B}_j}(t) = 
	\mathfrak{b}T\ln\left[\frac{k_j'(\boldsymbol{\xi})P_\mathrm{A}(\boldsymbol{\xi};t)}{k_j''(\boldsymbol{\xi})P_{\mathrm{B}_j}(\boldsymbol{\xi};t)}\right]
	,
\end{equation}
which with Eq.~\eqref{Rjpkin} and Eq.~\eqref{Rjpkin} allows us to obtain 

\begin{equation}\label{appmudiffyo}
	\mu_\mathrm{A}^*(\boldsymbol{\xi};t) - \mu_{\mathrm{B}_j}(t) = 
	\mathfrak{b}T\ln\left[\frac{\mathcal{R}_j'(\boldsymbol{\xi};t)}{\mathcal{R}_j''(\boldsymbol{\xi};t)}\right]
	.
\end{equation}
Combining Eqs.~\eqref{kkjnkwefsdf} and \eqref{appmudiffyo}, we can then write the total dissipation $\mathcal{D}_\mathrm{rxn}$ succinctly as

\begin{equation}\label{appDrxnttt}
	\mathcal{D}_\mathrm{rxn}(t) = 
	\sum_{j=1}^{M} \iiint \mathcal{D}_j^*(\boldsymbol{\xi};t) \,d^3\boldsymbol{\xi}
	,
\end{equation}
where $\mathcal{D}_j^*(\boldsymbol{\xi};t)$, the dissipation density for the $j$th reaction occurring at the end-to-end vector $\boldsymbol{\xi}$, is

\begin{equation}\label{appDjSrxnttt}
	\mathcal{D}_j^*(\boldsymbol{\xi};t) \equiv 
	n\mathfrak{b}T\left[\mathcal{R}_j'(\boldsymbol{\xi};t) - \mathcal{R}_j''(\boldsymbol{\xi};t)\right]\ln\left[\frac{\mathcal{R}_j'(\boldsymbol{\xi};t)}{\mathcal{R}_j''(\boldsymbol{\xi};t)}\right]
	.
\end{equation}
Eqs.~\eqref{appDrxnttt} and \eqref{appDjSrxnttt} are Eqs.~\eqref{Drxnttt} and \eqref{DjSrxnttt} in the manuscript, respectively. 
The right-hand side of Eq.~\eqref{appDrxnttt} is of the form $c(x-y)\ln(x/y)$, which is a quantity that is positive semidefinite for all $x>0$ and $y>0$ if $c\geq 0$.
Since the reactions rates $\mathcal{R}_j'(\boldsymbol{\xi};t)$ and $\mathcal{R}_j''(\boldsymbol{\xi};t)$ are both positive definite, we have $[\mathcal{R}_j'(\boldsymbol{\xi};t)-\mathcal{R}_j''(\boldsymbol{\xi};t)]\ln[\mathcal{R}_j'(\boldsymbol{\xi};t)/\mathcal{R}_j''(\boldsymbol{\xi};t)]\geq 0$.
For finite temperatures we also have $n\mathfrak{b}T>0$, which with Eq.~\eqref{appDjSrxnttt} then allows us to conclude that each $\mathcal{D}_j^*(\boldsymbol{\xi};t)\geq 0$, and therefore with Eq.~\eqref{appDrxnttt} that $\mathcal{D}_\mathrm{rxn}(t) \geq 0$.

\section{Extended details on implementing the $u$FJC model\label{appuFJC}}

\subsection{Asymptotic approximation for the $u$FJC model\label{appasymp}}

Here we obtain an asymptotic approximation for the single-chain mechanical response of the $u$FJC model.
For nondimensional end-to-end lengths $\gamma$ near and below unity, the mechanical response will closely match that of the EFJC (harmonic $u$) if $\kappa\gg 1$.
Physically, $\kappa\gg 1$ represents that thermal energy is much smaller than the characteristic energy to begin stretching the link, so thermal sampling will be effectively restricted to where $u$ is minimized, which is also where $u$ is harmonic.
If the force remains small enough, $\eta\ll 1$, nondimensional end-to-end lengths above unity will not be reached.
The analytic expression we use for the single-chain mechanical response of the EFJC in the Gibbs (isotensional) ensemble is \citep{fiasconaro2019analytical}

\begin{equation}\label{analyticresult}
	\gamma_\mathrm{EFJC}(\eta) = \mathcal{L}(\eta) + \frac{\eta}{\kappa}\left[1 + \frac{1 - \mathcal{L}(\eta)\coth(\eta)}{1 + (\eta/\kappa)\coth(\eta)}\right]
	,
\end{equation}
where $\mathcal{L}(\eta)=\coth(\eta)-1/\eta$ is the Langevin function.
Since we will utilize $\kappa\gg 1$ in order to asymptotically approximate the mechanical response of the $u$FJC, we will expand Eq.~\eqref{analyticresult} in a series that is valid as $\kappa\to\infty$.
Using the Maclaurin series for $1/(1+x)$, we then have

\begin{equation}
	\gamma_\mathrm{EFJC}(\eta) = 
	\mathcal{L}(\eta) + \frac{\eta}{\kappa}\left\{2 - \mathcal{L}(\eta)\coth(\eta) + \left[1 - \mathcal{L}(\eta)\coth(\eta)\right]\eta\coth(\eta)\sum_{n = 1}^\infty (-\kappa)^{-n}\right\}
	\quad\text{as }\kappa\to\infty
	.
\end{equation}
Now, we choose to make a first-order asymptotic approximation, only keeping $O(\kappa^{-1})$ terms and writing

\begin{equation}\label{fe34ere3r4tdfg}
	\gamma_\mathrm{EFJC}(\eta) \sim
	\mathcal{L}(\eta) + \frac{\eta}{\kappa}\left[2 - \mathcal{L}(\eta)\coth(\eta)\right]
	\quad\text{for }\kappa\gg 1
	.
\end{equation}
For sufficiently low forces, $\eta\ll 1$, the mechanical response of the $u$FJC matches that of the EFJC with the same stiffness since in either case the link stretching will be restricted to the harmonic regime.
Eq.~\eqref{fe34ere3r4tdfg} is then the $O(\kappa^{-1})$ asymptotic approximation of the single-chain mechanical response of the $u$FJC for $\eta\ll 1$, 

\begin{equation}\label{gammasymp1}
	\gamma(\eta) \sim 
	\mathcal{L}(\eta) + \frac{\eta}{\kappa}\left[2 - \mathcal{L}(\eta)\coth(\eta)\right] 
	\quad\text{for }\kappa\gg 1\text{ and }\eta\ll 1
	.
\end{equation}
Now, for $\kappa\gg 1$ and $\gamma\gtrsim 1$ the $u$FJC will remain aligned and the links will begin to be stretched directly.
As $\gamma$ continues to increase, the corresponding large forces required, $\eta\gg 1$, will be approximately due to stretching the links alone.
In other words, the forces required to significantly stretch the stiff links will eclipse the entropically-based forces.
In this limit the mechanical response of the chain is asymptotically given by that of the links,

\begin{equation}\label{gammasymp2}
	\gamma(\eta) \sim
	\lambda(\eta)
	\quad\text{for }\kappa\gg 1\text{ and }\eta\gg 1
	,
\end{equation}
where $\lambda(\eta)$ is the stretch $\ell/\ell_b$ of a link under the nondimensional force $\eta$, defined through

\begin{equation}\label{lambdadefyo}
	\eta = 
	\beta\ell_b\left.\frac{\partial u(\ell)}{\partial\ell}\right|_{\ell = \ell_b\lambda(\eta)}
	.
\end{equation}
We then have two asymptotic approximations when $\kappa\gg 1$ for the mechanical response in Eqs.~\eqref{gammasymp1} and \eqref{gammasymp2} that we must match.
We may do so using Prandtl's method of asymptotic matching \citep{powers2015mathematical,bender2013advanced}, which stipulates that the following must be true to obtain a composite approximation:

\begin{equation}\label{asdfwerw3}
	\lim_{\eta\to\infty}\gamma_{\eta\ll 1}(\eta) = 
	\lim_{\eta\to 0}\gamma_{\eta\gg 1}(\eta) 
	.
\end{equation}
Here the limits are $1+\eta/\kappa$ in either case, thus we satisfy Eq.~\eqref{asdfwerw3}.
Prandtl's method also stipulates that this limit must be subtracted from the composite solution obtained when adding the two approximations, otherwise it would be accounted for twice.
After subtracting this common part of $1+\eta/\kappa$, we obtain the composite first-order asymptotic approximation 

\begin{equation}\label{appgammasim}
	\gamma_1(\eta) \sim
	\mathcal{L}(\eta) + \frac{\eta}{\kappa}\left[1 - \mathcal{L}(\eta)\coth(\eta)\right] + \lambda(\eta) - 1
	\quad\text{for }\kappa\gg 1
	.
\end{equation}
This mechanical response has three distinct terms.
The first, $\mathcal{L}(\eta)$, is the entropic mechanical response of the FJC that dominates at low forces.
The third, $\lambda(\eta) - 1$, is based upon the mechanical response of the aligned chain $\lambda(\eta)$ that dominates at high forces.
After noting that

\begin{equation}
	\Big|\eta[1 - \mathcal{L}(\eta)\coth(\eta)]\Big| \leq 1
	\quad\text{for all }\eta
	,
\end{equation}
we see that the second term in Eq.~\eqref{appgammasim} is essentially an $O(\kappa^{-1})$ correction. 
If $\kappa$ is sufficiently high, this correction is negligible and we may take the even simpler (leading-order) approximation 

\begin{equation}\label{appgammasimplest}
	\gamma_0(\eta)\sim\mathcal{L}(\eta) + \lambda(\eta) - 1
	\quad\text{for }\kappa\gg 1
	,
\end{equation}
which is the asymptotic approximation utilized in Eq.~\eqref{gammasimplest} of the manuscript.

\begin{figure*}[t]
\begin{center}
$
\michaelarraysettings
\begin{array}{cc}
\includegraphics{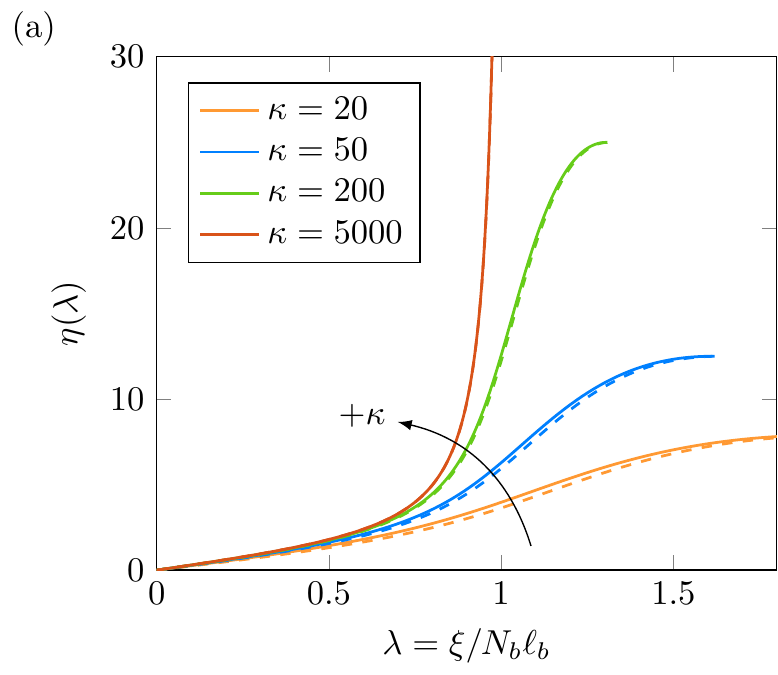}
& 
\includegraphics{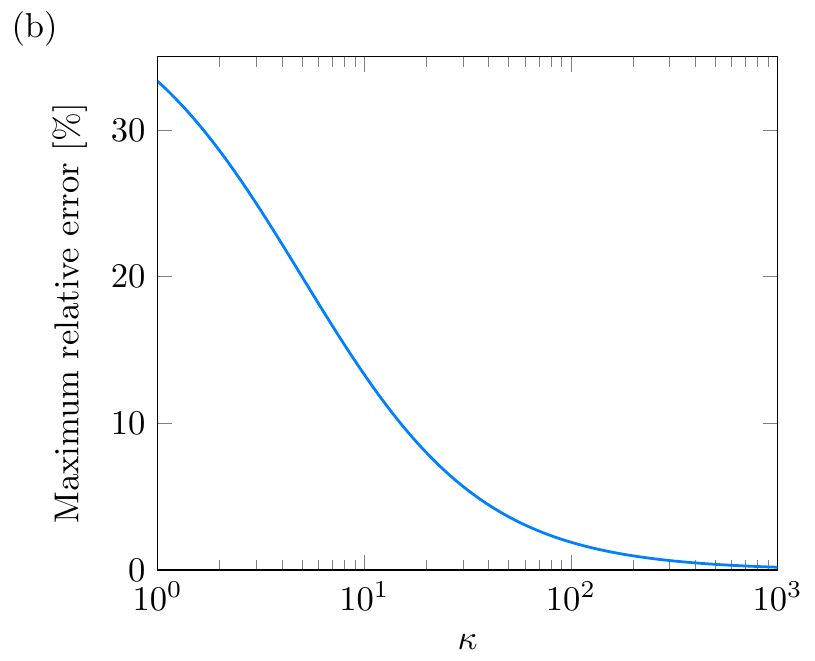}
\end{array}
$
\vspace{-5pt}
\end{center}
\caption{\label{appfigmech}
	(a) The nondimensional force $\eta=\beta f\ell_b$ as a function of the leading order (solid) and first-order corrected (dashed) asymptotic approximations of the corresponding nondimensional end-to-end length $\gamma=\xi/N_b\ell_b$ using the Morse-FJC model for $\beta u_b = 25$ and varying $\kappa$.
	(b) The maximum relative error $\mathrm{max}[e(\eta)]$ for $\beta u_b = 25$ and varying $\kappa$.
}
\end{figure*}

When applying the Morse potential to the $u$FJC model mechanical response in Eq.~\eqref{appgammasimplest}, with link potential energy $u(\ell)$ given by Eq.~\eqref{umorse}, we first compute the link force

\begin{equation}
	f = 
	\frac{\partial u(\ell)}{\partial\ell} = 
	\sqrt{2k_bu_b} \,e^{-\sqrt{k_b/2u_b}(\ell-\ell_b)} \left[1 - e^{-\sqrt{k_b/2u_b}(\ell-\ell_b)}\right]
	,
\end{equation}
and afterward nondimensionalize ($\eta\equiv\beta f\ell_b$, $\kappa\equiv \beta k_b\ell_b^2$, $\lambda\equiv \ell/\ell_b$) to obtain the nondimensional force

\begin{equation}\label{fewa3w3rwefd}
	\eta = 
	\sqrt{2\kappa\beta u_b} \,e^{-\sqrt{\kappa/2\beta u_b}[\lambda(\eta)-1]} \left\{1 - e^{-\sqrt{\kappa/2\beta u_b}[\lambda(\eta)-1]}\right\}
	.
\end{equation}
We then choose the transition state stretch $\lambda_\ddagger \equiv 1 + \ln(2)\sqrt{2\beta u_b/\kappa}$ corresponding to the maximum force $\eta_\mathrm{max} = \sqrt{\kappa\beta u_b/8}$.
Solving Eq.~\eqref{fewa3w3rwefd} for $\lambda(\eta)$, we then obtain

\begin{equation}\label{appMorselam}
	\lambda(\eta) = 
	1 + \sqrt{\frac{2\beta u_b}{\kappa}}\,\ln\left[\frac{2}{1+\displaystyle\sqrt{1 - \eta/\eta_\mathrm{max}}}\right]
	\quad\text{for } \eta \leq \eta_\mathrm{max} = \sqrt{\frac{\kappa\beta u_b}{8}}
	,
\end{equation}
which is Eq.~\eqref{Morselam} from the manuscript.
We plot the leading order $O(\kappa^0)$ and first-order corrected $O(\kappa^{-1})$ asymptotic approximations of the mechanical response from Eqs.~\eqref{appgammasim} and \eqref{appgammasimplest} for the Morse-FJC for varying $\kappa$ in Fig.~\ref{appfigmech}(a).
We found that varying $\beta u_b$ has little effect on the accuracy of the approximation of the mechanical response.
Fig.~\ref{appfigmech}(a) shows that the $O(\kappa^{-1})$ correction provides a small contribution for $\kappa=20$, a nearly negligible contribution for $\kappa=50$, and essentially no contribution for $\kappa=200$ and above.
Higher order corrections may be obtained, but if the $O(\kappa^{-1})$ correction is negligible, these will surely be negligible.
Going forward (and in the manuscript), we choose Eq.~\eqref{appgammasimplest} as the asymptotic approximation of the mechanical response $\gamma(\eta)$ and treat the $O(\kappa^{-1})$ correction from Eq.~\eqref{appgammasim} as an estimate of the error.
The relative error $e\equiv|\gamma_1-\gamma_0|/\gamma_1$ would then be

\begin{equation}
	e(\eta) \sim 
	\frac{(\eta/\kappa)\left[1 - \mathcal{L}(\eta)\coth(\eta)\right]}{\gamma_1(\eta)}
	\quad\text{for }\kappa\gg 1
	.
\end{equation}
We plot the maximum (over $\eta$) of this relative error in Fig.~\ref{appfigmech}(b) as a function of $\kappa$.
For smaller $\kappa$ the maximum relative error can be quite large, indicating that the $O(\kappa^{-1})$ correction is (and perhaps higher order corrections are) necessary.
As $\kappa$ increases the maximum relative error rapidly shrinks, showing that it indeed becomes accurate to ignore the $O(\kappa^{-1})$ correction (and all higher order corrections) and utilize the Eq.~\eqref{appgammasimplest} as the asymptotic approximation of the mechanical response $\gamma(\eta)$.

\subsection{Obtaining various single-chain quantities for the $u$FJC model\label{appuFJCsimp}}

Here we provide a full derivation of the single-chain quantities for the $u$FJC model leading up to the reaction rate coefficient function $k'(\gamma)$ in Eq.~\eqref{kBAfinal}.
We begin with the nondimensional configurational Helmholtz free energy per link $\vartheta_\mathrm{A,con}^*({\gamma})$, which under the Gibbs-Legendre approximation \citep{Buchestatistical2020} may be obtained from the mechanical response $\gamma(\eta)$ via

\begin{equation}
	\vartheta_\mathrm{A,con}^*({\gamma}) \sim
	\eta\gamma(\eta) - \int \gamma(\eta)\,d\eta
	\quad\text{for } N_b\gg 1
	.
\end{equation}
We then apply our asymptotic approximation for $\gamma(\eta)$ in Eq.~\eqref{gammasimplest} to obtain

\begin{equation}\label{apppsiconufjc}
	\vartheta_\mathrm{A,con}^*({\gamma}) = 
	\ln\left\{\frac{\eta\exp[\eta\mathcal{L}(\eta)]}{\sinh(\eta)}\right\} 
	+\beta u(\eta)
	-c_0
	,
\end{equation}
where $c_0=\ln(4\pi\ell_b^3\sqrt{2\pi/\kappa})$ is the constant of integration \citep{fiasconaro2019analytical}. 
This constant is part of the constant prefactor of the partition function, and does not appear to produce the correct units in Eq.~\eqref{psiconufjc} only because we have scaled away Planck's constant $h=1$. 
Starting from Eq.~\eqref{psiconufjc} and continuing in the following equations, note that functions of $\gamma$ are often written in terms of $\eta$, where $\eta=\eta(\gamma)$ then represents inverting the mechanical response $\gamma(\eta)$ in Eq.~\eqref{gammasimplest} for the $\eta$ corresponding to ${\gamma}$.
Using the principal thermodynamic connection formula, $\mathfrak{q}=e^{-N_b\vartheta}$, the configuration partition function is

\begin{equation}\label{appqAcon}
	\mathfrak{q}_\mathrm{A,con}^*({\gamma}) =
	\left[\ell_b\sqrt{\frac{2\pi}{\kappa}}\, e^{-\beta u(\eta)} \right]^{N_b} \mathfrak{q}_\mathrm{FJC,con}^*({\gamma}) 
	,
\end{equation}
where the configuration partition function of the FJC \citep{rubinstein2003polymer} is given by

\begin{equation}
	\mathfrak{q}_\mathrm{FJC,con}^*({\gamma}) = 
	\left\{\frac{4\pi\ell_b^2\sinh(\eta)}{\eta\exp[\eta\mathcal{L}(\eta)]}\right\}^{N_b}
	.
\end{equation}
After introducing the nondimensional equilibrium distribution $\mathscr{P}_\mathrm{A}^\mathrm{eq}({\gamma}) \equiv (N_b\ell_b)^3 P_\mathrm{A}^\mathrm{eq}({\xi})$, Eq.~\eqref{PAeq} gives Eq.~\eqref{PeqnicenoVb}.
The total probability that a chain is intact at equilibrium ${P}_\mathrm{A}^\mathrm{tot,eq}$ is the first factor in Eq.~\eqref{PeqnicenoVb}.
Note that the momentum contribution $\mathfrak{q}_\mathrm{A,mom}^*$ to the partition function $\mathfrak{q}_\mathrm{A}^*({\gamma})=\mathfrak{q}_\mathrm{A,mom}^*\mathfrak{q}_\mathrm{A,con}^*({\gamma})$ is independent of ${\gamma}$ and identical to $\mathfrak{q}_\mathrm{B,mom}^*$, causing it to vanish from Eq.~\eqref{PeqnicenoVb}.
Additionally, note that the constant $c_0$ from Eq.~\eqref{apppsiconufjc} cancels when substituting in $\vartheta_\mathrm{A,con}^*({\gamma})$, showing $c_0$ is irrelevant (only the relative free energy matters) when computing $\mathscr{P}_\mathrm{A}^\mathrm{eq}({\gamma})$, which is why $c_0$ does not appear in Eq.~\eqref{psiconufjc} from the manuscript.

We now turn to the partition function of the transition state $\mathfrak{q}_{\ddagger}^*({\gamma})$, which is a chain of $N_b-1$ intact links and a single link held rigidly in its transition state.
Since our Gibbs-Legendre method allows us to treat each link independently, we may compute these partition function separately and multiply them.
The transition state link partition function is that of the FJC with $\ell_b\mapsto\ell_\ddagger$, which also takes $\eta\mapsto\lambda_\ddagger\eta$ with $\lambda_\ddagger\equiv\ell_\ddagger/\ell_b$ due to the nondimensionalization.
We then have

\begin{equation}\label{appqTScon}
	\mathfrak{q}_{\ddagger,\mathrm{con}}^*({\gamma}) =
	e^{-\beta u(\ell_\ddagger)}
	\left\{\left.\mathfrak{q}_\mathrm{FJC,con}^*({\gamma})\right|_{\ell_b=\ell_\ddagger}
	\mathfrak{q}_\mathrm{A,con}^*({\gamma})^{(N_b-1)}\right\}^{1/N_b}
	.
\end{equation}
We then compute the ratio of $\mathfrak{q}_{\ddagger}^*({\gamma})$ to $\mathfrak{q}_\mathrm{A}^*({\gamma})$ in order to retrieve the reaction rate coefficient function $k'({\gamma})$ from Eq.~\eqref{kBjA}. 
The ratio of the configuration contributions is found using Eqs.~\eqref{appqAcon} and \eqref{appqTScon}, where the contributions from the $N_b-1$ non-transition state links cancel due to the link-independence that the Gibbs-Legendre method permits.
The ratio of the momentum contributions similarly cancels except for the momentum degree of freedom from stretching the link.
The contribution from this stretching degree of freedom is equivalent to that from the one-dimensional translation of the reduced mass $\nu=m/2$ of the link \citep{mcq}, so we then have

\begin{equation}\label{appqmomratio}
	\frac{\mathfrak{q}_{\ddagger,\mathrm{mom}}^*({\gamma})}{\mathfrak{q}_\mathrm{A,mom}^*({\gamma})} = 
	(2\pi \nu\mathfrak{b}T)^{-1/2}
	.
\end{equation}
To obtain $k'(\gamma)$ from Eq.~\eqref{kBjA}, we now multiply the ratio of $\mathfrak{q}_{\ddagger,\mathrm{con}}^*({\gamma})$ in Eq.~\eqref{appqTScon} to $\mathfrak{q}_\mathrm{A,con}^*({\gamma})$ in Eq.~\eqref{appqAcon} by $\mathfrak{q}_{\ddagger,\mathrm{mom}}^*({\gamma})/\mathfrak{q}_\mathrm{A,mom}^*({\gamma})$ in Eq.~\eqref{appqmomratio} and simplify.
When taking the ratio of $\mathfrak{q}_{\ddagger,\mathrm{con}}^*({\gamma})$ to $\mathfrak{q}_\mathrm{A,con}^*({\gamma})$, all contributions from all links cancel except that from the link that may be in the transition state (analogous to the momentum contribution).
After simplifying the ratio of $\mathfrak{q}_{\ddagger,\mathrm{con}}^*({\gamma})$ to $\mathfrak{q}_\mathrm{A,con}^*({\gamma})$ through powers of $N_b$, we have

\begin{equation}
	\frac{ \mathfrak{q}_{\ddagger,\mathrm{con}}^*({\gamma}) }{ \mathfrak{q}_\mathrm{A,con}^*({\gamma}) } = 
	\frac{ e^{-\beta u(\ell_\ddagger)} \left[\mathfrak{q}_\mathrm{FJC,con}^*({\gamma})|_{\ell_b=\ell_\ddagger}\right]^{1/N_b} }{ \mathfrak{q}_\mathrm{A,con}^*({\gamma})^{1/N_b}}
	.
\end{equation}
The denominator is the partition function of a single $u$FJC link, while the numerator is the partition function of a single $u$FJC link in its transition state, which would then be that of a single FJC link of length $\ell_\ddagger$ (and correspondingly higher potential energy $u(\ell_\ddagger)$ due to stretching beyond the rest length $\ell_b$).
Expanding this result using Eq.~\eqref{appqAcon}, we obtain

\begin{equation}\label{ljknwedf}
	\frac{ \mathfrak{q}_\mathrm{A,con}^*({\gamma}) }{ \mathfrak{q}_{\ddagger,\mathrm{con}}^*({\gamma}) } = 
	\ell_b \sqrt{\frac{2\pi}{\kappa}}\, e^{-\beta[u(\ell_\ddagger) - u(\eta)]} \left[\frac{\mathfrak{q}_\mathrm{FJC,con}^*({\gamma})|_{\ell_b=\ell_\ddagger}}{\mathfrak{q}_\mathrm{FJC,con}^*({\gamma})}\right]^{1/N_b}
	,
\end{equation}
which consists of three parts. 
The first part carries units of length with $\ell_b$ multiplying a unitless $\sqrt{2\pi/\kappa}$, where the units of length are a direct result of the single additional configurational degree of freedom (stretching) that a $u$FJC link has as opposed to the transition state link.
The second part is the exponential function of the potential energy differences.
The third part is the ratio of the FJC configurational partition functions of different lengths ($\ell_\ddagger$ and $\ell_b$), purely entropic in nature.
Multiplying Eq.~\eqref{ljknwedf} by Eq.~\eqref{appqmomratio}, scaling by $\beta=1/\mathfrak{b}T$ and simplifying, we receive the forward reaction rate coefficient function from Eq.~\eqref{kBjA},

\begin{equation}\label{wedsffvgt}
	k'({\gamma}) = 
	\frac{1}{2\pi}\sqrt{\frac{\kappa}{\nu\beta\ell_b^2}}\, e^{-\beta[u(\ell_\ddagger) - u(\eta)]} \left[\frac{\mathfrak{q}_\mathrm{FJC,con}^*({\gamma})|_{\ell_b=\ell_\ddagger}}{\mathfrak{q}_\mathrm{FJC,con}^*({\gamma})}\right]^{1/N_b}
	.
\end{equation}
Next, we reference classical transition state theory \citep{zwanzig2001nonequilibrium} in order to make sense of Eq.~\eqref{wedsffvgt}. 
The quantity resulting from the square root function has units of frequency, so we define the attempt frequency $\omega_\ddagger$ as

\begin{equation}\label{appomegab}
	\omega_\ddagger \equiv 
	\sqrt{\frac{1}{\nu}\left.\frac{\partial^2u}{\partial\ell^2}\right|_{\ell=\ell_b}} = 
	\sqrt{\frac{\kappa}{\nu\beta\ell_b^2}}
	.
\end{equation}
We simplify the last term in Eq.~\eqref{wedsffvgt}, which is proportional to the logarithm of the entropy barrier,

\begin{equation}\label{frew43er}
	\left[\frac{\mathfrak{q}_\mathrm{FJC,con}^*({\gamma})|_{\ell_b=\ell_\ddagger}}{\mathfrak{q}_\mathrm{FJC,con}^*({\gamma})}\right]^{1/N_b} = 
	\frac{\lambda_\ddagger\sinh(\lambda_\ddagger\eta)\exp[\eta\mathcal{L}(\eta)]}{\sinh(\eta)\exp[\lambda_\ddagger\eta\mathcal{L}(\lambda_\ddagger\eta)]}
	.
\end{equation}
Combining the entropic and potential energy contributions in Eqs.~\eqref{wedsffvgt} and \eqref{frew43er}, we obtain the Helmholtz free energy barrier $\Delta\Psi_\ddagger^*({\gamma})$ as
\begin{equation}\label{appDeltaPsiTS}
	\Delta\Psi_\ddagger^*({\gamma}) \equiv 
	u(\ell_\ddagger) - u(\eta)
	- \mathfrak{b}T\ln\left\{\frac{\lambda_\ddagger\sinh(\lambda_\ddagger\eta)\exp[\eta\mathcal{L}(\eta)]}{\sinh(\eta)\exp[\lambda_\ddagger\eta\mathcal{L}(\lambda_\ddagger\eta)]}\right\} 
	.
\end{equation}
Writing $k'(\gamma)$ in Eq.~\eqref{wedsffvgt} in terms of $\omega_\ddagger$ and $\Delta\Psi_\ddagger^*({\gamma})$, we obtain 

\begin{equation}\label{appkBAfinal}
	k'({\gamma}) = 
	\frac{\omega_\ddagger}{2\pi}\,e^{-\beta\Delta\Psi_\ddagger^*({\gamma})}
	.
\end{equation}
Eqs.~\eqref{appomegab}--\eqref{appkBAfinal} are equivalent to Eqs.~\eqref{kBAfinal}--\eqref{DeltaPsiTS} from the manuscript.

\section{Solving for the distribution of intact chains\label{appexact}}

\subsection{Obtaining and verifying the exact solution}

Here we exactly solve Eq.~\eqref{PAevokinVBfinalsinglek}, the evolution equation for the probability density distribution of intact chains $P_\mathrm{A}(\boldsymbol{\xi};t)$.
Our solution is valid for all chain models with a single reaction coordinate or any number of identical reaction coordinates.
Eq.~\eqref{PAevokinVBfinalsinglek} in terms of the chain end-to-end vector $\boldsymbol{\xi}$ is

\begin{equation}\label{PAevokinVBfinalsinglekintermsofxi}
	\dfrac{\partial P_\mathrm{A}(\boldsymbol{\xi};t)}{\partial t} = 
	- \left[\dfrac{\partial P_\mathrm{A}(\boldsymbol{\xi};t)}{\partial\boldsymbol{\xi}}\,\boldsymbol{\xi}\right]:\mathbf{L}(t) 
	-k(\boldsymbol{\xi})\left\{P_\mathrm{A}(\boldsymbol{\xi};t) - \dfrac{P^\mathrm{eq}_\mathrm{A}(\boldsymbol{\xi})}{P_{\mathrm{B}}^\mathrm{tot,eq}}\left[1 - \iiint P_\mathrm{A}(\tilde{\boldsymbol{\xi}};t)\,d^3\tilde{\boldsymbol{\xi}}\right]\right\}
	.
\end{equation}
We begin solving Eq.~\eqref{PAevokinVBfinalsinglekintermsofxi} by defining the new variable, 

\begin{equation}\label{Htrans}
	H(\boldsymbol{\xi};t) \equiv 
	P_\mathrm{A}\left[\mathbf{F}(t)\cdot\boldsymbol{\xi};t\right] e^{\int_0^t k\left[\mathbf{F}(s)\cdot\boldsymbol{\xi}\right]\,ds}
	,
\end{equation}
where $H(\boldsymbol{\xi};0) = P_\mathrm{A}(\boldsymbol{\xi};0)$ if $\mathbf{F}(0)=\mathbf{1}$.
Substitution of Eq.~\eqref{Htrans} into Eq.~\eqref{PAevokinVBfinalsinglekintermsofxi} produces

\begin{equation}\label{Hevo}
	\frac{\partial H(\boldsymbol{\xi};t)}{\partial t}\,e^{-\int_0^t k\left[\mathbf{F}(s)\cdot\boldsymbol{\xi}\right]\,ds}
	=
	\dfrac{k\left[\mathbf{F}(t)\cdot\boldsymbol{\xi}\right]P^\mathrm{eq}_\mathrm{A}\left[\mathbf{F}(t)\cdot\boldsymbol{\xi}\right]}{P_{\mathrm{B}}^\mathrm{tot,eq}}
	\left\{1 - \iiint H(\tilde{\boldsymbol{\xi}};t)\,e^{-\int_0^t k\left[\mathbf{F}(s)\cdot\tilde{\boldsymbol{\xi}}\right]\,ds}\,d^3\tilde{\boldsymbol{\xi}}\right\}
	.
\end{equation}
Eq.~\eqref{Hevo} may be rearranged to cause the right-hand side to become a function of time only,

\begin{align}
\label{dHdtrearr}
	 \frac{\partial H(\boldsymbol{\xi};t)}{\partial t}\,\dfrac{ e^{-\int_0^t k\left[\mathbf{F}(s)\cdot\boldsymbol{\xi}\right]\,ds}}{k\left[\mathbf{F}(t)\cdot\boldsymbol{\xi}\right]P^\mathrm{eq}_\mathrm{A}\left[\mathbf{F}(t)\cdot\boldsymbol{\xi}\right]}
	= &
	\frac{1}{P_{\mathrm{B}}^\mathrm{tot,eq}}\left\{1 - \iiint H(\tilde{\boldsymbol{\xi}};t)\,e^{-\int_0^t k\left[\mathbf{F}(s)\cdot\tilde{\boldsymbol{\xi}}\right]\,ds}\,d^3\tilde{\boldsymbol{\xi}}\right\}
	,\\ \equiv & 
	{\rho}(t)
	,
\end{align}
which we have now defined as ${\rho}(t)$. 
We see from the right-hand side of Eq.~\eqref{dHdtrearr} that this function happens to be ${\rho}(t)=P_\mathrm{B}^\mathrm{tot}(t)/P_\mathrm{B}^\mathrm{tot,eq}$.
Rearranging and integrating Eq.~\eqref{dHdtrearr}, we then retrieve the solution

\begin{equation}\label{Hsol}
	H(\boldsymbol{\xi};t) = 
	H(\boldsymbol{\xi};0)
	+ \int_0^t \dfrac{k\left[\mathbf{F}(\tau)\cdot\boldsymbol{\xi}\right]P^\mathrm{eq}_\mathrm{A}\left[\mathbf{F}(\tau)\cdot\boldsymbol{\xi}\right]}{ e^{-\int_0^\tau k\left[\mathbf{F}(s)\cdot\boldsymbol{\xi}\right]\,ds} }  \,{\rho}(\tau)\,d\tau
	,
\end{equation}
where we now must determine a solution for ${\rho}(t)$ that is consistent with the solution in Eq.~\eqref{Hsol} by substituting it back into Eq.~\eqref{dHdtrearr}. 
Our results are simplified by introducing the reaction propagator $\Xi(\boldsymbol{\xi};t,\tau)$, which we define as

\begin{equation}\label{XiApp}
	\Xi(\boldsymbol{\xi};t,\tau) \equiv
	\exp\left\{-\int_\tau^t k\left[{}_{(t)}\mathbf{F}(s)\cdot\boldsymbol{\xi}\right]\,ds\right\}
	,
\end{equation}
where the relative deformation \citep{paolucci2016continuum} is defined as

\begin{equation}\label{reldef}
	{}_{(t)}\mathbf{F}(\tau) \equiv
	\mathbf{F}(\tau)\cdot\mathbf{F}^{-1}(t)
	.
\end{equation}
Next we define the kernel function $K(t,\tau)$, which we may rewrite in terms of the reaction propagator $\Xi(\boldsymbol{\xi};t,\tau)$ using the invariance (due to incompressibility) of $d^3\boldsymbol{\xi}$ to the transformation $\boldsymbol{\xi}\mapsto\mathbf{F}^{-1}(t)\cdot\boldsymbol{\xi}$, as

\begin{align}
	K(t,\tau) \equiv &
	\iiint \dfrac{k\left[\mathbf{F}(\tau)\cdot\boldsymbol{\xi}\right]P^\mathrm{eq}_\mathrm{A}\left[\mathbf{F}(\tau)\cdot\boldsymbol{\xi}\right]}{P_\mathrm{B}^\mathrm{tot,eq} e^{\int_\tau^t k\left[\mathbf{F}(s)\cdot\boldsymbol{\xi}\right]\,ds}} \,d^3\boldsymbol{\xi}
	,\\ = &
	\frac{1}{P_\mathrm{B}^\mathrm{tot,eq}} \iiint P^\mathrm{eq}_\mathrm{A}\left[{}_{(t)}\mathbf{F}(\tau)\cdot\boldsymbol{\xi}\right] \,\frac{\partial\Xi(\boldsymbol{\xi};t,\tau)}{\partial\tau} \,d^3\boldsymbol{\xi}
\label{Kttaudef}
	,
\end{align}
where $K(t,\tau>t)\equiv 0$.
Then we similarly define the right-hand side function

\begin{align}
	b(t) \equiv &
	\frac{1}{P_\mathrm{B}^\mathrm{tot,eq}}\left\{1 - \iiint H(\boldsymbol{\xi};0) e^{-\int_0^t k\left[\mathbf{F}(s)\cdot\tilde{\boldsymbol{\xi}}\right]\,ds} \,d^3\boldsymbol{\xi}\right\}
	,\\ = &
	\frac{1}{P_\mathrm{B}^\mathrm{tot,eq}}\left\{1 - \iiint P_\mathrm{A}\left[\mathbf{F}^{-1}(t)\cdot\boldsymbol{\xi};0\right] \, \Xi(\boldsymbol{\xi};t,0) \,d^3\boldsymbol{\xi}\right\}
\label{beqnappen}
	.
\end{align}
Now, when we substitute Eq.~\eqref{Hsol} into Eq.~\eqref{dHdtrearr} and simplify, we obtain the consistency condition

\begin{equation}\label{Volterraeqn}
	{\rho}(t) + \int_0^t K(t,\tau) {\rho}(\tau)\,d\tau = 
	b(t)
	.
\end{equation}
Eq.~\eqref{Volterraeqn} is a linear Volterra integral equation of the second kind with eigenvalue $-1$.

We first consider the special case of motions with constant stretch history \citep{truesdell2004non,paolucci2016continuum} that have $\mathbf{L}=$~constant, which with $\mathbf{F}(0)=\mathbf{1}$ allow $\mathbf{F}(t)=e^{t\mathbf{L}}$ and therefore ${}_{(t_2)}\mathbf{F}(t_1)=\mathbf{F}(t_1-t_2)=\mathbf{F}^{-1}(t_2-t_1)$.
This allows us to rewrite the kernel $K(t,\tau)$ in Eq.~\eqref{Kttaudef} to be of the convolution type \citep{kanwal2013linear} by depending only on the difference $t-\tau$,

\begin{equation}
	K(t-\tau) =
	\frac{1}{P_\mathrm{B}^\mathrm{tot,eq}} \iiint P^\mathrm{eq}_\mathrm{A}(\boldsymbol{\xi}) k'(\boldsymbol{\xi}) e^{-\int_0^{t-\tau} k\left[\mathbf{F}(s)\cdot\boldsymbol{\xi}\right]\,ds} \,d^3\boldsymbol{\xi}
	.
\end{equation}
Kernels of the convolution type allow Eq.~\eqref{Volterraeqn} to be solved using the Laplace transform $\mathfrak{L}$. 
After utilizing the convolution theorem twice \citep{rahman2007integral}, the solution may be written as

\begin{equation}
	{\rho}(t) = 
	\int_0^t W(t-\tau) b(\tau) \,d\tau
	,
\end{equation}
where the solution kernel (also of the convolution type) is given by

\begin{equation}
	W(t) = 
	\mathfrak{L}^{-1}\left\{\frac{1}{1+\mathfrak{L}[K(t)]}\right\}
	.
\end{equation}
The solution as $t\to\infty$ may be obtained without need for the inverse Laplace transform $\mathfrak{L}^{-1}$ and can be utilized to study the steady-state mechanical response under these special deformations \citep{tanaka1992viscoelastic}.

For arbitrary incompressible deformation histories, the solution to Eq.~\eqref{Volterraeqn}, obtained using Picard's method of successive approximations \citep{cochran1972analysis,kanwal2013linear}, is the Liouville-Neumann series

\begin{equation}\label{rhosol}
	{\rho}(t) = 
	b(t) + \sum_{m=1}^\infty (-1)^m \int_0^t K_m(t,\tau) b(\tau)\,d\tau
	.
\end{equation}
The functions $K_m(t,\tau)$, where we begin with $K_1(t,\tau) \equiv K(t,\tau)$, are defined as

\begin{equation}
	K_m(t,\tau) \equiv 
	\int_\tau^t K(t,s) K_{m-1}(s,\tau)\,ds
	.
\end{equation}
This series solution for ${\rho}(t)$ allows the solution for $P_\mathrm{A}(\boldsymbol{\xi};t)$ and subsequently $\boldsymbol{\sigma}(t)$ in Eq.~\eqref{sigmafinal} to be written as series.
This resulting series for $\boldsymbol{\sigma}(t)$ resembles the general viscoelastic constitutive equation for the stress of the integral type \citep{truesdell2004non,paolucci2016continuum}.
However, in our case it is much more practical to construct $P_\mathrm{A}(\boldsymbol{\xi};t)$ and integrate the single term for $\boldsymbol{\sigma}(t)$ afterward.

The Liouville-Neumann series in Eq.~\eqref{rhosol} converges for some total time interval $t\in[0,\mathcal{T}]$ if the kernel function $K(t,\tau)$ is square-integrable \citep{cochran1972analysis}, which requires

\begin{equation}\label{Ksqint}
	\|K\|^2 \equiv \int_0^{\mathcal{T}} \int_0^t \big|K(t,\tau)\big|^2 \,d\tau \,dt < \infty
	.
\end{equation}
Since $0\leq\Xi(\boldsymbol{\xi};t,\tau)\leq 1$ in Eq.~\eqref{Kttaudef}, we have $K\leq{\hat{K}}$ and thus $\|K\|^2\leq\|{\hat{K}}\|^2$, where 

\begin{equation}
	{\hat{K}}(t,\tau) \equiv 
	\frac{1}{P_\mathrm{B}^\mathrm{tot,eq}}\iiint k\left[{}_{(t)}\mathbf{F}(\tau)\cdot\boldsymbol{\xi}\right] P^\mathrm{eq}_\mathrm{A}\left[{}_{(t)}\mathbf{F}(\tau)\cdot\boldsymbol{\xi}\right] \,d^3\boldsymbol{\xi}
	,
\end{equation}
so we may prove Eq.~\eqref{Ksqint} by proving $\|{\hat{K}}\|^2 < \infty$. 
We transform $\boldsymbol{\xi}\mapsto {}_{(\tau)}\mathbf{F}(t)\cdot\boldsymbol{\xi}$ in order to see that ${\hat{K}}$ is not truly a function of $t$ and $\tau$ and is more simply

\begin{equation}\label{faew3rwefdr543w}
	{\hat{K}} =
	\frac{1}{P_\mathrm{B}^\mathrm{tot,eq}}\iiint k(\boldsymbol{\xi}) P_\mathrm{A}^\mathrm{eq}(\boldsymbol{\xi})\,d^3\boldsymbol{\xi}
	,
\end{equation}
which means $\|{\hat{K}}\|^2 < \infty$ is proven for a finite time interval, where $0<P_\mathrm{B}^\mathrm{tot,eq}<1$ and $k(\boldsymbol{\xi})\propto k'(\boldsymbol{\xi})$, if

\begin{equation}\label{kPeqAPP}
	\iiint k'(\boldsymbol{\xi}) P_\mathrm{A}^\mathrm{eq}(\boldsymbol{\xi})\,d^3\boldsymbol{\xi} < \infty
	.
\end{equation}
If the positive semidefinite functions $k'(\boldsymbol{\xi})$ and $P^\mathrm{eq}(\boldsymbol{\xi})$ are prescribed without considering a chain model, this relation provides a constraint.
On the other hand, if we utilize our statistical mechanical framework -- namely Eqs.~\eqref{kBjA}, \eqref{PAeqgen}, and \eqref{qABjTS} -- we see Eq.~\eqref{faew3rwefdr543w} is a requirement that $\mathfrak{b}T\mathfrak{q}_{\ddagger}/\mathfrak{q}_\mathrm{B}<\infty$.
This requirement is generally true for finite, nonzero temperatures because of the following.
First, the momentum portions of the partition functions are known exactly and are finite.
Second, the configuration portions in general have an upper bound that is powers of the volume (zero potential case), which is then finite if the volume is finite.
Since the partition functions are positive-definite for nonzero temperatures and an appropriate Hamiltonian, the proof is complete.
For the specific case of the Morse-FJC we have approximately considered in Sec.~\ref{subsecuFJC}, we can immediately see that it is explicitly true: the integral of $\mathfrak{q}_{\ddagger}^*(\boldsymbol{\xi})$ over all end-to-end lengths exists and is finite and $\mathfrak{q}_\mathrm{B}=\mathfrak{q} - \mathfrak{q}_\mathrm{A}$ is a nonzero finite positive number.

While we have obtained the solution for ${\rho}(t)$ in Eq.~\eqref{rhosol} and have proven its convergence, we must now consider error estimates since we cannot sum to infinity in practice, and since it is unlikely that we will recognize what the series converges to for a given single-chain model.  
Including $M$ terms in Eq.~\eqref{rhosol},

\begin{equation}
	{\rho}(t) = 
	b(t) + \sum_{m=1}^M (-1)^m \int_0^t K_m(t,\tau) b(\tau)\,d\tau + R_{M+1}(t)
	,
\end{equation}
where $R_{M+1}(t)$ is then the residual function.
It can be shown \citep{lovitt1924linear} that the absolute value of the residual function in our case is bound by the inequality

\begin{equation}
	\left|R_{M+1}(t)\right| \leq 
	\frac{\big[\mathrm{max}|K(t,\tau)|\mathcal{T}\big]^{M+1}\mathrm{max}|{\rho}(t)|}{(M+1)!}
	.
\end{equation}
The limit of $\left|R_{M+1}(t)\right|$ as $M\to\infty$ is then zero, further verifying the convergence of our series solution in Eq.~\eqref{rhosol}.
The maximum of $K(t,\tau)>0$ over all $t$ and $\tau\leq t$ is $\hat{K}$ in Eq.~\eqref{faew3rwefdr543w}, and the maximum of ${\rho}(t)>0$ is $1/P_\mathrm{B}^\mathrm{tot,eq}$ since the maximum of $P_\mathrm{B}^\mathrm{tot}(t)$ is unity.
Therefore, we may bound the residual as

\begin{equation}\label{fewa234r4er}
	\left|R_{M+1}\right| \leq 
	\frac{({\hat{K}}\mathcal{T})^{M+1}}{P_\mathrm{B}^\mathrm{tot,eq}(M+1)!}
	.
\end{equation}
The right-hand side of Eq.~\eqref{fewa234r4er} is then our $t$-independent estimate for the residual which is computed once after the full time interval of interest $\mathcal{T}$ is specified.
This estimate can be scaled by ${\rho}(t)$ for an estimate of the relative error at time $t$. 
We see two effective timescales -- the total time interval $\mathcal{T}$ and the timescale $1/{\hat{K}}$, where ${\hat{K}}$ is actually the total reverse reaction rate coefficient.
For $\mathcal{T}<1/{\hat{K}}$ (short total time), the residual estimate in Eq.~\eqref{fewa234r4er} rapidly becomes small as $M$ increases, while for $\mathcal{T}>1/{\hat{K}}$ (long total time) the residual estimate may require considerably large $M$ to become small.

Now that our solution for ${\rho}(t)$ has been proven and error estimates in summing for it have been considered, we can go back to Eq.~\eqref{Hsol} and transform back using Eq.~\eqref{Htrans} to finally write the solution

\begin{equation}\label{PAsolP0}
	P_\mathrm{A}(\boldsymbol{\xi};t) = 
	P_\mathrm{A}\left[\mathbf{F}^{-1}(t)\cdot\boldsymbol{\xi};0\right]\Xi(\boldsymbol{\xi};t,0)
	+ \int_0^t P_\mathrm{A}^\mathrm{eq}\left[{}_{(t)}\mathbf{F}(\tau)\cdot\boldsymbol{\xi}\right] \,\frac{\partial\Xi(\boldsymbol{\xi};t,\tau)}{\partial\tau}\,{\rho}(\tau)\,d\tau
	.
\end{equation}
Substitution of this solution back into Eq.~\eqref{PAevokinVBfinalsinglekintermsofxi} for verification shows that the first term in Eq.~\eqref{PAsolP0} is the homogeneous solution and the second term is the particular solution.
For further verification, substitution of the solution Eq.~\eqref{PAsolP0} into the conservation requirement Eq.~\eqref{q3rewrgt3wedf} retrieves again the integral equation for ${\rho}(t)=P_\mathrm{B}^\mathrm{tot}(t)/P_\mathrm{B}^\mathrm{tot,eq}$ from Eq.~\eqref{Volterraeqn}.
As a final check, substitution of the solution Eq.~\eqref{PAsolP0} into the evolution equation for broken chains Eq.~\eqref{PBjtotevokin} retrieves (after simplifying and integrating the resulting ordinary integro-differential equation) the integral equation from Eq.~\eqref{Volterraeqn} yet again.
The homogeneous solution accounts for the decay of the initial distribution of chains as it is deformed and chains break.
The particular solution accounts for broken chains reforming in time throughout the deformation history, which is why its integrand (a rate of reforming) is proportional to the total probability of broken chains through ${\rho}(t)=P_\mathrm{B}^\mathrm{tot}(t)/P_\mathrm{B}^\mathrm{tot,eq}$.
If the reaction propagator $\Xi(\boldsymbol{\xi};t,\tau)$ is treated as independent of end-to-end vector $\Xi(t,\tau)$ and instead constitutively prescribed, the integral of Eq.~\eqref{PAsolP0} over all end-to-end vectors for $P_\mathrm{A}^\mathrm{tot}(t)$ resembles many recent models for the total number of intact bonds in a dynamic network \citep{long2013modeling,sun2016thermomechanics,meng2016stress,hui2012constitutive,long2014time,guo2016mechanics}.

If the system is initially equilibrated before $t=0$, which is $P_\mathrm{A}(\boldsymbol{\xi};t\leq 0)=P_\mathrm{A}^\mathrm{eq}(\boldsymbol{\xi})$ and $\mathbf{F}(t\leq 0)=\mathbf{1}$, we may use integration by parts and ${\rho}(t\leq 0)=1$ to rewrite Eq.~\eqref{PAsolP0} as a single term,

\begin{equation}\label{PAsolApp}
	P_\mathrm{A}(\boldsymbol{\xi};t) = 
	\int_{-\infty}^t P_\mathrm{A}^\mathrm{eq}\left[{}_{(t)}\mathbf{F}(\tau)\cdot\boldsymbol{\xi}\right] \,\frac{\partial\Xi(\boldsymbol{\xi};t,\tau)}{\partial\tau}\,{\rho}(\tau)\,d\tau
	,
\end{equation}
which is equivalent to the solution from Eq.~\eqref{PAsol} of the manuscript.

\subsection{Computational considerations\label{appcomut}}

We recognize that as the current time $t$ grows, it becomes computationally prohibitive to store certain quantities (such as the reaction propagator) over the entire history.
Fortunately, we are able to rewrite our solution at $t$ in terms of the solution at any previous time, which allows us to periodically reset and resolve in order to satisfy memory requirements.
For an intermediate time $t_i$ obeying $0\leq \tau\leq t_i\leq t$, the reaction propagator has the property

\begin{equation}\label{propertysavestheday}
	\Xi(\boldsymbol{\xi};t,\tau) = 
	\Xi(\boldsymbol{\xi};t,t_i) \, \Xi\left[{}_{(t)}\mathbf{F}(t_i)\cdot\boldsymbol{\xi};t_i,\tau\right]
	.
\end{equation}
This property allows the solution in Eq.~\eqref{PAsolP0} at time $t$ to be written in terms of the solution at an intermediate time $0\leq t_i\leq t$ and the history from $t_i$ to $t$,

\begin{equation}
	P_\mathrm{A}(\boldsymbol{\xi};t) = 
	P_\mathrm{A}\left[{}_{(t)}\mathbf{F}(t_i)\cdot\boldsymbol{\xi};t_i\right]\Xi(\boldsymbol{\xi};t,t_i)
	+ \int_{t_i}^t P_\mathrm{A}^\mathrm{eq}\left[{}_{(t)}\mathbf{F}(\tau)\cdot\boldsymbol{\xi}\right] \,\frac{\partial\Xi(\boldsymbol{\xi};t,\tau)}{\partial\tau}\,{\rho}(\tau)\,d\tau
	.
\end{equation}
When applying Eq.~\eqref{propertysavestheday} to the solution for $\rho(t)$ in Eq.~\eqref{rhosol}, two adjustments must be made.
First, the bounds of integration in Eq.~\eqref{rhosol} must be $\tau\in[t_i,t]$.
Second, $b(t)$ in Eq.~\eqref{beqnappen} must be rewritten as

\begin{equation}
	b(t) = 
	\frac{1}{P_\mathrm{B}^\mathrm{tot,eq}}\left\{1 - \iiint P_\mathrm{A}\left[{}_{(t)}\mathbf{F}(t_i)\cdot\boldsymbol{\xi};t_i\right] \, \Xi(\boldsymbol{\xi};t,t_i) \,d^3\boldsymbol{\xi}\right\}
	.
\end{equation}
Next, we note that it is most computationally expedient to compute $\rho(t)$ in Eq.~\eqref{rhosol} by performing successive approximations \citep{lovitt1924linear}.
This is as opposed to computing, storing, integrating, and summing each of the $M$ functions $K_m(t,\tau)b(\tau)$, which is more computationally expensive.
After starting with $\rho_0(t)\equiv b(t)$ and computing $K(t,\tau)$, storing both $b(t)$ and $K(t,\tau)$, we successively approximate the solution $\rho(t)$ by iterating

\begin{equation}
	\rho_M(t) = b(t) - \int_0^t K(t,\tau) \rho_{M - 1}(\tau)\,d\tau
	.
\end{equation}
To computationally obtain the solution as $\rho(t)=\lim_{M\to\infty} \rho_M(t)$, we take finite $M$ such that the residual in Eq.~\eqref{fewa234r4er} falls below some specified tolerance.
Finally, we note that exploiting the symmetries of $P_\mathrm{A}(\boldsymbol{\xi};t)$ that are preserved over the deformation history additionally serves to alleviate computation expense when computing the integrals over $\boldsymbol{\xi}$.
The $u$FJC model we utilize here is spherically-symmetric (only depend on $\|\boldsymbol{\xi}\|_2 = \xi$), allowing many integrals to be reduced to one-dimensional integrals over the scalar end-to-end length $\xi$. 
For example, Eq.~\eqref{kPeqAPP} for spherically-symmetric single-chain models is reducible to 

\begin{equation}
	\iiint k'(\boldsymbol{\xi}) P_\mathrm{A}^\mathrm{eq}(\boldsymbol{\xi}) \,d^3\boldsymbol{\xi} =
	4\pi \int_0^\infty k'({\xi}) P_\mathrm{A}^\mathrm{eq}({\xi})\,\xi^2\,d\xi
	.
\end{equation}
While the time-independent single-chain functions will always retain their symmetry, the distribution $P_\mathrm{A}(\boldsymbol{\xi};t)$ will in general not.
Fortunately, many deformation histories of interest will preserve a portion of the original spherical symmetry: when applying uniaxial tension (not necessarily monotonic), where the relative deformation is given by

\begin{equation}
	{}_{(t)}\mathbf{F}(\tau) = 
	\left(\begin{array}{ccc}
	\tfrac{F_{11}(\tau)}{F_{11}(t)} & 0 & 0 \\
	0 & \left[\tfrac{F_{11}(t)}{F_{11}(\tau)}\right]^{1/2} & 0 \\
	0 & 0 & \left[\tfrac{F_{11}(t)}{F_{11}(\tau)}\right]^{1/2}
	\end{array}\right)
	,
\end{equation}
the angular symmetry about the $\xi_1$-axis is preserved.
This is made clear after writing

\begin{align}
	\left\| {}_{(t)}\mathbf{F}(\tau)\cdot\boldsymbol{\xi} \right\|_2 = &
	\left\{\left[\frac{F_{11}(\tau)}{F_{11}(t)}\right]^2\xi_1^2 + \frac{F_{11}(t)}{F_{11}(\tau)}\,\left(\xi_2^2 + \xi_3^2\right)\right\}^{1/2}
	,\\  = &
	\left\{\left[\frac{F_{11}(\tau)}{F_{11}(t)}\right]^2 z^2 + \frac{F_{11}(t)}{F_{11}(\tau)}\, r^2\right\}^{1/2}
	,
\end{align}
where $z\equiv\xi_1$ and $r\equiv (\xi_2^2 + \xi_3^2)^{1/2}$ are the height and radius in cylindrical coordinates. 
Physically, the deformation stretches spheres into spheroids, which is symmetric about the $z$-axis along which the deformation is directed.
We then write each single chain function in terms of $z$ and $r$ rather than simply $\xi$, such as $P_\mathrm{A}^\mathrm{eq}(z,r)\equiv P_\mathrm{A}^\mathrm{eq}[(z^2 + r^2)^{1/2}]$.
After doing so we simplify each of the three-dimensional integrals over $\boldsymbol{\xi}$ into two-dimensional integrals over $z$ and $r$; note that we also exploit the symmetry about the $z$-plane.
For example: for an initially-equilibrated system, the right-hand side function $b(t)$ from Eq.~\eqref{beqnappen} is

\begin{align}
	b(t) \equiv &
	\frac{1}{P_\mathrm{B}^\mathrm{tot,eq}}\left\{1 - \iiint P_\mathrm{A}^\mathrm{eq}\left[\mathbf{F}^{-1}(t)\cdot\boldsymbol{\xi}\right] \, \Xi(\boldsymbol{\xi};t,0) \,d^3\boldsymbol{\xi}\right\}
	, \\ = &
	\frac{1}{P_\mathrm{B}^\mathrm{tot,eq}}\left\{1 - 4\pi\int_0^\infty r\,dr \int_0^\infty\,dz\, P_\mathrm{A}^\mathrm{eq}\left[F_{11}^{-1}(t)z,\, F_{11}^{1/2}(t) r\right] \, e^{-\int_0^t\,ds\, k\left[\tfrac{F_{11}(s)}{F_{11}(t)}\,z,\,\left[\tfrac{F_{11}(t)}{F_{11}(s)}\right]^{1/2} r\right]} \right\}
	.
\end{align}
The kernel function $K(t, \tau)$ simplifies similarly, allowing $\rho(t)$ and afterward $P_\mathrm{A}(\boldsymbol{\xi};t)$ to be evaluated with less computational expense.
Further, $P_\mathrm{A}(\boldsymbol{\xi};t)$ too retains a symmetry about $\xi_1$, which allows the stress in Eq.~\eqref{sigmafinal} to also be written in terms of a two-dimensional integral over $z$ and $r$:

\begin{equation}\label{fe23rf23213}
	\sigma_{11}(t) = 
	2\pi n\int_0^\infty r\,dr \int_0^\infty\,dz\, P_\mathrm{A}(z, r; t) \,f(z, r)\, \frac{2z^2 - r^2}{\left(z^2 + r^2\right)^{1/2}}
	.
\end{equation}
Note that we have also applied the traction-free boundary conditions and correspondingly solved for the pressure $p(t)$ in order to obtain Eq.~\eqref{fe23rf23213}.

\subsection{Special cases\label{appspeccases}}

As we discussed in Sec.~\ref{speccaseirrev} and observed in Figs.~\ref{figk}--\ref{figkappaH}, the reaction rate coefficient function $k'(\xi)$ often behaves as being constant at its initial value $k'(0)$ before suddenly becoming infinite beyond some critical extension $\xi_c$.
For the net reaction rate coefficient function $k(\xi)=N_bk'(\xi)$, this is more specifically

\begin{equation}
	k(\xi) \sim 
	\begin{cases}
	k_0, & \xi \leq \xi_c, \\
	\infty, & \xi > \xi_c.
	\end{cases}
\end{equation}
Applying this approximation to the reaction propagator $\Xi(\boldsymbol{\xi};t,\tau)$ in Eq.~\eqref{XiApp}, we obtain

\begin{equation}\label{appxiapproxgen}
	\Xi(\boldsymbol{\xi};t,\tau) \sim {\Theta(\boldsymbol{\xi};t,\tau)} e^{-k_0(t-\tau)}
	\quad\text{and}\quad
	\frac{\partial\Xi(\boldsymbol{\xi};t,\tau)}{\partial\tau} \sim {\Theta(\boldsymbol{\xi};t,\tau)} k_0 e^{-k_0(t-\tau)}
	,
\end{equation}
where the yield function $\Theta(\boldsymbol{\xi};t,\tau)$ is defined as

\begin{equation}
	{\Theta(\boldsymbol{\xi};t,\tau)} \equiv 
	\begin{cases}
	1, & \left\|{}_{(t)}\mathbf{F}(s)\cdot\boldsymbol{\xi}\right\|_2 \leq \xi_c ~~\forall s\in[\tau,t]
	,\\
	0, & \text{otherwise}
	.
	\end{cases}
\end{equation}
The yield function accounts for chains that have been broken via extension past $\xi_c$ by assigning zero to any vector $\mathbf{F}^{-1}(t)\cdot\boldsymbol{\xi}$ being sampled in ${P}_\mathrm{A}^\mathrm{eq}\left[\mathbf{F}^{-1}(t)\cdot\boldsymbol{\xi}\right]$ that was outside or deformed outside the yield surface at $\|\boldsymbol{\xi}\|_2=\xi_c$ during the deformation history.
With Eq.~\eqref{appxiapproxgen}, our solution for the distribution of intact chains ${P}_\mathrm{A}(\boldsymbol{\xi};t)$ from Eq.~\eqref{PAsolApp} becomes

\begin{equation}\label{PAsolspecorig}
	{P}_\mathrm{A}(\boldsymbol{\xi};t) = 
	\int_{-\infty}^t {P}_\mathrm{A}^\mathrm{eq}\left[{}_{(t)}\mathbf{F}(\tau)\cdot\boldsymbol{\xi}\right] \Theta(\boldsymbol{\xi};t,\tau) k_0 e^{-k_0(t-\tau)} \,{\rho}(\tau)\,d\tau
	.
\end{equation}
This special case is transient chain breaking combined with a finite critical extension.

\subsubsection*{Rate-independent irreversible breaking}

Applying Eq.~\eqref{appxiapproxgen} to our solution for the distribution of intact chains ${P}_\mathrm{A}(\boldsymbol{\xi};t)$ from Eq.~\eqref{PAsolP0},

\begin{equation}\label{PAsol0specorig}
	P_\mathrm{A}(\boldsymbol{\xi};t) = 
	P_\mathrm{A}\left[\mathbf{F}^{-1}(t)\cdot\boldsymbol{\xi};0\right]\Theta(\boldsymbol{\xi};t,0) e^{-k_0(t-\tau)}
	+ \int_0^t {P}_\mathrm{A}^\mathrm{eq}\left[{}_{(t)}\mathbf{F}(\tau)\cdot\boldsymbol{\xi}\right] \Theta(\boldsymbol{\xi};t,\tau) k_0 e^{-k_0(t-\tau)} \,{\rho}(\tau)\,d\tau
	.
\end{equation}
We assume that the distribution is initially equilibrated, and then Eq.~\eqref{PAsol0specorig} is equivalent to Eq.~\eqref{PAsolspecorig}.
Now, we consider the special case where $k_0\approx 0$, which greatly simplifies Eq.~\eqref{PAsol0specorig} to

\begin{equation}\label{appPAsolspecirrev}
	{P}_\mathrm{A}(\boldsymbol{\xi};t) = 
	{P}_\mathrm{A}^\mathrm{eq}\left[\mathbf{F}^{-1}(t)\cdot\boldsymbol{\xi}\right] \Theta(\boldsymbol{\xi};t,0)
	.
\end{equation}
Eq.~\eqref{appPAsolspecirrev} is equivalent to Eq.~\eqref{PAsolspecirrev} from the manuscript and is the special case of the rate-independent irreversible breaking of chains.

\subsubsection*{Transient breaking}

Next, we consider the special case where $\xi_c\to\infty$ but $k_0$ remains appreciable, where Eq.~\eqref{PAsolspecorig} becomes

\begin{equation}\label{transPapp}
	{P}_\mathrm{A}(\boldsymbol{\xi};t) = 
	\int_{-\infty}^t {P}_\mathrm{A}^\mathrm{eq}\left[{}_{(t)}\mathbf{F}(\tau)\cdot\boldsymbol{\xi}\right] k_0 e^{-k_0(t-\tau)} \,d\tau
	.
\end{equation}
Eq.~\eqref{transPapp} is equivalent to Eq.~\eqref{PAsoltransient} from the manuscript and is the special case where chains constantly break and reform, i.e. the transient network model.
Note that an infinitely-extensible single-chain model (such as the ideal or EFJC models) must be utilized since $\xi_c\to\infty$.
Also note that we have taken $\rho(t)=1$ within the solution in Eq.~\eqref{transPapp}, which means that the total number of intact chains $P_\mathrm{A}^\mathrm{tot}(t)$ remains constant at $P_\mathrm{A}^\mathrm{tot,eq}$ for all time.
This can be verified by integrating Eq.~\eqref{PAevokinVBfinalsinglek} over $\boldsymbol{\xi}$ with $k(\boldsymbol{\xi}) = k_0$ for an initially-equilibrated system.
Integrating Eq.~\eqref{transPapp} over $\boldsymbol{\xi}$ additionally will show that $P_\mathrm{A}^\mathrm{tot}(t) = P_\mathrm{A}^\mathrm{tot,eq}$ for all time as long as $P_\mathrm{A}^\mathrm{tot}(0) = P_\mathrm{A}^\mathrm{tot,eq}$.
Further, we point out that $k(\boldsymbol{\xi}) = k_0$ is likely the only way to guarantee that the fraction of intact chains remains constant at its equilibrium value. 
For this to be true, we must have ${\rho}(t) = 1$ as well as $P_\mathrm{A}^\mathrm{tot}(t) = P_\mathrm{A}^\mathrm{tot,eq}$.
Considering Eq.~\eqref{Volterraeqn} with ${\rho}(t) = 1$, we reference Eq.~\eqref{Kttaudef} for $K(t,\tau)$ and Eq.~\eqref{beqnappen} for $b(t)$ in simplifying the following:

\begin{align}
	\int_0^t K(t,\tau)\,d\tau = 
	\frac{1}{P_\mathrm{B}^\mathrm{tot,eq}} \int_0^t \iiint P^\mathrm{eq}_\mathrm{A}\left[{}_{(t)}\mathbf{F}(\tau)\cdot\boldsymbol{\xi}\right] \,\frac{\partial\Xi(\boldsymbol{\xi};t,\tau)}{\partial\tau} \,d^3\boldsymbol{\xi} \,d\tau
	,\\ = 
	\frac{1}{P_\mathrm{B}^\mathrm{tot,eq}} \int_0^t \iiint \left( \frac{\partial}{\partial\tau} \Big\{ P^\mathrm{eq}_\mathrm{A}\left[{}_{(t)}\mathbf{F}(\tau)\cdot\boldsymbol{\xi}\right] \Xi(\boldsymbol{\xi};t,\tau) \Big\} - \Xi(\boldsymbol{\xi};t,\tau) \,\frac{\partial}{\partial\tau}\,P^\mathrm{eq}_\mathrm{A}\left[{}_{(t)}\mathbf{F}(\tau)\cdot\boldsymbol{\xi}\right] \right) \,d^3\boldsymbol{\xi} \,d\tau
	,\\ = 
	\frac{1}{P_\mathrm{B}^\mathrm{tot,eq}} \iiint \left( P^\mathrm{eq}_\mathrm{A}(\boldsymbol{\xi}) - P^\mathrm{eq}_\mathrm{A}\left[\mathbf{F}^{-1}(t)\cdot\boldsymbol{\xi}\right] \Xi(\boldsymbol{\xi};t,0) - \int_0^t \Xi(\boldsymbol{\xi};t,\tau) \,\frac{\partial}{\partial\tau}\,P^\mathrm{eq}_\mathrm{A}\left[{}_{(t)}\mathbf{F}(\tau)\cdot\boldsymbol{\xi}\right] \,d\tau \right) \,d^3\boldsymbol{\xi}
	,\\ = 
	b(t) - 1  - \frac{1}{P_\mathrm{B}^\mathrm{tot,eq}} \int_0^t \iiint \Xi(\boldsymbol{\xi};t,\tau) \left[\left.\frac{\partial P_\mathrm{A}^\mathrm{eq}(\boldsymbol{\xi})}{\partial\boldsymbol{\xi}}\right|_{\boldsymbol{\xi}={}_{(t)}\mathbf{F}(\tau)\cdot\boldsymbol{\xi}}\cdot\mathbf{L}(\tau)\cdot {}_{(t)}\mathbf{F}(\tau)\cdot\boldsymbol{\xi}\right] \,d^3\boldsymbol{\xi}\,d\tau
\label{few3r4wefd34werd}
	.
\end{align}
We can alternatively arrive at this result by integrating Eq.~\eqref{PAsolP0} over all $\boldsymbol{\xi}$, setting both ${\rho}(t) = 1$ and $P_\mathrm{A}^\mathrm{tot}(t) = P_\mathrm{A}^\mathrm{tot,eq}$, and similarly simplifying.
Substituting Eq.~\eqref{few3r4wefd34werd} into Eq.~\eqref{Volterraeqn}, we find in general that $P_\mathrm{A}^\mathrm{tot}(t) = P_\mathrm{A}^\mathrm{tot,eq}$ is only satisfied when

\begin{equation}\label{whenamitrue}
	\int_0^t \iiint \exp\left\{-\int_\tau^t k\left[{}_{(\tau)}\mathbf{F}(s)\cdot\boldsymbol{\xi}\right]\,ds\right\} \left[\frac{\partial P_\mathrm{A}^\mathrm{eq}(\boldsymbol{\xi})}{\partial\boldsymbol{\xi}}\cdot\mathbf{L}(\tau)\cdot\boldsymbol{\xi}\right]\,d^3\boldsymbol{\xi}\,d\tau
	= 0
	.
\end{equation}
This condition may also be retrieved through simplifying the spatial integral of Eq.~\eqref{PAsolApp} with ${\rho}(t)=1$ and setting the result equal to $P_\mathrm{A}^\mathrm{tot,eq}$.
While Eq.~\eqref{whenamitrue} is indeed satisfied for the transient network model where $k(\boldsymbol{\xi}) = k_0$, it is unlikely to be satisfied in all the other cases where $k(\boldsymbol{\xi})$ is not constant.
We therefore find that the transient network model is the only case of our model where $P_\mathrm{A}^\mathrm{tot}(t)$ can be guaranteed to remain constant at its equilibrium value $P_\mathrm{A}^\mathrm{tot,eq}$.

\bibliography{main}

\end{document}